\documentclass{iopjournal}
\usepackage{multirow}
\usepackage{xspace}
\usepackage{comment}
\usepackage{setspace}
\usepackage{appendix}
\usepackage[T1]{fontenc}
\usepackage[scaled=0.9]{inconsolata} 
\usepackage{xcolor}
\usepackage{listings}
\usepackage{natbib}
\usepackage{amsmath}
\usepackage{amssymb}
\usepackage{graphicx}
\usepackage{subcaption}
\begin{document}

\title{Black Hole Binary Detection Landscape for the Laser Interferometer Lunar Antenna (LILA): Signal-to-Noise Calculations $\&$ Science Cases}

\author{Tintin Nguyen$^{1, 2, *}$\orcid{0000-0002-7524-5219}, Anjali Yelikar$^3$\orcid{0000-0002-8065-1174}, Ryan Nowicki$^{3}$\orcid{0009-0008-6626-0725}, Karan Jani$^{3}$\orcid{0000-0003-1007-8912}, Angelo Ricarte$^{1, 2}$\orcid{0000-0001-5287-0452}}

\affil{$^1$Center for Astrophysics $\vert$ Harvard \& Smithsonian, Cambridge, MA 02138, USA}

\affil{$^2$Black Hole Initiative, Harvard University, Cambridge, MA 02138, USA}

\affil{$^3$Department of Physics and Astronomy, Vanderbilt University, Nashville, TN 37235, USA}

\affil{$^*$Author to whom any correspondence should be addressed.}

\email{tintin.nguyen@cfa.harvard.edu}


\begin{abstract}
The Laser Interferometer Lunar Antenna (LILA) is a proposed gravitational-wave project aiming to take full advantage of the Moon's environment to access the deci-Hz band and detect intermediate-mass black hole (IMBH) binaries of mass $\sim 10^2-10^6 \, M_{\odot}$. With an observational period of 4 years, LILA can extend its IMBH detection horizon to the very early Universe, directly probing the first population of massive black holes ($z \sim 20-30$). LILA could also detect intermediate-mass-ratio inspiral systems with a total mass of $\sim 10^4 - 10^6 \, M_{\odot}$ and a mass ratio of $\sim 10^{-4} - 10^{-2}$. LILA can discover IMBH binaries months to years before merger with measurable eccentricity residuals retained from their formation, providing crucial early warning for multi-messenger and multi-band follow-up. The high SNR ($\gtrsim 100$) events detectable with LILA would enable strong-field tests of gravity. With these capabilities, LILA will provide important insights into the formation and evolution of massive black holes, as well as the astrophysical environments and evolutionary pathways of black hole binaries. LILA will also complement current LIGO/Virgo/KAGRA detections of pair-instability mass gap events, hierarchical merger candidates, and light IMBH mergers, while expanding the upper envelope of discovered black holes with stellar origin to masses of $\gtrsim 250 \, M_{\odot}$.

\end{abstract}

\section{Introduction}
Gravitational-wave (GW) astronomy just celebrated its ten-year anniversary of the first stellar-mass black hole merger event GW150914 \citep{GW150914}, discovered by the Laser Interferometer GW Observatory (LIGO). Since then, ground GW detectors (LIGO, Virgo and KAGRA, hereafter LVK) have detected hundreds of black hole merger events \citep{LIGO_O3_population} with final black hole masses ranging from as small as $\sim 5 M_{\odot}$ (GW230529\_181500; 
\citealt{GW230529_mass_gap_and_NS}) to as massive as $\sim 225 \, M_{\odot}$ (GW231123; 
\citealt{GW231123_total_225_Msol, Chatterjee_2025_ML_GW231123_lite_IMBH}), complementing previous and ongoing X-ray binary observations \citep{McClintock_Remillard_2006_X-ray_compact_binary}. Since GWs preserve spacetime information from the sources, they provide stringent tests of general relativity (GR) \citep{GW150914_GR, LIGO_GWTC3_GR} and precise measurements of black hole mass and spin \citep{LIGO_O3_population}. The spin distributions of discovered binary stellar-mass black holes in gravitational waves (predominantly slowly-spinning) and X-ray (predominantly rapidly-spinning) detections are significantly different \citep{Reynolds_2021_BH_spin_review, Zdziarski_2025_BH_spin_Xray_GW}. Though there are major uncertainties in X-ray reflection spectroscopy spin measurements subject to astrophysical modeling assumptions \citep{Dauser_2013_irradiation_disk_jet, Mall_2024_BH_spin_Xray_precession, Shashank_2025_BH_spin_reflection_spectroscopy_GRMHD}, this still demonstrates the potential of multi-messenger probes in unveiling different stellar-mass black hole populations. Additionally, LVK compact binary merger events provide rich insights into the formation mechanisms and evolution of their progenitor stellar binary systems \citep{Kruckow_2018_GW_merger_progenitor, Giacobbo_2018_compact_binary_progenitors, Belczynski_2020_binary_black_hole_evolution, Fishbach_2020_GW190521_straddling_binary, van_Son_2022_mass_transfer_NSBH_gap, Broekgaarden_2022}, which could leave an imprint on the redshift evolution of the black hole merger rate \citep{Fishbach_2018_BH_merger_rate_redshift, van_Son_2022_BH_merger_rate_evolution}. As LVK detectors are approaching their design sensitivity limits, third-generation ground detectors, such as Einstein Telescope (ET; \citealt{Abac_2025_Einstein_Telescope}) and Cosmic Explorer (CE; \citealt{Abbott_2017_Cosmic_Explorer}), are expected to be launched in the next decade, improving the sensitivity in the $\sim 10-10^4$ Hz frequency band by at least an order of magnitude.

Another exciting development is the strong evidence for the stochastic GW background detection made by the North American Nanohertz Observatory for GW (NANOGrav; \citealt{NANOGrav_15_year}), which can be attributed to the population of supermassive black hole (SMBH) binaries and mergers over cosmic time \citep{NANOGrav_15_year_BH_binary, Ellis_2024a}. Unlike ground GW detectors in the $\sim 10-10^4$ Hz band, pulsar timing arrays like NANOGrav work in a much lower frequency regime of $\sim 10^{-9}-10^{-7}$ Hz. Interestingly, models found that NANOGrav results are consistent with the black hole to host galaxy stellar mass ratio suggested by high-redshift overmassive SMBHs \citep{Pacucci_2023, Pacucci_2024, Ellis_2024b} in active galactic nuclei (AGN) discovered by the James Webb Space Telescope (JWST) and AGN X-ray luminosity function \citep{Kusakabe_2026_SMBH_binary_demographics_Xray_GW_background}, highlighting the importance of multi-messenger astronomy in uncovering the universal presence of SMBHs and their mergers. Future next-generation radio telescopes like the Square Kilometer Array (SKA; \citealt{SKA}) will enable more sensitive pulsar timing arrays that could potentially detect individual black hole binaries, in addition to an improved measurement of the stochastic GW background \citep{Lazio_2013_SKA_PTA, Shannon_2025_SKA_PTA}.

In the next decade, the space-based Laser Interferometer Space Antenna (LISA) will probe an entire new frequency band of $\sim 10^{-4}-10^{-1}$ Hz, which will primarily target SMBH inspirals and mergers \citep{LISA_2017, Amaro-Seoane_2023_LISA_astrophysics}. Existing observational evidence for dual AGN and binary SMBH systems is circumstantial, many of which are unresolved or biased towards radio-bright compact jetted systems (see, e.g., review by \citealt{Krause_2025_SMBH_binary_review}). Many SMBH binary candidates have been identified by their periodic variability in the Palomar Transient Factory quasar sample \citep{Charisi_2016_SMBHB_candidates_PTF_quasars}. There is limited evidence for sub-parsec SMBH binaries, among the strongest being OJ 287 \citep{Valtonen_2009_OJ287_outbursts, Britzen_2018_OJ287, Spitzer_2020_OJ287_Spitzer, Gomez_2026_OJ287_polarization_EHT} and PKS 1302-102 \citep{Graham_2015_SMBH_binary_PKS1302_102} with quasi-periodic optical outbursts, hindering our understanding of how SMBHs merge within the age of the Universe when they are a few parsecs of each other (the ``final parsec problem"; \citealt{Milosavljevic_Merritt_2003_final_parsec_problem}). Hence, LISA and other gravitational-wave detectors covering a similar frequency band will significantly improve observational evidence for SMBH binary systems and our theoretical understanding of their dynamics. The advent of the Rubin Observatory and the Legacy Survey of Space and Time (LSST) will revolutionize time-domain astrophysics and provide complementary probes in identifying SMBH binary systems with light curve variability \citep{Witt_2022_SMBH_binary_time_domain, Xin_2026_SMBH_binary_LSST_Bayesian}. LSST, Roman, LISA, and pulsar timing arrays will take full advantage of multi-messenger astrophysics in unveiling massive black hole binaries in the next decade \citep{De_Rosa_2019_SMBHB_multi_messenger, Xin_Haiman_2021_short_period_SMBHB_LSST_LISA, Charisi_2022_multimessenger_time_domain_SMBHB, Haiman_2023_Roman_LISA_precursor_SMBHB, DOrazio_Charisi_2023_observational_signature_SMBHB, Xin_Haiman_2024_EM_counterparts_LISA_LSST}.

\begin{figure}
    \centering
    \includegraphics[width=0.9\linewidth]{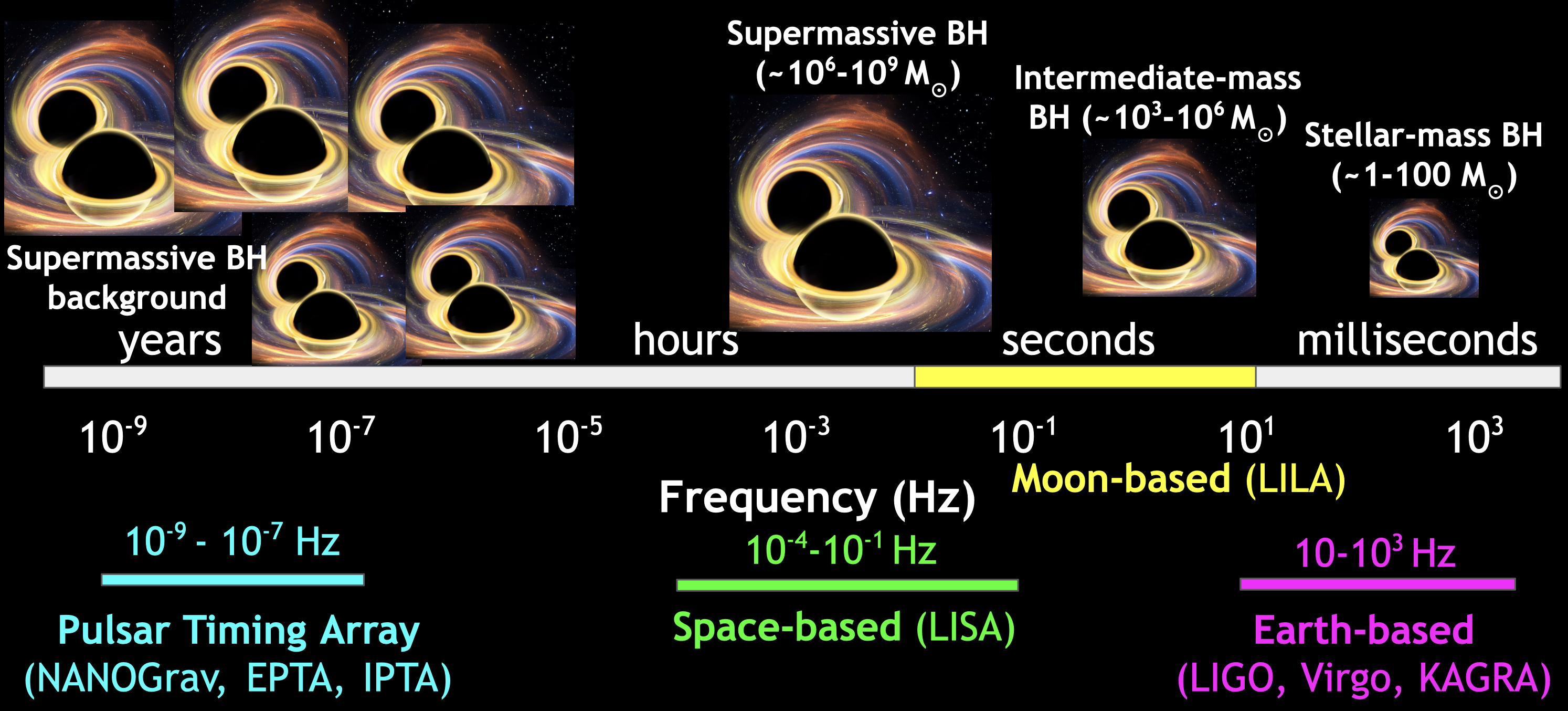}
    \caption{A schematic diagram describing the GW spectrum with the corresponding detector types and black hole targets for different frequency bands}
    \label{fig:gw_spectrum}
\end{figure}

Another proposed future GW detector mission is the Laser Interferometer Lunar Antenna (LILA; \citealt{Jani_2025_LILA}), which is targeting two phases called LILA-Pioneer and LILA-Horizon. LILA aims to fill in the deci-Hz frequency band between LISA and LVK of $\sim 10^{-1}-10$ Hz \citep{Creighton_2025_LILA_noise_sensitivity, Shapiro_2025_LILA_vibration_isolation}, extending sensitivity to intermediate-mass black holes (IMBHs) with masses of $\sim 10^3-10^6 \, M_{\odot}$ at earlier cosmic times \citep{Jani_Loeb_2020_GLOC}. LILA-Horizon could also reach LISA-like sensitivity in the $\sim 10^{-4}-10^{-1}$ Hz band with arcminute sky localization, compared with a few degrees for LIGO \citep{Jani_2025_LILA}. Figure \ref{fig:gw_spectrum} provides an overview of the GW spectrum and how LILA targets the band gap between ground detectors and LISA. The Moon provides two decisive advantages to LILA: (1) The lack of seismic and other dynamical noise allows it to access the deci-Hz band gap and (2) The lunar normal modes provide a natural amplifier to reach higher sensitivity at the desired frequency band \citep{Creighton_2025_LILA_noise_sensitivity, Trippe_2026_LILA_site_selection}. Other astrophysical sources not discovered thus far by LVK but could be detected by LILA in the milli-to-deci-Hz band include white dwarf binary mergers \citep{Pichardo_Marcano_2025_WD_merger_Moon, Das_2026_supersoft_Xray_deciHz_GW}, type Ia supernovae \citep{Gill_2024_deciHz_GW_CCSNe, Murphy_2025_GW_CCSNe_structure, Richardson_2025_GW_3D_CCSNe}, and tidal disruption events around IMBHs \citep{Mandel_2018_deciHz_GW_science_case}. As a lunar interferometer, LILA can also be sensitive to cosmological horizon-scale gravitational potentials, providing direct constraints on dark energy properties \citep{Gurrola_2026_dark_energy_sound_speed_LILA, Gurrola_2026_dark_energy_Moon}. 

Other proposed GW experiments in the deci-Hz band include the Deci-hertz Interferometer GW Observatory (DECIGO; \citealt{Kawamura_2021_DECIGO}), the Lunar GW Antenna (LGWA; \citealt{Harms_2021_LGWA}), TianQin \citep{Luo_2016_TianQin}, TianGO \citep{Kuns_2020_TianGO}, Indian Initiative in GW Observation detector (IndIGO-D; \citealt{Sharma_2026_IndIGO-D}), and the Artificial Precision Time Array (APTA; \citealt{Alves_2024_APTA}). Our work on the black hole binary detection landscape for LILA is relevant for the general deci-Hz GW forecast (see \citealt{Mandel_2018_deciHz_GW_science_case, Arca_Sedda_2020_deciHz_GW_discoveries, Izumi_Jani_2021_deciHz_GW_landscape, deciHz_JHU_workshop} for comprehensive reviews on deci-Hz GW astrophysics). The exciting opportunities for multi-band GW detections with ground and space detectors open novel prospects for tests of gravity \citep{Vitale_2016_multiband_GW, Carson_Yagi_2020_multiband_GW_IMR_gravity_test, Datta_multiband_GW_IMBH_GR_test}, parameter estimation improvement \citep{Curt_2019_multiband_GW, Grimm_2020_multiband_GW, Nakano_2021_multiband_GW_B-DECIGO, Multiband_Ranjan_2025}, searches for IMBHs \citep{Jani_2020_IMBH_multiband_GW}, early warning for multi-messenger follow-up \citep{Sesana_2016_multiband_GW, Multiband_Ruiz_Rocha_2025, Yelikar_Jani_2025_NS_binary_lunar_detector, Pichardo_Marcano_2025_WD_merger_Moon}, and early black hole growth \citep{Valiante_2021_multiband_GW_BH_growth}.

While the search for IMBHs has remained mostly elusive so far (see, e.g., \citealt{Greene_2020_review} for a comprehensive review), there is considerable tentative evidence for the existence of IMBHs with masses of $10^3-10^6 \, M_{\odot}$ through indirect means, such as reverberation mapping \citep{Peterson_2005_IMBH_reverberation_mapping}, X-ray variability modeling \citep{Maccarone_2007_IMBH_globular_cluster}, hyperluminous X-ray source spectral modeling \citep{Straub_2014_HLX-1_IMBH, Kaaret_2017_ultraluminous_Xray_review, Barrows_2019_hyperluminous_Xray_IMBH}, single-epoch virial estimators \citep{Chilingarian_2018, Fei_2025_JWST_GLIMPSE_IMBH_direct_collapse}, tidal disruption events \citep{Lin_2020_IMBH_candidate_3XMM, Angus_2022_IMBH_TDE}, black hole mass scaling relations \citep{Graham_2013_Mbh_Lspheroid_relation_IMBH_candidates, Baldassare_2015}, and stellar dynamical mass measurements \citep{Lanzoni_2013_IMBH_candidate_globular_cluster, Lutzgendorf_2015_IMBH_candidate_globular_cluster, Nguyen_2017_NGC404_dynamical_measurement, Nguyen_2018_dynamical_measurement_BH_NSC, Nguyen_2019, Pechetti_2022_M31_GC_IMBH, Vitral_2023_IMBH_globular_cluster_M4, Haberle_2024_IMBH_candidate_fast_stars}. All of these findings are subject to uncertain modeling assumptions and inconclusive interpretations, e.g., in some studies, alternative physical models, such as dense nuclear clusters, can fit equally well to observational data without a massive central object. In the near future, the advent of LISA and deci-Hz GW detectors like LILA will provide a robust pathway towards conclusively detecting IMBHs if they exist \citep{Jani_2020_IMBH_multiband_GW, Arca_Sedda_2020_deciHz_GW_discoveries, Izumi_Jani_2021_deciHz_GW_landscape, Amaro-Seoane_2023_LISA_astrophysics, Fragione_Loeb_2023_IMBH_merger_history, Song_2025_IMBH_binary_LGWA, Kusakabe_2026_SMBH_binary_forecast_GW}. These IMBH searches with GW events will be complemented by time-domain variability studies with the Rubin Observatory \citep{Burke_Natarajan_2026_IMBH_time_domain_variability} and future high-resolution integral field spectrographs, such as the HARMONI instrument on the Extremely Large Telescope \citep{Nguyen_2025_IMBH_simulation_mass_measurement_ELT/HARMONI, Ngo_2025_IMBH_detection_ELT/HARMONI, Ngo_2025_IMBH_simulation_ELT/HARMONI}. 

From a theoretical standpoint, IMBHs could form in dense stellar clusters \citep{Gurkan_2004_BH_star_clusters, Griersz_2015_IMBH_formation_globular_clusters, Arca_Sedda_2018_BH_globular_clusters, Gonzalez_2021_IMBH_star_clusters}, possibly through runaway stellar collisions \citep{Miller_Hamilton_2002_IMBH_formation_GC, Zwart_McMillan_2002_IMBH_runaway_collision, Gurkan_2006_BH_binary_collisional_runaway, Rodriguez_2019_repeated_BH_mergers, Di_Carlo_2021_IMBH_star_clusters, Purohit_2024_IMBH_collisional_runaway, Gonzalez_Prieto_2024_IMBH_stellar_collisions, Vergara_2025_IMBH_runaway_collisions, Rantala_2026_IMBH_formation_runaway_metallicity,Mestichelli_2026_IMBH_seeding_stellar_collisions}, and grow through tidal capture and tidal disruption events \citep{Sakurai_2018_IMBH_growth_TDE, Rizzuto_2023_IMBH_growth_tidal_capture_TDE, Chang_2025_stellar_TDE_IMBH_rate}, as well as hierarchical mergers \citep{Christian_2018_BHMF_hierarchical_star_clusters, Rodriguez_2019_repeated_BH_mergers, Antonini_2019_BH_merger_star_clusters, Fragione_Rasio_2023_hierarchical_merger_star_cluster, Kritos_2025_BH_growth_merging_star_clusters} and efficient accretion \citep{Natarajan_2021_IMBH_accretion_formation, Kritos_2024_SMBH_NSC} in stellar clusters. The predictions for IMBH populations in realistic astrophysical environments and cosmological contexts \citep{Boekholt_2018, Tagawa_2020_compact_binary_AGN_disks} are very uncertain because of the difficulties in resolving stellar clusters and their coevolution with IMBHs inside their systems. Despite the theoretical and observational challenges so far, IMBHs are vital for understanding the cosmic assembly of SMBHs (e.g., \citealt{Volonteri_2005, Shapiro_2005, King_2006, Ricarte_Natarajan_2018_SEROTINA_SMBH_assembly}) and the coevolution with their host galaxies \citep{Kormendy_Ho_2013, McConnell_Ma_2013}, as well as the dynamical evolution of dense stellar systems. The widespread presence of heavy ($\gtrsim 10^7 \, M_{\odot}$) black holes in the early Universe ($z \gtrsim 3$) suggested by observations \citep{Lyke_2020_SDSS_quasar_16th_DR, Fan_2023, Pacucci_2023, Taylor_2025_JWST_AGN_BHMF} further motivate the need to search for IMBHs at earlier epochs to bridge the gap between the first seeding populations and the observed demographics \citep{BL01, Madau_Rees_2001, Bromm_Loeb_2003, Volonteri_2010, Ferrara_2014, Pacucci_2022_search}. This highlights the importance of forecasting the IMBH detection landscape for deci-Hz GW detectors like LILA.

This paper details the signal-to-noise calculation framework used to characterize the detection landscape and develop science cases for LILA, with an emphasis on IMBH binaries. Section \ref{sec:strain_model} explains the characteristic strain evolution model for binary systems, accounting for inspiral-merger-ringdown phases and eccentricity. Section \ref{sec:SNR} outlines the signal-to-noise calculations for LILA given the characteristic strain of a binary system and accounting for the compact galactic binary foreground. Section \ref{sec:detection_landscape} demonstrates the LILA detection horizon and early-warning potential of black hole binaries across different initial eccentricities, mass ranges, mass ratios, and redshifts, with implications for massive black hole formation and growth modeling, multi-band GW astronomy, and GR tests. Section \ref{sec:conclusion} summarizes the main findings of the paper.

\section{Characteristic Strain Model for Binary Systems}
\label{sec:strain_model}

\subsection{Characteristic Strain Evolution of Non-Relativistic Circular Binary Systems}
\label{subsec:strain}
We consider a binary system consisting of two masses $M_1$ and $M_2$ in a circular orbit of radius $R$ and angular frequency $\Omega$, observed at an inclination angle $\iota$ with respect to the binary rotational axis by a detector located at a distance $D$ from the binary system. The distance is denoted as the comoving distance $D(z)$ as a function of cosmological redshift $z$, where the flat $\Lambda$CDM cosmological parameters constrained by the Planck measurements of the cosmic microwave background \citep{Planck_2018} are used throughout this study. In the weak-field (Newtonian) limit, the binary system can be treated as an effective one-body system with the reduced mass $\mu \equiv M_1 M_2/M$ orbiting the central mass $M \equiv M_1 + M_2$. We introduce the ``chirp mass" $M_c \equiv \mu^{3/5} M^{2/5}$. From the post-Newtonian multipole expansion, the GW polarizations to the lowest order are given by \citep{Sathyaprakash_Schutz_2009_GW_review}:
\begin{subequations}
\begin{align}
    h_{+}(t) &= h_0 \frac{1 + \cos^2 \iota}{2} \cos(2 \Omega t) \\
    h_{\times}(t) &= h_0 \cos \iota \sin(2\Omega t) \\
    h_0 &\equiv \frac{4G \mu R^2 \Omega^2}{D(z) \, c^4} = \frac{4 (GM_c)^{5/3} \Omega^{2/3}}{D(z) \, c^4}
\end{align}
\end{subequations}
Here, we set the angular frequency to be $\Omega = \pi f_r$, where $f_r$ is the rest-frame GW frequency, which is twice the orbital frequency. The root-mean-square of the GW strain, averaging over polarization modes, time (one wave period), and inclination (assuming isotropy) is:
\begin{equation}
    h(f) = \langle h_{+}^2 + h_{\times}^2 \rangle_{\iota, t}^{1/2} = \left( \frac{\Omega}{2 \pi} \int_{0}^{\pi/\Omega} dt \int_{-1}^{1} d(\cos \iota) \, [h_{+}^2 + h_{\times}^2]\right)^{1/2} =
    \sqrt{\frac{2}{5}} h_0 = \frac{8 \pi^{2/3}}{\sqrt{10}} \frac{(G M_c)^{5/3}}{D(z) \, c^4} [f(1+z)]^{2/3}
    \label{eq:strain}
\end{equation}
where the observed frequency $f$ is related to the rest-frame frequency $f_r$ by $f = f_r (1+z)^{-1}$. Based on the average strain above, the characteristic strain of an orbiting and inspiraling black hole binary over an observation period of $t_{\rm obs}$ can be modeled by the following piecewise function \citep{Sesana_2005}:
\begin{equation}
h_c(f) \approx \begin{cases}
h(f) \, \sqrt{f t_{\rm obs}} \propto f^{7/6} & \text{if } n > ft_{\rm obs} \\
h(f) \, \sqrt{n} \propto f^{-1/6} & \text{if } n < ft_{\rm obs},
\label{eq:characteristic_strain}
\end{cases}
\end{equation}
where $n$ is the number of cycles spent during the frequency interval $\Delta f \sim f$ around an observed frequency f:
\begin{equation}
    n(f) \approx \frac{f_r^2}{df_r/dt} = \frac{5}{96 \pi^{8/3}} \frac{c^5}{(GM_c)^{5/3}} [f(1+z)]^{-5/3},
    \label{eq:cycle}
\end{equation}
with the rest-frame GW frequency evolution as
\begin{equation}
    \frac{df_r}{dt} = \frac{96 \pi^{8/3} (G M_c)^{5/3}}{5c^5} f_r^{11/3}.
    \label{eq:rest_frame_dfr_dt}
\end{equation}
The conceptual idea behind the piecewise function is that during the early phase when the orbital separation is still large enough and stable, the source spends many cycles at a roughly constant frequency $f$, so $h_c (f) \approx h\sqrt{ft_{\rm obs}}$. At the later inspiral phase when the orbital separation rapidly shrinks, the source evolves quickly through frequency $f$ and only stays for a few cycles $n(f)$, hence $h_c(f) = h\sqrt{n}$. The characteristic strain physically characterizes how much power the strain contributes to different GW frequency bands. Hence, the power law is positive ($h_c \propto f^{7/6}$) during the early orbiting phase when the strain becomes stronger as the binary separation decreases while the binary system still spends significant time in the lower-frequency regime. On the contrary, the power law is negative ($h_c \propto f^{-1/6}$) during the late inspiral phase since the binary system rapidly evolves in frequency space so that it no longer contributes much power to the high frequency band in spite of the increase in GW strain. \\

\subsection{Inspiral-Merger-Ringdown Model}
\label{subsec:IMR}
Because the derivation in the previous subsection \ref{subsec:strain} relies on the weak-field, non-relativistic source regime, it only describes the system accurately up until the black holes accelerate to some speed comparable to the speed of light. Characterizing the strong-field regime in detail requires full numerical relativity calculations, but for the purposes of this paper, we approximate using an inspiral-merger-ringdown (IMR) phenomenological template derived from numerical relativity waveforms, written in the form of the waveform amplitude in the frequency domain (PhenomA model; \citealt{Ajith_2008_phenomelogical_GW_template_bank}):
\begin{equation}
A \propto \begin{cases}
f^{-7/6} & \text{if } f < f_{\rm merge} \\
f^{-2/3} & \text{if } f_{\rm merge} \leq f < f_{\rm ring} \\
\mathcal{L}(f; f_{\rm ring}, \sigma) & \text{if } f_{\rm ring} \leq f < f_{\rm cut},
\label{eq:phenom}
\end{cases}
\end{equation}
where
\begin{subequations}
\begin{align}
    f_{\rm merge} &= \mathcal{A}_0; \, \, f_{\rm ring} = \mathcal{A}_1; \, \, \sigma = \mathcal{A}_2; \, \, f_{\rm cut} = \mathcal{A}_3 \\
    \mathcal{A}_n &\equiv \frac{a_n \eta^2 + b_n \eta + c_n}{\pi(GM/c^3) (1+z)} \\
    \mathcal{L}(f; f_{\rm ring}, \sigma) &\equiv \frac{1}{2 \pi} \frac{\sigma}{(f-f_{\rm ring})^2 + \sigma^2/4}
\end{align}
\label{eq:frequencies}
\end{subequations} 
with $\eta \equiv \mu/M$ and the parameters $\mathcal{A}_n$ are derived from numerical relativity template fits reported in Table 1 of \citealt{Ajith_2008_phenomelogical_GW_template_bank}. The frequencies above are defined in the observed frame. Additionally, $\mathcal{L}(f; f_{\rm ring}, \sigma)$ denotes the Lorentzian function of width $\sigma$ centered around $f_{\rm ring}$.
According to the Parseval theorem, the characteristic strain $h_c$ is related to the Fourier transform of the strain (Equation \ref{eq:strain}) $\tilde{h}$ as $h_c(f) = \sqrt{2} \, f_r \, | \tilde{h}(f)|$ \citep{Sesana_2005}. Using the Parseval theorem relation, we can then append the IMR model in Equation \ref{eq:phenom} to the piecewise function in Equation \ref{eq:characteristic_strain} to fully model the characteristic strain as the following:
\begin{equation}
h_c(f) = \begin{cases}
\displaystyle h(f) \sqrt{f t_{\rm obs}} \propto f^{7/6} & \text{if } f < f_{\rm knee} \\
\displaystyle h(f) \sqrt{n(f)} \propto f^{-1/6} & \text{if } f_{\rm knee} \leq f < f_{\rm merge} \\
\displaystyle h(f_{\rm merge}) \sqrt{n(f_{\rm merge})} \left(\frac{f}{f_{\rm merge}}\right)^{1/3} \propto f^{1/3} & \text{if } f_{\rm merge} \leq f < f_{\rm ring} \\
\displaystyle h(f_{\rm merge}) \sqrt{n(f_{\rm merge})} \left( \frac{f_{\rm ring}}{f_{\rm merge}}\right)^{1/3} \frac{f}{f_{\rm ring}} \frac{\mathcal{L}(f; f_{\rm ring}, \sigma)}{\mathcal{L}(f_{\rm ring}; f_{\rm ring}, \sigma)} & \text{if } f_{\rm ring} \leq f < f_{\rm cut}
\end{cases}
\label{eq:characteristic_strain_imr}
\end{equation}
where we add an additional transitional frequency $f_{\rm knee}$ based on Equation \ref{eq:characteristic_strain} and \ref{eq:cycle}:
\begin{equation}
    f_{\rm knee} \equiv n(f) \, t_{\rm obs} = \frac{5}{96 \pi^{8/3}} \frac{c^5}{(GM_c)^{5/3}t_{\rm obs}} [f(1+z)]^{-5/3}
    \label{eq:f_knee}
\end{equation}
The coefficients in the third and fourth branches of the piecewise function in Equation \ref{eq:characteristic_strain_imr} are defined that way to ensure continuity while following the scaling relations from the IMR model in Equation \ref{eq:phenom} and the Parseval relation. Equations \ref{eq:strain}, \ref{eq:cycle}, and \ref{eq:frequencies}-\ref{eq:f_knee} fully model the characteristic strain evolution for circular binary systems, accounting for early orbital, inspiral, merger, and post-merger ringdown phases.


\subsection{Effects of Eccentricity on Binary Dynamics}
\label{subsec:eccentricity}
Many studies have shown that significant eccentricity can naturally arise in black hole binary systems due to effects like stellar scattering \citep{Sesana_2010_eccentric_SMBHB_stellar_envi}, galaxy mergers \citep{Iwasawa_2011_SMBHB_eccentric_evolution}, and interactions with the circumbinary gas disk \citep{Armitage_Natarajan_2005_SMBHB_eccentricity}. Dynamically active environments, such as dense stellar clusters \citep{Rodriguez_2018_eccentric_BH_mergers_dense_star_clusters, Samsing_2018_eccentric_BH_merger_globular_clusters, Zevin_2019_eccentric_BH_mergers_dense_star_clusters, Fragione_2019_eccentric_BH_merger_globular_clusters}, galactic nuclei \citep{Gondan_2018_eccentric_BH_merger_galactic_nuclei, Tagawa_2021_eccentric_BH_mergers_AGN, Samsing_2022_eccentric_BH_mergers_AGN}, and triples \citep{Arca_Sedda_2021_eccentric_BH_mergers_triples}, are expected to produce highly eccentric black hole binary systems. Therefore, it is important to account for eccentric binaries and their effects on the GW signals. When the binary system has a non-zero eccentricity $e \neq 0$, its dynamics contribute to multiple GW harmonic modes at the same time. Hence, the characteristic strain needs to be summed over all harmonic modes, leading to the following modified expression \citep{Huerta_2015_eccentric_SMBHB_SNR}:
\begin{equation}
    h(f_r, e) = \left(\frac{4 - \sqrt{1-e^2}}{\sqrt{1-e^2}}\right)^{1/2} \frac{8 \, \pi^{2/3}}{\sqrt{30}} \frac{(GM_c)^{5/3}}{D(z) \, c^4} f_r^{2/3}
    \label{eq:strain_eccentricity}
\end{equation}
Note that in the circular orbit limit ($e \rightarrow 0$), the equation above reproduces Equation \ref{eq:strain}. The rate of rest-frame gravitational wave frequency change in Equation \ref{eq:rest_frame_dfr_dt} is also modified into \citep{Peters_Mathews_1963_GW, Enoki_Nagashima_GW_eccentric_SMBHB}:
\begin{equation}
    \frac{df_r}{dt} = \frac{96 \pi^{8/3} (G M_c)^{5/3}}{5c^5} F(e) \, f_r^{11/3},
    \label{eq:dfr_dt_ecc}
\end{equation}
with the eccentricity factor
\begin{equation}
    F(e) = \frac{1 + 73/24 e^2 + 37/96 e^4}{(1-e^2)^{7/2}}.
    \label{eq:eccentricity_factor}
\end{equation}
The frequency rate becomes faster due to the additional energy loss during pericentric approach for elliptical orbits. Hence, the number of orbits spent during the frequency interval $\Delta f \sim f$ becomes:
\begin{equation}
    n \approx \frac{f_r^2}{df_r/dt} = \frac{5}{96 \pi^{8/3}} \frac{c^5}{(GM_c)^{5/3}} F(e)^{-1 \,} f_r^{-5/3},
    \label{eq:cycle_eccentricity}
\end{equation}
The only physical parameter left to model the time evolution is eccentricity
\citep{Peters_1964_GW_binary}:
\begin{equation}
    \frac{de}{dt} = \frac{-304}{15} \frac{G^3 M_1 M_2 M}{c^5 a^4} E(e),
\end{equation}
with the eccentricity factor
\begin{equation}
    E(e) = \frac{e(1 + 121/304 \, e^2)}{(1-e^2)^{5/2}}.
\end{equation}
The orbital semimajor axis $a$ can be replaced by Kepler's third law
\begin{equation}
    a = \left[ \frac{GM}{(\pi f_r)^{2}} \right]^{1/3}
\end{equation}
and using the definition of chirp mass $M_c = \mu^{3/5} M^{2/5}$, we obtain the orbital eccentricity rate of
\begin{equation}
    \frac{de}{dt} = \frac{-304 \pi^{8/3}}{15} \frac{(GM_c)^{5/3}}{c^5} E(e) f_r^{-8/3}.
    \label{eq:de_dt}
\end{equation}
Recall that we can convert between rest-frame and observed frequency as $f_r = f(1+z)$. Then, by combining Equation \ref{eq:dfr_dt_ecc} and \ref{eq:de_dt}, we can evolve the eccentricity in observed frequency space by the following differential equation:
\begin{equation}
    \frac{de}{df} = \frac{de}{dt} \left(\frac{df_r}{dt}\right)^{-1} \frac{df_r}{df} = \frac{-19}{18}\frac{E(e)}{F(e)} f^{-1} \Rightarrow \frac{de}{d(\ln f)} = \frac{-19}{18} \frac{E(e)}{F(e)} 
    \label{eq:de_df}
\end{equation}

\subsection{Full Characteristic Strain Evolution Model for Binary Systems}
\label{subsec:full_strain_model}
We now combine the elements from the previous Sections \ref{subsec:strain}-\ref{subsec:eccentricity} to fully model the characteristic strain evolution for a binary system with given masses $M_1$ and $M_2$, with the chirp mass defined as $M_c \equiv (M_1 M_2)^{3/5}/(M_1 + M_2)^{1/5}$, and an initial eccentricity $e_0$ at time $\tau$ before merger at redshift $z$. First, we find the observed frequency associated with time $\tau$ before merger at frequency $f_{\rm merge}$ (defined in Equation \ref{eq:frequencies}). From Equation \ref{eq:dfr_dt_ecc}, the time to reach the merger frequency $f_{\rm merge, \, r} = f_{\rm merge} (1+z)$ from the initial rest-frame frequency $f_{r}$ is:
\begin{subequations}
\begin{align}
    \tau &= \int_{f_r}^{f_{\rm merge}} \frac{df_r}{df_r/dt} = C(f_r^{{-8/3}} - f_{\rm merge, \, r}^{-8/3}) \\
    C &\equiv \frac{5}{256 \pi^{8/3}} \frac{c^5}{(GM_c)^{5/3}} F(e_0)^{-1}
\end{align}
\end{subequations}
with the eccentricity factor $F(e)$ defined in Equation \ref{eq:eccentricity_factor}. Inverting the above relation, we obtain:
\begin{align}
    f_r (\tau) = \left( \frac{\tau}{C} + f_{\rm merge, r}^{-8/3} \right)^{-3/8}
\end{align}
And hence, the observed frequency at time $\tau$ before merger is:
\begin{align}
    f (\tau) = \frac{1}{1+z} \left[ \frac{\tau}{C} + f_{\rm merge}^{-8/3} \, (1+z)^{-8/3} \right]^{-3/8}
    \label{eq:frequency_time_before_merger}
\end{align}
Then, given the initial observed frequency $f(\tau)$ above (Equation \ref{eq:frequency_time_before_merger}) and eccentricity $e_0$, we can evolve the characteristic strain (Equation \ref{eq:characteristic_strain_imr} using the eccentric strain factor in Equation \ref{eq:strain_eccentricity}) and the eccentricity (Equation \ref{eq:de_dt}) simultaneously in observed frequency space until the final frequency $f_{\rm cut}$ (as defined in Equation \ref{eq:frequencies}). We note that the characteristic strain evolution model in Section \ref{sec:strain_model} is only strictly correct for isolated binary systems with no other dynamical effects besides inspiral due to gravitational radiation. Dynamical interactions on galactic \citep{Izquierdo-Villalba_2020_SMBH_galactic_nuclei_halo_outskirt, Ricarte_2021_wandering_BH, Di_Matteo_2023_wandering_merging_IMBH_cosmic_noon} and sub-kiloparsec scales \citep{Milosavljevic_Merritt_2003_final_parsec_problem, Dosopoulou_Antonini_2017_final_hundred_parsec_problem} can substantially delay black hole mergers. Preferential accretion onto the secondary in moderately eccentric binaries in circumbinary disks can drive the mass ratio towards unity \citep{Duffell_2020_circumbinary_disks_accretion_torque, Munoz_2020_circumbinary_accretion, Siwek_2023_binary_evolution_circumbinary_disks}. Another factor we do not account for is black hole spin, which has minimal impact on the farthest detectable distance for masses of $\lesssim 10^3 \, M_{\odot}$ but could increase (or decrease) the detection distance by a factor of 2 if the individual black hole spins are aligned (or misaligned) with the orbital angular momentum for equal-mass binaries of $\gtrsim 10^4 \, M_{\odot}$ \citep{Jani_2020_IMBH_multiband_GW}.

\section{Signal-to-Noise Calculations}
\label{sec:SNR}
Given the characteristic noise $h_{\rm noise}(f)$ and the characteristic strain $h_c(f)$ for a binary system as evolved in Subsection \ref{subsec:full_strain_model}, the signal-to-noise ratio (SNR) is:
\begin{equation}
    \rm{S/N} = \left[ \int_{f_1}^{f_2} d(\ln f) \, \left( \frac{h_c(f)}{h_{\rm noise}(f)} \right)^2  \right]^{1/2},
    \label{eq:SNR}
\end{equation}
where the SNR is integrated over the frequency band [$f_1, f_2$], representing the full observation period, in quadrature. The black hole binary is considered detected if the integrated SNR is above a certain threshold. The characteristic noise in Equation \ref{eq:SNR} accounts for the LILA instrumental noise and the compact galactic binary (CGB) foreground. We neglect the stellar-mass black hole binary contribution to the background noise, as recently constrained by the LVK fourth observing run \citep{LVK_2025_O4a_persistent_GW}, because it is orders of magnitude below the LILA sensitivity curve at its peak band of $\sim 1-100$ Hz. We also neglect the SMBH binary contribution to the background noise, since it peaks at frequencies that are orders of magnitude lower than the LISA band \citep{NANOGrav_15_year_BH_binary}. Other cosmological sources of GW background such as inflation \citep{Maggiore_2000_GW_early_cosmology}, cosmic strings \citep{Auclair_2020_cosmic_string_GW_background}, and primordial black holes \citep{Bagui_2025_primordial_BH_GW} are not accounted for. For the instrumental noise $h_{\rm LILA} (f)$, we consider the sensitivity curves of two planned LILA detectors, LILA-Pioneer and LILA-Horizon \citep{Creighton_2025_LILA_noise_sensitivity}. LILA-Horizon is a dual-band detector, extending the LILA-Pioneer sensitivity down by about an order of magnitude in the deci-Hz band and to higher frequencies of $\sim 10-100$ Hz, i.e., that of the ground detectors \citep{Shapiro_2025_LILA_vibration_isolation}.

The CGB foreground is dominated by white dwarf binaries \citep{Ruiter_2010}, in which a small fraction has been identified electromagnetically \citep{Stroeer_Vecchio_2006_LISA_verification_binary, Marsh_2011_WDB_LISA, Kupfer_Korol_2024_LISA_galactic_binary_Gaia}. Population models have predicted up to $\sim 60$ million CGBs in the LISA band \citep{Nelemans_2004_AMCVn_LISA, Korol_2017_WDB_detectability_Gaia_LSST_LISA, Korol_2018_LISA_detectability_WD}, with only $\sim 0.01 \, \%$ of them being individually resolvable by LISA \citep{Amaro-Seoane_2023_LISA_astrophysics}. Thus, CGB foreground noise could play an important role for LILA, since LILA and LISA share strong overlap in their frequency bands. To model CGB noise, we employ the following empirical power spectral density (PSD) model:
\begin{align}
    S_{\rm CGB}(f) &= \frac{A}{2} f^{-7/3} e^{-(f/f_1)^{\alpha}} \left[ 1 + \tanh \left( \frac{f_k - f}{f_2} \right) \right] \\
    \log_{10} (f_1) &= a_1 \log_{10} (t_{\rm obs}) + b_1 \\
    \log_{10} (f_k) &= a_k \log_{10} (t_{\rm obs}) + b_k,
\end{align}
with chosen parameters $A = 1.14 \times 10^{-44}$, $f_2 = 3.1 \times 10^{-4}$, $\alpha = 1.8$, $a_1 = -0.25$, $b_1 = -2.7$, $a_k = -0.27$, and $b_k = -2.47$, which are calibrated to Galactic white dwarf binary population models \citep{Korol_2020_WDB_galactic_population} assuming a combined SNR threshold of 7 and the mean values for smoothing PSD estimation \citep{Karnesis_2021_compact_binary_background}. To convert the CGB foreground PSD into characteristic strain, we take $h_{\rm CGB} = \sqrt{f S_{\rm CGB}}$. Then, the characteristic noise in Equation \ref{eq:SNR} is taken to be the maximum of the LILA instrumental noise and the CGB foreground noise: $h_{\rm noise}(f) = \max[h_{\rm LILA}(f), h_{\rm CGB} (f)]$. In this paper, the SNR calculations carried out for other current and future detectors or pulsar timing arrays, including LIGO (assuming the A+ design in the fourth observing run; \citealt{Barsotti_2018_Advanced_LIGO_sensitivity}), ET \citep{Hild_2011_ET_sensitivity}, CE \citep{Abbott_2017_Cosmic_Explorer}, LISA \citep{Robson_2019_LISA_sensitivity}, and NANOGrav (15 years; \citealt{NANOGrav_15yr_detector_characterization_noise}), use the sensitivity curves in corresponding references therein. The 15-year SKA sensitivity curve is simulated using the software \texttt{gwent} \citep{Kaiser_McWilliams_2011_sensitivity_GW_spectrum} with the fiducial parameter estimates from \citealt{Sesana_2008_GW_background_BH_binary_PTA} and the methods from \citealt{Hazboun_2019_PTA_sensitivity_curves}, assuming the white noise and stochastic GW background model. \\

\begin{figure}[ht]
  \centering
  \begin{subfigure}{0.5\linewidth}
    \centering
    \includegraphics[width=\linewidth]{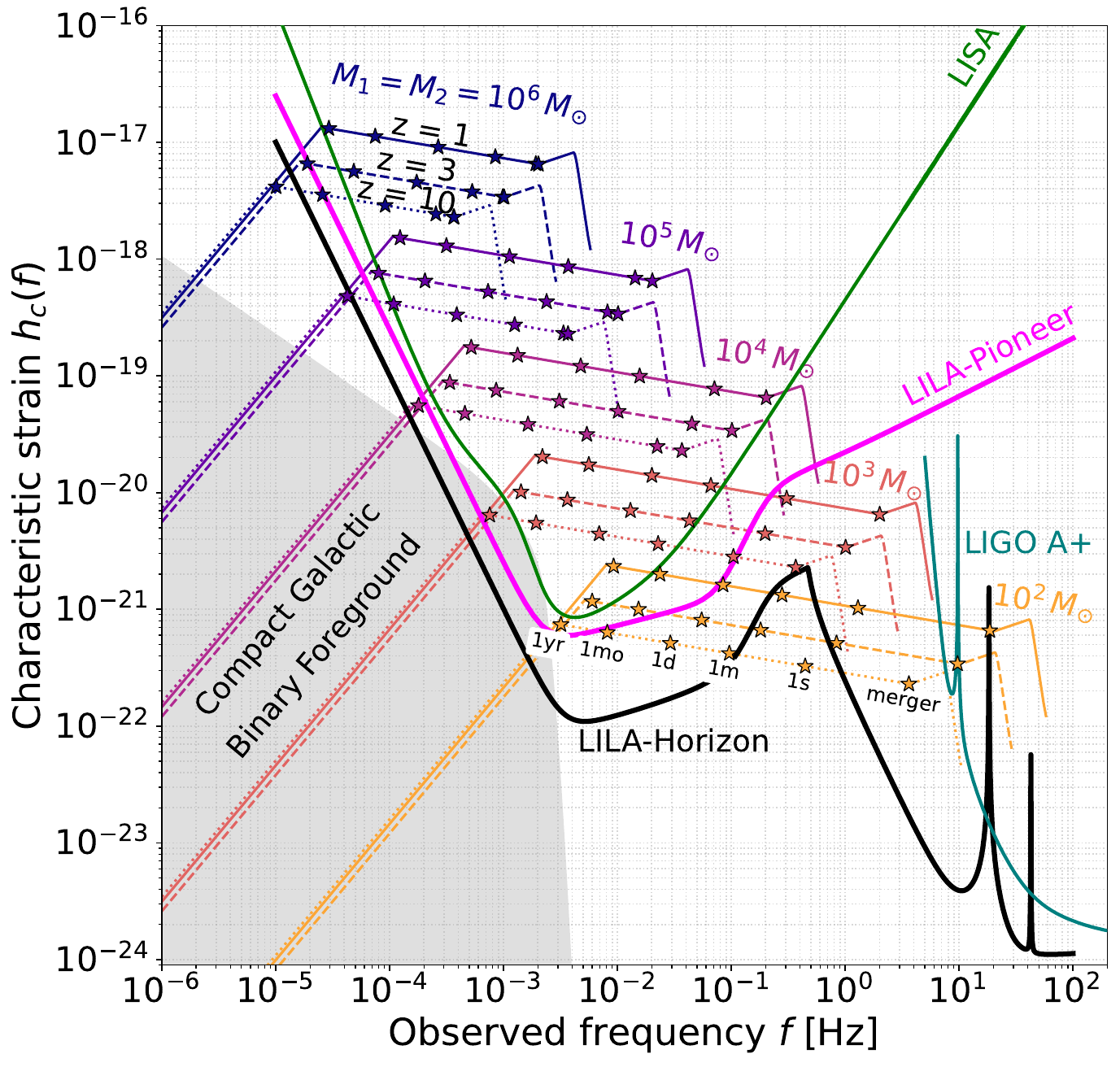}
  \end{subfigure}\hfill
  \begin{subfigure}{0.5\linewidth}
    \centering
    \includegraphics[width=\linewidth]{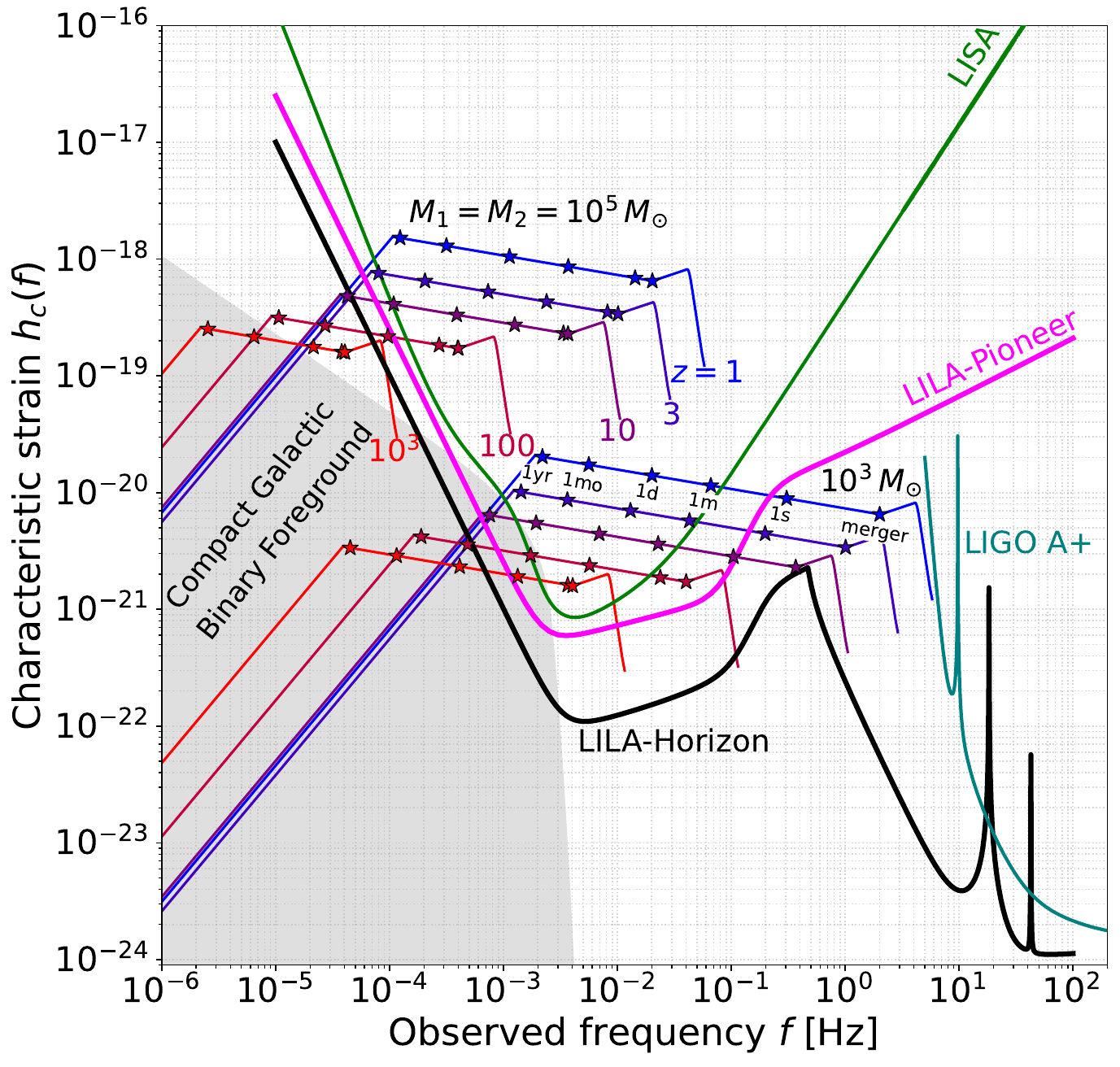}
  \end{subfigure}
  \caption{Characteristic strain evolution of equal-mass, non-spinning, circular black hole binary systems with different masses and at different redshifts, in comparison to the instrumental sensitivity of LILA \citep{Creighton_2025_LILA_noise_sensitivity}, LISA \citep{Robson_2019_LISA_sensitivity}, and LIGO A+ \citep{Barsotti_2018_Advanced_LIGO_sensitivity}, as well as the compact Galactic binary foreground \citep{Karnesis_2021_compact_binary_background} expected to be quantified by LISA for a 4-year observational period with a combined SNR threshold of 7. On each evolutionary track for a given black hole mass and redshift, in the order of increasing frequency, the stars represent 1 year, 1 month (30 days), 1 day, 1 hour, 1 second before merger, and merger, respectively. The left panel illustrates the evolution for masses of $10^2-10^6 \, M_{\odot}$ observed at redshift $z = 1-10$. The right panel illustrates the evolution for masses of $10^3-10^5 \, M_{\odot}$ observed at redshift $z = 1-1000$.}
  \label{fig:strain_evolution_circular}
\end{figure}

\begin{figure}[ht]
  \centering
  \begin{subfigure}{0.5\linewidth}
    \centering
    \includegraphics[width=\linewidth]{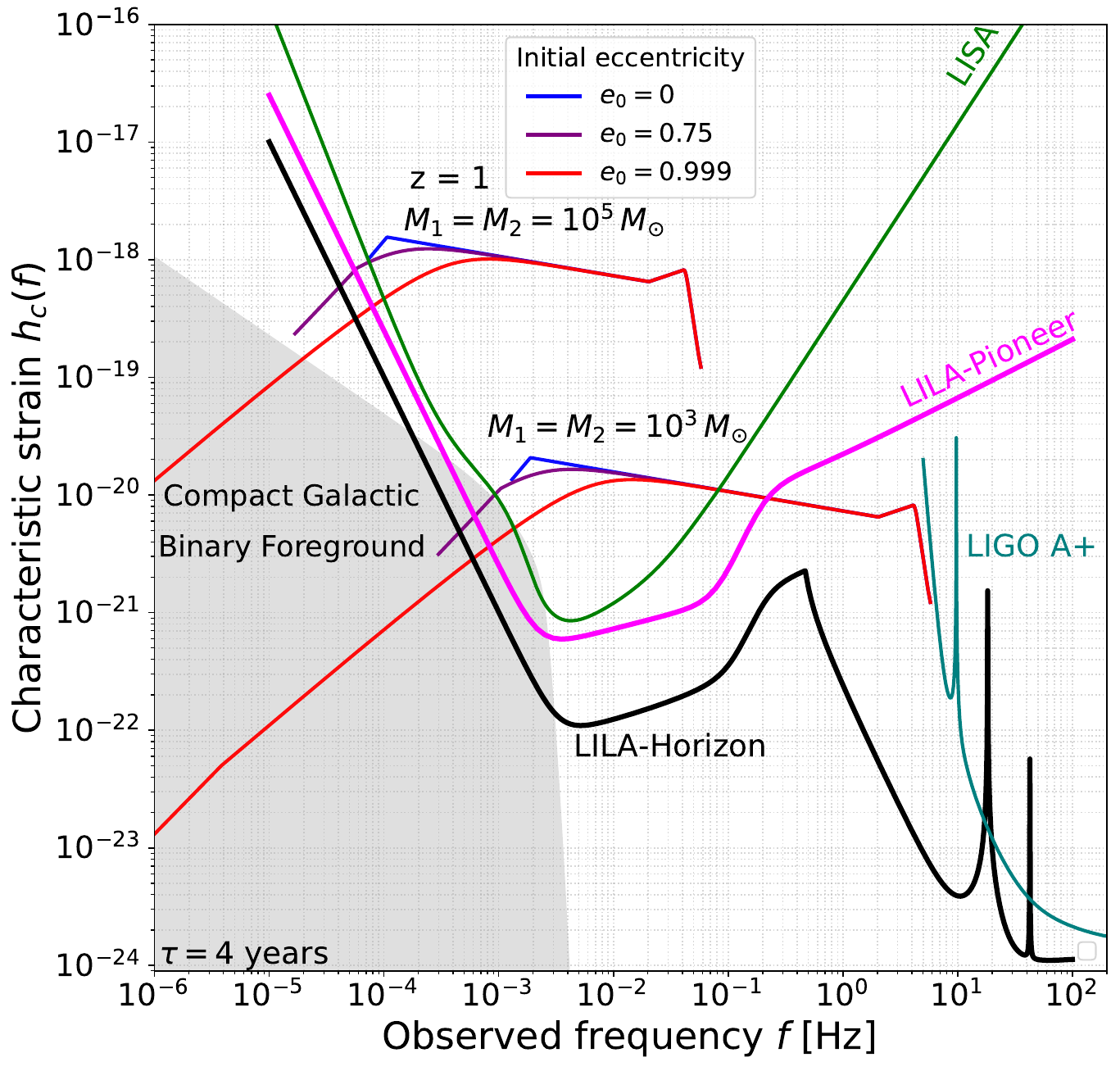}
  \end{subfigure}\hfill
  \begin{subfigure}{0.5\linewidth}
    \centering
    \includegraphics[width=\linewidth]{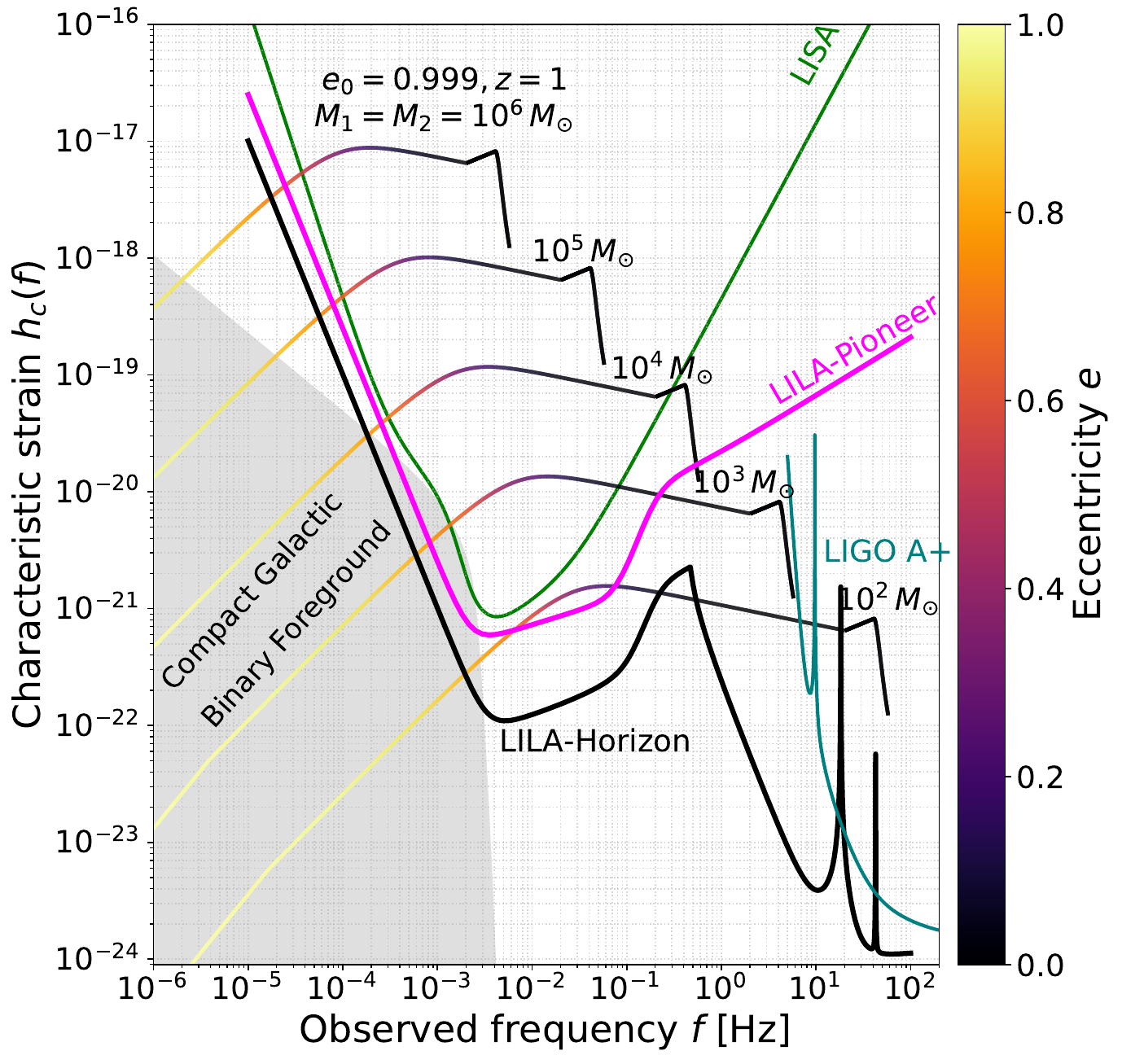}
  \end{subfigure}
  \caption{Characteristic strain evolution of equal-mass, non-spinning black hole binary systems with different masses and initial eccentricities observed at redshift $z = 1$, in comparison to the instrumental sensitivities and Galactic binary foreground noise as mentioned in Figure \ref{fig:strain_evolution_circular}. The left panel illustrates the evolution for black hole masses $10^3 \, M_{\odot}$ and $10^5 \, M_{\odot}$ with initial eccentricities of $0, 0.75$, and $0.999$, observed at redshift $z = 1$, from 4 years before merger until merger. The right panel illustrates the evolution for masses of $10^2-10^6 \, M_{\odot}$ with an initial eccentricity of $0.999$, observed at redshift $z = 1$, where the eccentricity is color-coded throughout each evolutionary track.}
  \label{fig:strain_evolution_eccentricity}
\end{figure}

\section{Detection Landscape and Science Cases for LILA}
\label{sec:detection_landscape}

\subsection{LILA Observational Horizon: Implications for Massive Black Hole Seeding \& Growth Models}
\label{subsec:LILA_observational_horizon}
\begin{figure}[ht]
  \centering
  \begin{subfigure}{0.5\linewidth}
    \centering
    \includegraphics[width=\linewidth]{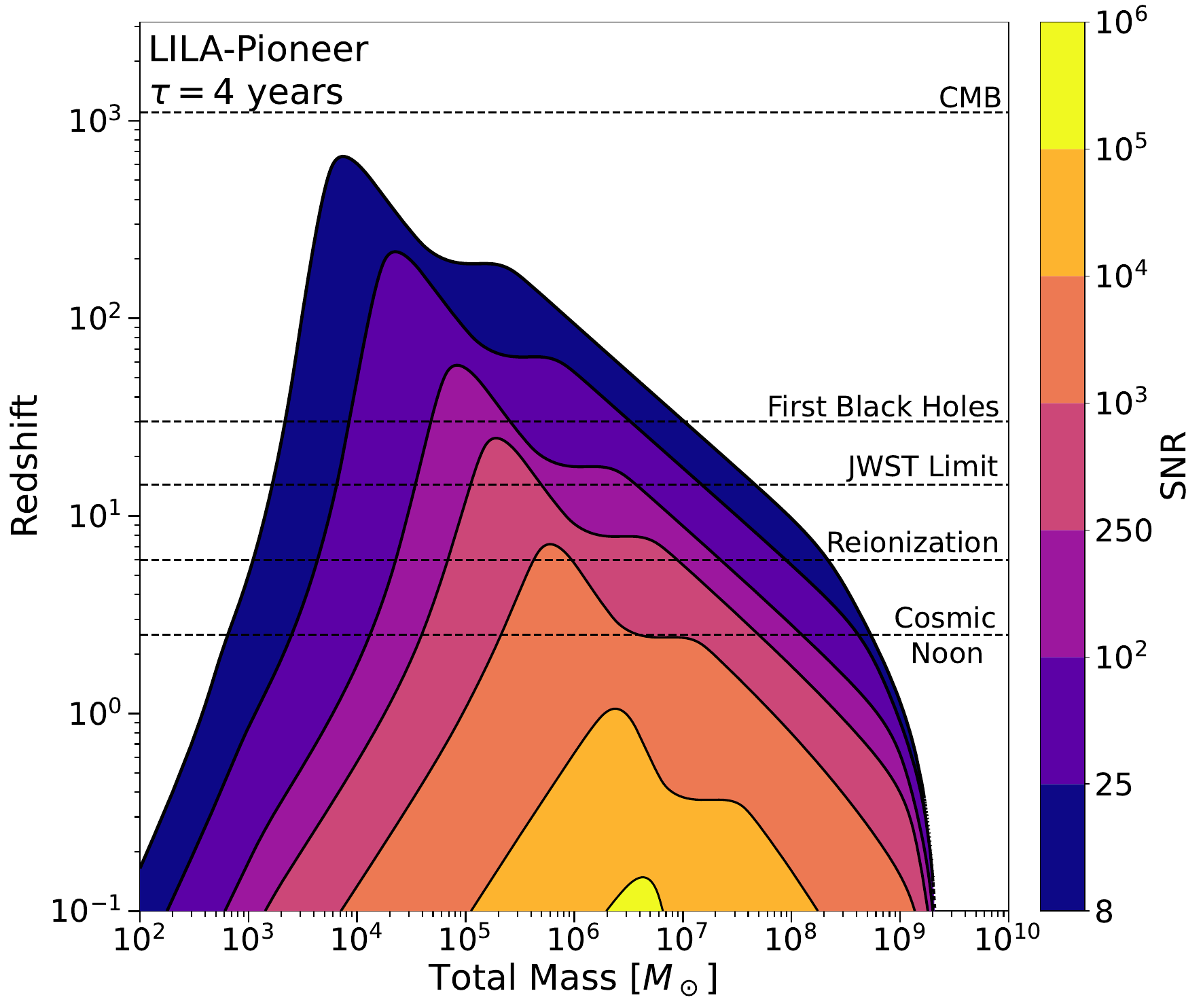}
  \end{subfigure}\hfill
  \begin{subfigure}{0.5\linewidth}
    \centering
    \includegraphics[width=\linewidth]{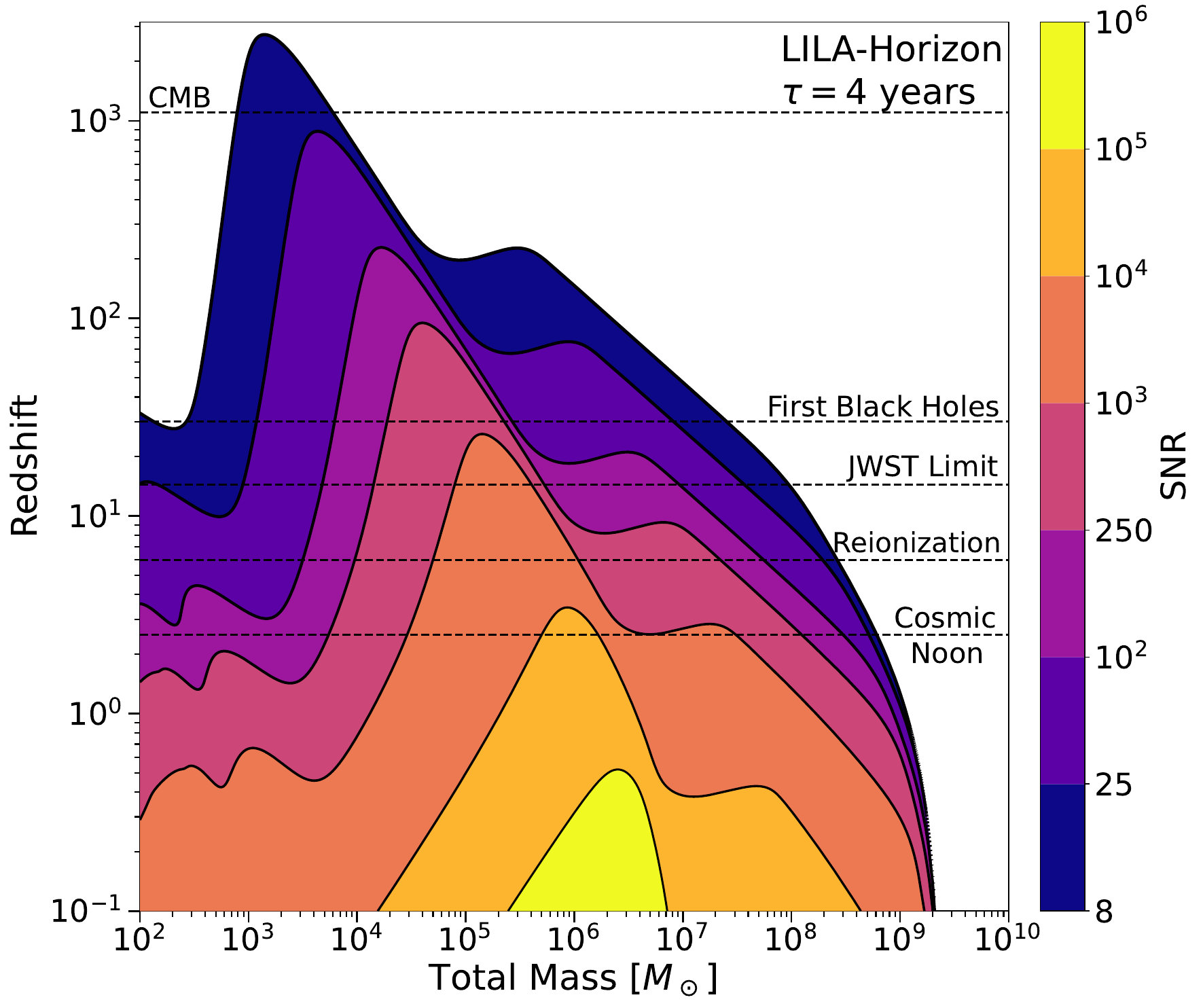}
  \end{subfigure}
  \caption{Detection horizon curves of LILA-Pioneer (left) and LILA-Horizon (right) instruments for SNR $= 8-10^6$, which illustrate the maximum detectable redshift $z$ for each equal-mass, non-spinning, circular binary system of total mass $M$ to reach a certain SNR threshold, assuming the systems are observed from four years before merger until merger. The horizontal dashed lines indicate important cosmic epochs, such as cosmic noon ($z \sim 2.5$; \citealt{Madau_2014}), end of reionization ($z \sim 7$; \citealt{BL01}), highest redshift detection by JWST ($z \sim 14.4$; \citealt{Naidu_2026_JWST_galaxy_z14.44}), first sources of light in the universe ($z \sim 30$; \citealt{BL01}), and the cosmic microwave background ($z \sim 1100$; \citealt{Planck_2018_overview})}
  \label{fig:snr_horizon}
\end{figure}

Based on the gravitational wave strain model in Section \ref{sec:strain_model}, Figures \ref{fig:strain_evolution_circular} and \ref{fig:strain_evolution_eccentricity} illustrate the characteristic strain evolution of equal-mass IMBH binaries for different masses, initial orbital eccentricities, and redshifts, against the instrumental sensitivity curves of LILA-Pioneer and LILA-Horizon. These figures highlight how IMBH binaries extending back to the black hole seeding epoch ($z \gtrsim 20$) can become detectable with LILA months to years in advance, which is crucial for SMBH formation and growth modeling and early-warning capabilities. To expand on observational prospects of LILA, using the SNR calculations outlined in Section \ref{sec:SNR}, we calculate the detection redshift for a given SNR threshold in Figure \ref{fig:snr_horizon}. With a SNR threshold of 8 and an observational time of 4 years, LILA-Pioneer and LILA-Horizon can extend the detection horizon towards most of the observable Universe at maximum redshifts of $z \sim 600$ and $z \sim 2500$, respectively. LILA sensitivity allows direct probes into the black hole seeding epochs, when light seeds of $\sim 10 - 10^3 \, M_{\odot}$ from the collapse of Pop III stars \citep{Madau_Rees_2001} or heavy seeds of $\sim 10^4 - 10^5 \, M_{\odot}$ from the direct collapse of primordial gas clouds \citep{Loeb_Rasio_1994, Bromm_Loeb_2003} could form as early as $z \sim 20 - 30$ \citep{BL_2000, Volonteri_2010, Ferrara_2014, Ricarte_Natarajan_2018_signatures_BH_seeding, Pacucci_2022_search}. The existence of direct collapse black holes remains an open question, motivated by recent JWST observations, such as the overmassive black hole galaxy UHZ1 \citep{Natarajan_2024_UHZ1} and ``Little Red Dot" galaxies \citep{Jeon_2026_LRD_progenitor_DCBH, Pacucci_2026_LRD_DCBH}. Direct collapse black holes are expected to produce gravitational wave signals in the deci-Hz band \citep{Pacucci_2015_GW, Kelly_2025_DCBH_GW_signatures}, offering new opportunities for LILA to test the existence of direct collapse as a black hole formation mechanism. Observational evidence such as the NANOGrav GW background amplitude \citep{NANOGrav_15_year_BH_binary} and overmassive high-redshift SMBHs recently discovered by JWST \citep{Pacucci_2023, Harikane_2023, Kocevski_2023, Maiolino_2024_JADES_BH_z4_11, Jones_2025_LRD_M_Mstar_evolution} suggest evolution models with efficient black hole merging in the early Universe \citep{Izquierdo-Villalba_2022_GW_background_BH_evolution, Bhowmick_2026_BRAHMA_heavy_seeds, Zhou_2026_AMBRA_seeds_mergers}. Therefore, we could expect promising detection rates for LILA across different redshifts, a calculation we defer to future studies. 

Gravitational waveforms provide reliable mass measurements of the initial and final black holes, and the detection horizon in Figure \ref{fig:snr_horizon} suggests LILA's potential to map out the black hole mass distribution across cosmic time all the way back to the seeding epoch. The differences in the black hole mass function inferred from light and heavy seeding scenarios are typically only significant for masses of $\lesssim 10^6 \, M_{\odot}$ \citep{Ricarte_2025_SEROTINA_spin_evolution, Jeon_2025_BHMF_high_z, Fei_2025_JWST_GLIMPSE_IMBH_direct_collapse}, so LILA discoveries of IMBH binaries across different redshifts would help distinguish different seeding mechanisms. In fact, recent studies have inferred the black hole mass function from JWST high-redshift AGN observations \citep{Mathee_2024_JWST_LRD, Taylor_2025_JWST_AGN_BHMF}, revealing that the number density of observed faint AGNs is an order of magnitude higher than previously estimated. The number density of black holes with masses of $\sim 10^{5.5}-10^7 \, M_{\odot}$ inferred in these studies has favored heavy seeding scenarios or light seeding scenarios with episodic super-Eddington accretion \citep{Fei_2025_JWST_GLIMPSE_IMBH_direct_collapse, Geris_2026_JADES_high_z_BH}. In light of new JWST constraints on high-redshift black hole populations, recent studies have forecasted up to tens of mergers with masses of $\gtrsim 10^3-10^4 \, M_{\odot}$ every year \citep{Liu_Inayoshi_2025_GW_forecast_JWST_constraints, Caceres-Burgos_2026_BH_merger_rate_JWST_constraints}, which is promising for future IMBH searches of LILA in the early Universe. However, JWST analysis of the high-redshift black hole mass function has major limitations due to the small-number statistics, systematic uncertainties in broad-line virial mass measurements, and selection effects driving biases towards discovering more massive and active black holes \citep{TRINITY_2, Li_2024, Silverman_2025_M_Mstar_SHELLQS-JWST, Ziparo_2026_M_Mstar_selection_effect, Roberts_2026_M_Mstar_comparison}. In addition, the inherent degeneracy between the initial seeding mass, accretion rate, and duty cycle has made it difficult to draw robust inferences about the nature of SMBH formation and growth with JWST AGN observations \citep{Fragione_2023}. Hence, by pushing the detection horizon towards higher redshifts, less active black holes, and the intermediate-mass range, LILA can potentially improve the determination of the black hole mass function in these regimes where the differences between light and heavy seeding scenarios become more prominent. Moreover, probing the low-mass end of the massive black hole mass function is crucial to galaxy evolution and cosmological simulation models \citep{Pacucci_2018, Habouzit_2021}. Mapping out the IMBH binary inspiral and merger population throughout cosmic time with LILA will also allow us to examine its contribution to the stochastic GW background and aid our physical interpretations of recent NANOGrav results \citep{NANOGrav_15_year, NANOGrav_15_year_BH_binary, Ellis_2024a, Izquierdo-Villalba_2024_NANOGrav}.

In addition to mass, the spin-orbit coupling during compact binary inspirals and mergers enables highly sensitive black hole spin measurements with gravitational wave detection, as demonstrated at sub-percent precision for LISA \citep{Berti_2005_LISA_parameter_inference_gravity_test, Lang_Hughes_2006, Lang_Hughes_2007, Lang_Hughes_2008, Lang_2011, Baibhav_2020_LISA_parameter_inference_localization, Burke_2020, Bhagwat_2022_LISA_ET_BH_spectroscopy, Watarai_2026_LISA_mass_spin_inference_ringdown}. By determining the spin distribution of black holes across different redshifts from detected inspirals and mergers, LILA can also investigate other aspects of cosmic black hole assembly, such as merger, accretion, and magnetic field dynamics \citep{Volonteri_2005, Shapiro_2005, King_2006, Ricarte_2025_SEROTINA_spin_evolution}.
Near-horizon physics, such as accretion-driven spin up \citep{Novikov_Thorne_1973, Shapiro_2005} and jet-driven spin down
\citep{Narayan_2022_jet_MAD_geometry_power_spin_down, Ricarte_2023_jet_spin_MAD}, impacts the cosmological spin distribution of black holes. The relative contribution and transition between accretion-dominated and merger-driven phases in cosmic black hole assembly \citep{Pacucci_Loeb_2020_accretion_merger_BH_growth, Izquierdo-Villalba_2026_overmassive_undermassive_BH} will be further elucidated by comparing redshift-dependent black hole mass and spin distributions with predictions from semi-analytical models  \citep{Sesana_2014_BH_spin_evolution_galaxy_kinematics, Ricarte_Natarajan_2018_SEROTINA_SMBH_assembly, Ricarte_2018, Ricarte_2019_SEROTINA_undetected_highz_BH, Trinca_2023, Porras_Valverde_SMBH_assembly_SAM} and cosmological simulations \citep{Dubois_2014, Bustamante_2019_SMBH_spin_evolution_feedback, Sala_2024_SMBH_spin_evolution_cosmo_sim, Bhowmick_2024}. In comparison to mass, the black hole spin distribution across cosmic time is a much more robust discriminant of different black hole accretion histories \citep{Ricarte_2025_SEROTINA_spin_evolution}. Black hole spin is also essential to galaxy evolution models since it influences radiative efficiency and jet power in black hole feedback, as shown by cosmological \citep{Bustamante_2019_SMBH_spin_evolution_feedback, Bollati_2024_AGN_feedback_BH_spin} and general relativistic magnetohydrodynamical (GRMHD; \citealt{Tchekhovskoy_2010_BH_spin_AGN_radio_dichotomy, Cho_2025_bridging_scales, Cho_2026_bridging_scales_subgrid_prescription}) simulations. Thus, LILA's capability to detect black hole binaries across a wide range of masses and redshifts will contribute to a broader understanding of black hole growth and galaxy evolution. On cosmological scales, mapping out the black hole mergers across different masses and redshifts with LILA will help probe the entropy budget of the universe \citep{Siyuan_2026_entropy_budget_BH_mergers} and different formation pathways when compared with the underlying large-scale structure \citep{Smith_2026_binary_BH_large_scale_structure}. Additionally, precise black hole spin measurements from GW events are also crucial for beyond-standard-model physics studies, such as black hole superradiance \citep{Arvanitaki_2011_string_axiverse_BH, Arvanitaki_2015_axion_BH_GW} where LILA sensitivity to IMBHs would complement recent ultralight boson constraints from LVK events (e.g., \citealt{Brito_2017_GW_searches_ultralight_bosons, Ng_2021_ultralight_boson_BH_mergers, Ng_2021_ultralight_boson_GWTC-2, Ghosh_Sachdeva_2021_light_dark_photon_constraints, Aswathi_2025_ultralight_boson_BH_binary}) in probing a new region of ultralight-boson parameter space at the corresponding energy scales.

\subsection{Eccentric IMBH Binaries}
\label{subsec:eccentric}

Because of circularization effects of gravitational radiation \citep{Peters_1964_GW_binary} as demonstrated by Equation \ref{eq:de_dt}, we expect the final orbital eccentricity of compact binary mergers detected by LVK and accessible by ground-based detector frequency range in general to be too small to be measured \citep{LIGO_GW150914_astrophysical_implications}. In fact, only a very small fraction of LVK black hole merger detections so far such as GW190521 \citep{GW190521} has evidence for significant eccentricity and dynamical evolution \citep{Romero-Shaw_2020_GW190521_eccentricity, Romero-Shaw_2022_eccentric_mergers_LVK, Gamba_2023_GW190521_dynamical_capture}, though limited eccentric waveform templates pose major challenges in LVK searches for eccentric black hole mergers \citep{GWTC3_search_eccentric_binary}. Black hole binary formation in dynamically active environments, such as dense stellar clusters \citep{Rodriguez_2016, Rodriguez_2018_eccentric_BH_mergers_dense_star_clusters, Arca_Sedda_2018_BH_globular_clusters,  Samsing_2018_eccentric_BH_merger_globular_clusters, Zevin_2019_eccentric_BH_mergers_dense_star_clusters, Fragione_2019_eccentric_BH_merger_globular_clusters}, galactic nuclei \citep{Antonini_Rasio_2016_BH_binary_galactic_nuclei, Gondan_2018_eccentric_BH_merger_galactic_nuclei, Zhang_2019_BH_mergers_galactic_nuclei, Tagawa_2021_eccentric_BH_mergers_AGN, Samsing_2022_eccentric_BH_mergers_AGN}, and triples \citep{Antonini_2017_BH_merger_triples, Arca_Sedda_2021_eccentric_BH_mergers_triples} via the Kozai-Lidov mechanism \citep{Kozai_1952, Lidov_1962, Naoz_2016_Kozai_Lidov_effect}, could imprint their signatures on orbital eccentricity. In addition, moderately eccentric binaries formed in circumbinary disks would not circularize and instead retain much eccentricity as they are driven towards inspirals and mergers \citep{Siwek_2023_binary_evolution_circumbinary_disks}. 

\begin{figure}[hbt!]
  \centering
  \begin{subfigure}{0.5\linewidth}
    \centering
    \includegraphics[width=\linewidth]{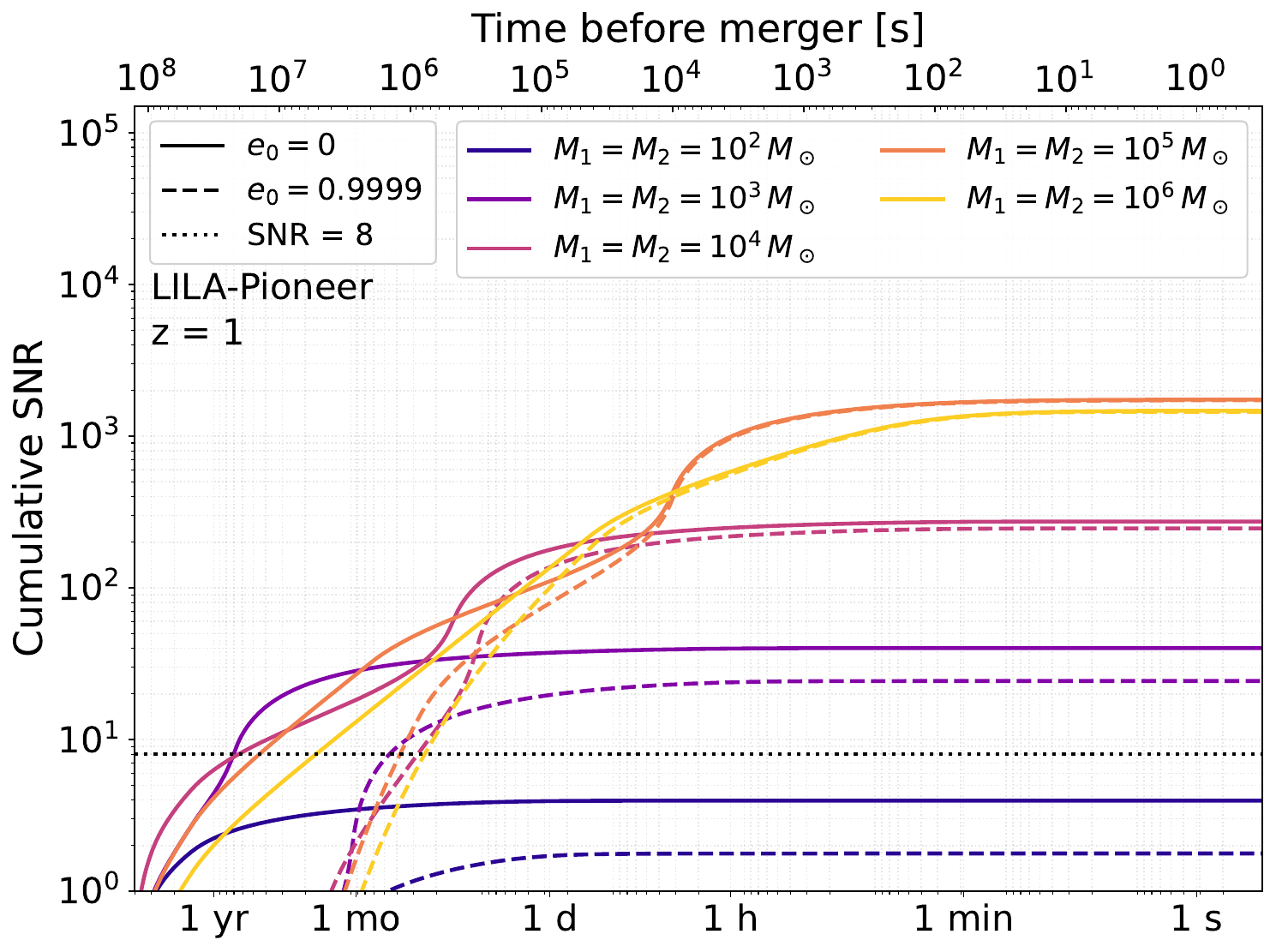}
  \end{subfigure}\hfill
  \begin{subfigure}{0.5\linewidth}
    \centering
    \includegraphics[width=\linewidth]{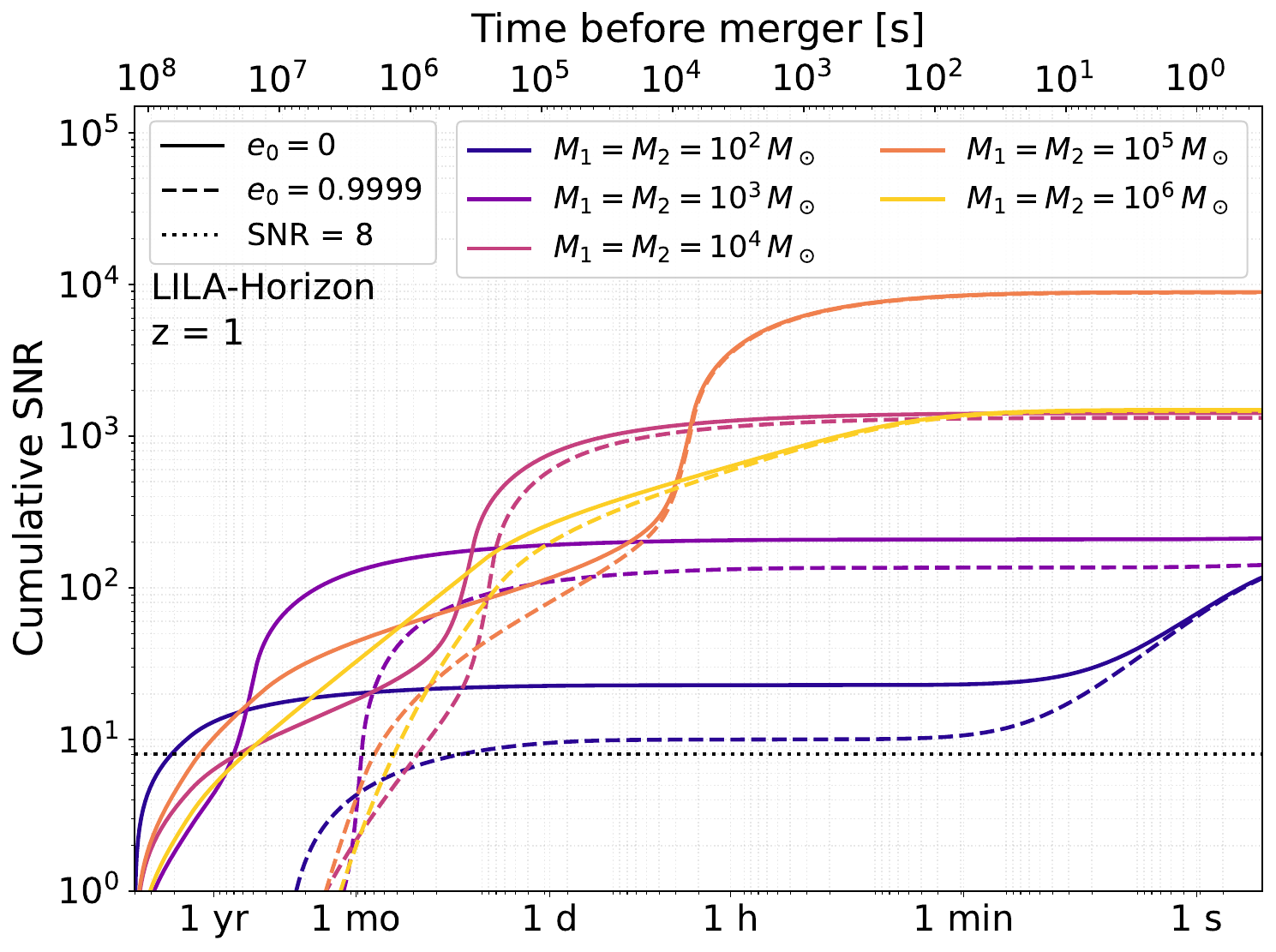}
  \end{subfigure}
    \begin{subfigure}{0.5\linewidth}
    \centering
    \includegraphics[width=\linewidth]{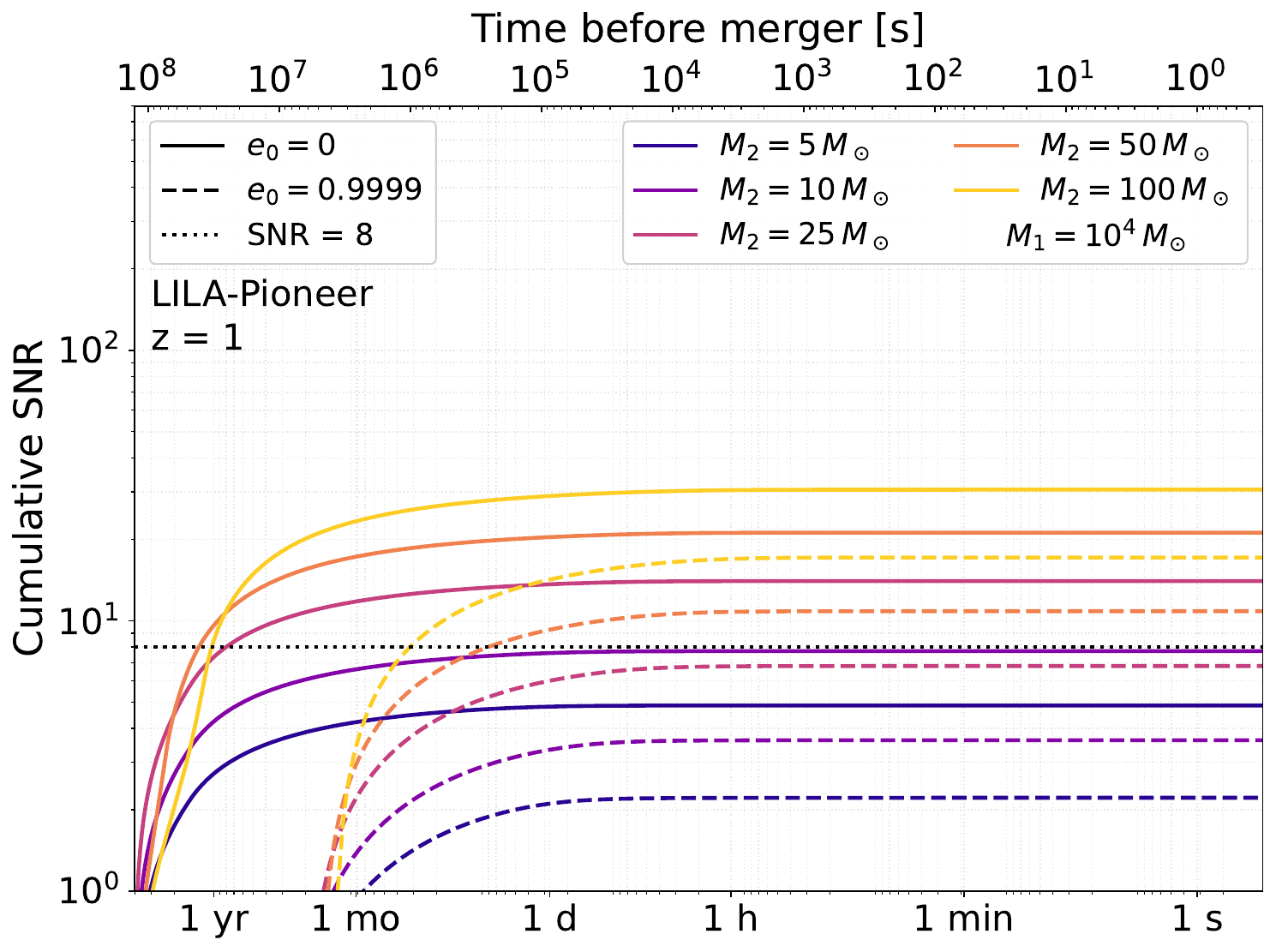}
  \end{subfigure}\hfill
  \begin{subfigure}{0.5\linewidth}
    \centering
    \includegraphics[width=\linewidth]{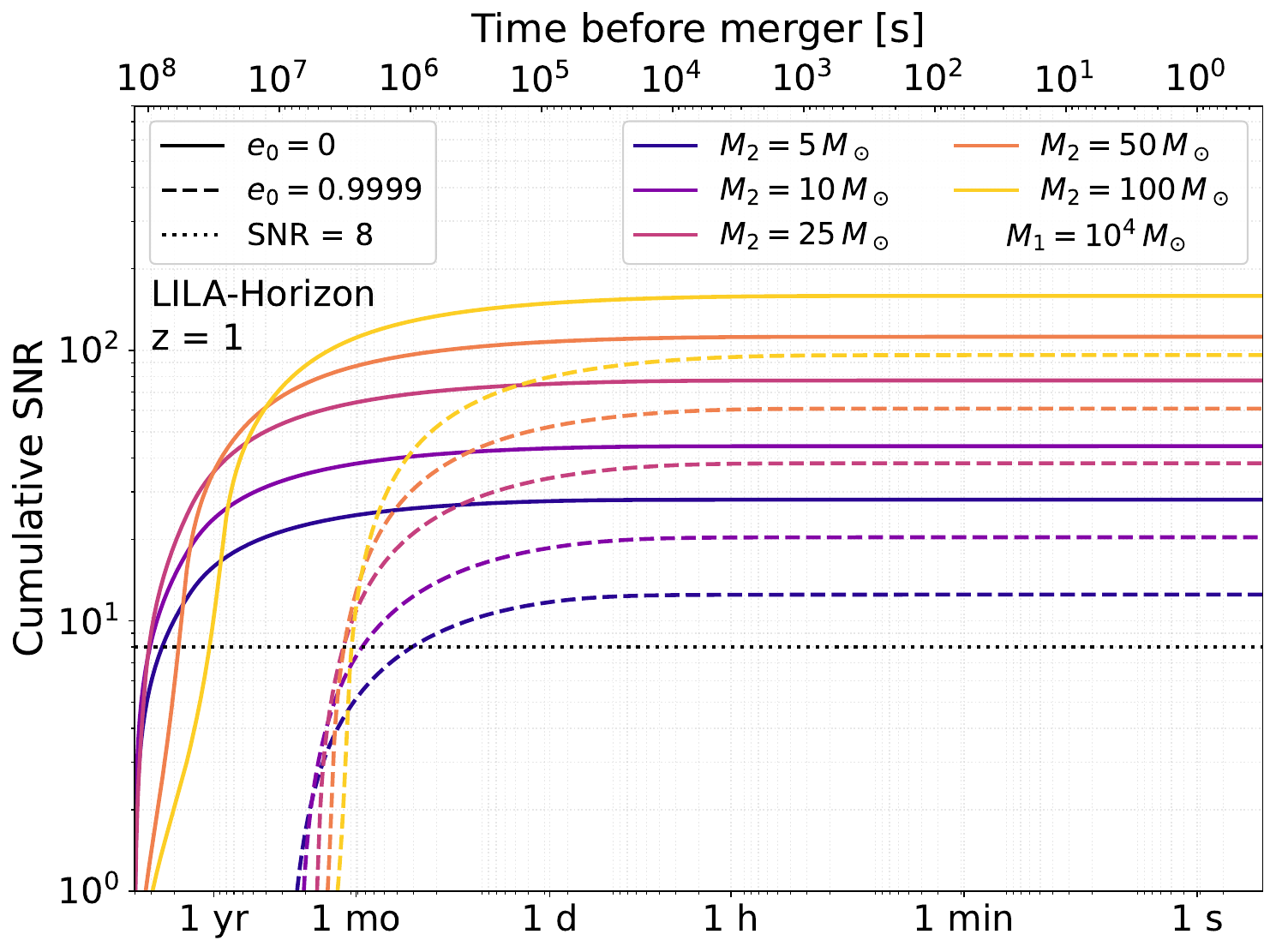}
  \end{subfigure}
  \caption{The cumulative SNR of equal-mass (top row) and intermediate-mass ratio (bottom row) inspirals as a function of time before merger, assuming an initial observation time of 4 years before merger and initial eccentricities of 0 (solid) and 0.9999 (dashed), for LILA-Pioneer (left) and LILA-Horizon (right). The horizon dotted line in each panel denotes SNR = 8, a typical detection threshold for GW events.}
  \label{fig:snr_vs_time}
\end{figure}

LISA is expected to distinguish black hole binaries formed in isolation from those in more dynamical scenarios in the milli-Hz frequency band for some nearby systems \citep{Breivik_2016_LISA_BH_binary_formation_channels, Nishizawa_2016_LISA_eccentricity_BH_binary_formation, Nishizawa_2017_LISA_eccentricity_BH_binary_formation, Kremer_2018_LISA_sources_globular_clusters}, but for binaries with very high initial eccentricities, the dominant gravitational wave signals are in the deci-Hz regime outside of the LISA band \citep{Chen_2017_deciHz_GW_BH_binary_formation, D_Orazio_2018_LISA_BH_mergers_globular_clusters, Kremer_2019_LISA_BH_binaries_dense_star_clusters}. In particular, simulations of dense star cluster populations suggest that 15\% of dynamical binaries will have eccentricities $e \gtrsim 0.1$ at 0.1 Hz \citep{Kritos_2024_BH_binary_star_cluster_evolution}. Thus, as a deci-Hz detector, LILA is well-suited for discovering systems some time before mergers with measurable eccentricity residuals retained from their formation mechanisms, potentially probing astrophysical environments and evolutionary pathways of black hole binaries. Figure \ref{fig:strain_evolution_eccentricity} shows that equal-mass IMBH binaries with masses of $10^2-10^6 \, M_{\odot}$ and initial orbital eccentricity of $e_0 = 0.999$, observed four years before merger, enter the LILA sensitivity band with significant eccentricity residuals of $e \sim 0.6$. Figure \ref{fig:snr_vs_time} illustrates that binary systems with masses $10^3-10^5 \, M_{\odot}$ and an initial eccentricity of $e = 0.9999$ are detectable (with SNR = 8 threshold) by both LILA-Pioneer and LILA-Horizon more than a month before merger, highlighting LILA's prospects for discovering highly eccentric binaries and their evolutionary signatures as products of dynamically active environments.

\subsection{Intermediate-Mass Ratio Inspirals (IMRIs)}
\label{subsec:IMRIs}

\begin{figure}[hbt!]
  \centering
  \begin{subfigure}{0.99\linewidth}
    \centering
    \includegraphics[width=\linewidth]{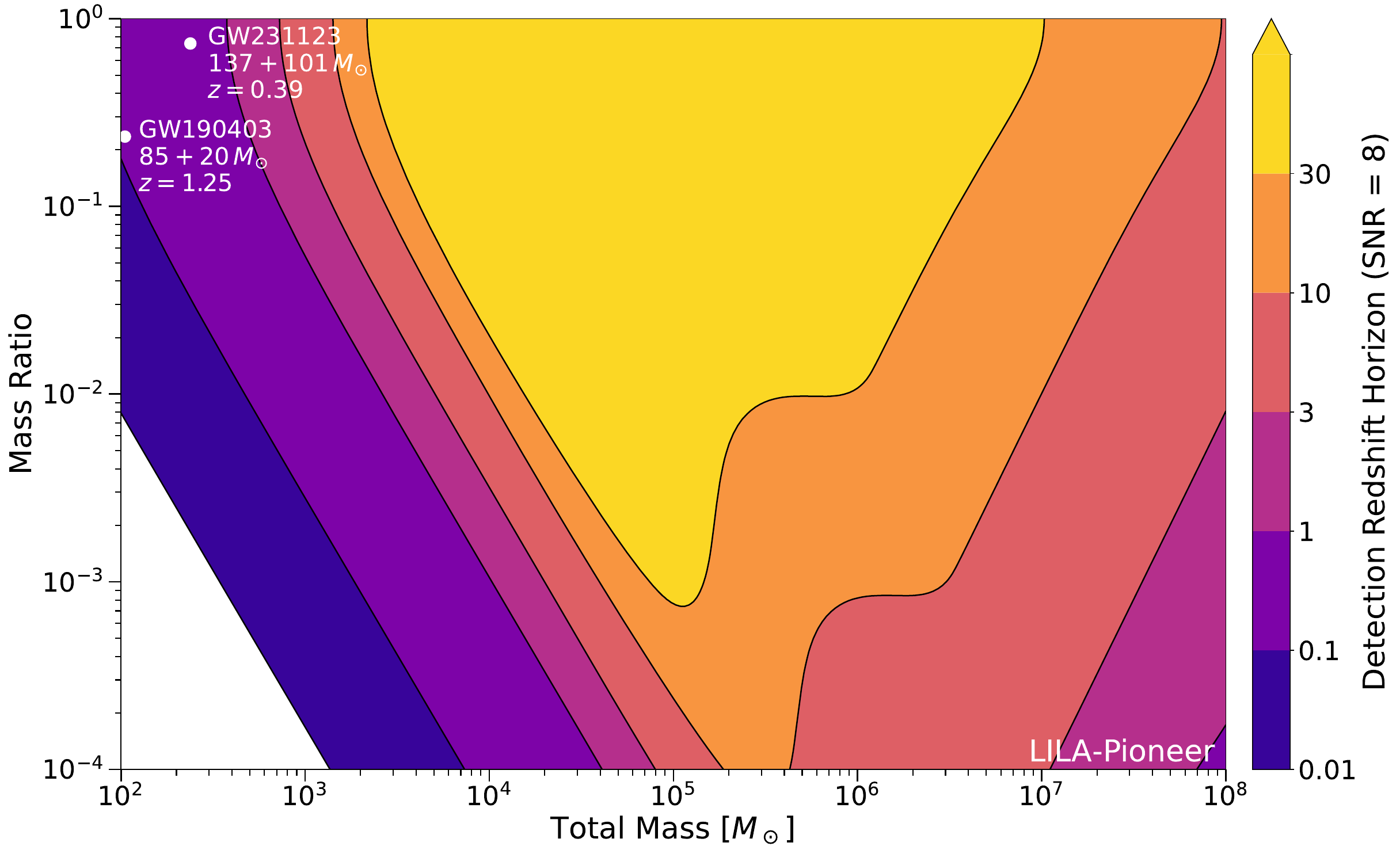}
  \end{subfigure}\hfill
  \begin{subfigure}{0.99\linewidth}
    \centering
    \includegraphics[width=\linewidth]{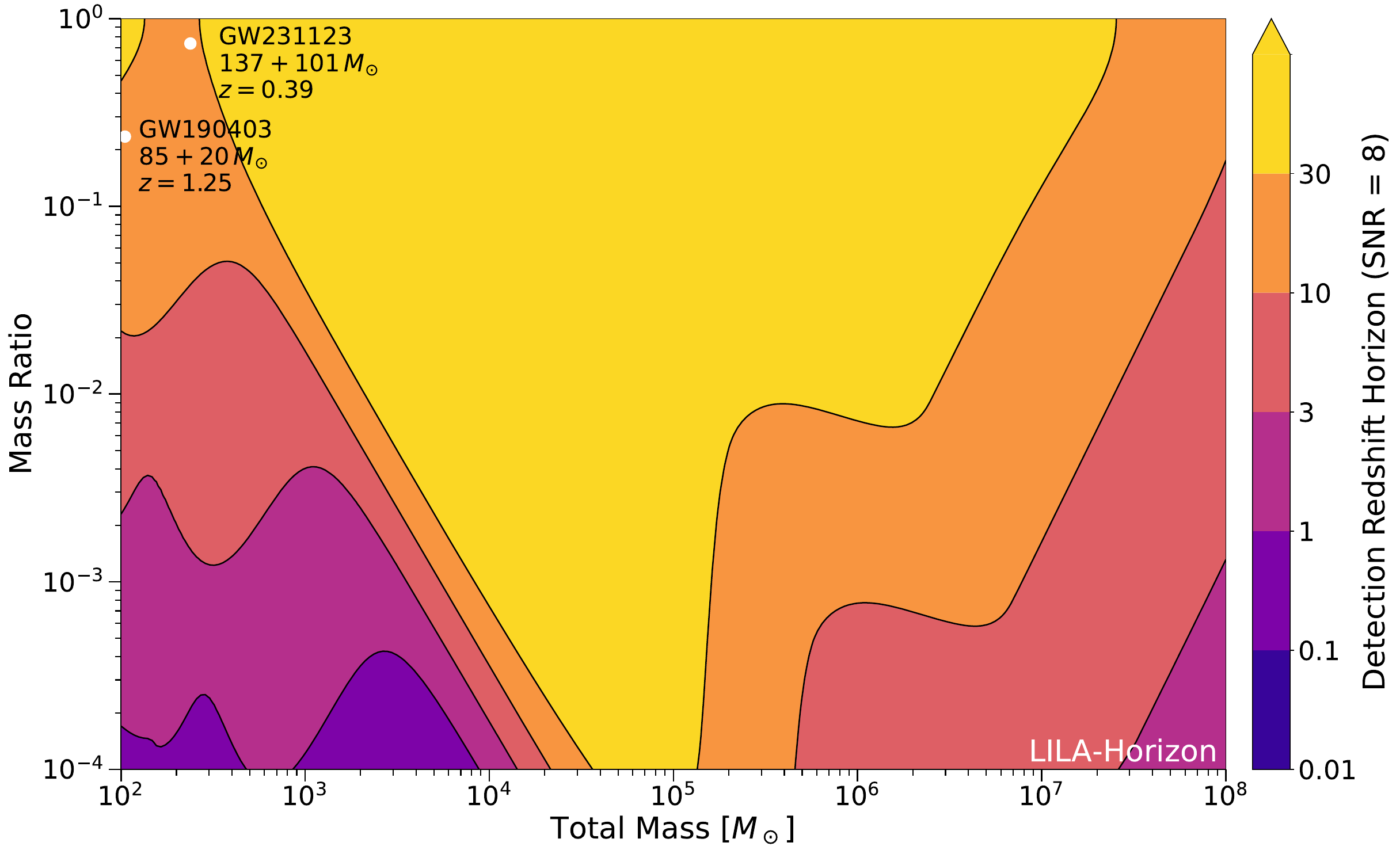}
  \end{subfigure}
  \caption{SNR = 8 detection redshift contours, defined as the farthest redshift for a black hole binary to achieve a detection SNR $>$ 8, of LILA-Pioneer (top) and LILA-Horizon (bottom) with different total masses and mass ratios ($M_2/M_1$), assuming non-spinning, circular binary systems initially observed 4 years before merger. For reference, GW231123, the most massive black hole binary detected thus far \citep{GW231123_total_225_Msol}, and GW190403$\_$051519, a high-significance candidate event with a mass ratio of $\sim 0.24$ \citep{GWTC2.1_extended_catalog}, are labeled on the mass ratio - total mass diagram. }
  \label{fig:mass_ratio_horizon}
\end{figure}

IMBHs are likely to form within dense stellar cluster environments based on our theoretical models and simulations \citep{Griersz_2015_IMBH_formation_globular_clusters, Rodriguez_2016, Arca_Sedda_2018_BH_globular_clusters, Antonini_2019_BH_merger_star_clusters}, so they can potentially form binary systems with compact stellar remnants \citep{MacLeod_2016_IMBH_stellar_companions, Arca_Sedda_2021_stellar_IMBHs_dense_clusters, Ye_2023_IMBH_WD_dense_stellar, Lee_2025_IMBH_star_cluster_formation_evolution}. This leads to intermediate-mass ratio inspirals (IMRIs) with mass ratio $M_2/M_1 \sim 10^{-2}-10^{-4}$ \citep{Haster_2016_IMRI_star_clusters, Amaro-Seoane_2018_IMRI_ground_space}, which could have extremely high initial eccentricities of $e \sim 0.999$ and small AU-scale separation \citep{Konstantinidis_2013_IMBH_star_clusters_simulations, Leigh_2014_stellar_IMBH_globular_clusters}. Given the number of IMBHs forming IMRI systems $n_{\rm rep}$, the fraction of clusters hosting an IMBH $p_{\rm IMBH}$, the fraction of clusters forming an IMRI system $p_{\rm IMRI}$, the fraction of compact objects in the cluster $f_{\rm com}$, the star cluster formation rate per unit mass $\rho_{\rm SFR}$, the average cluster mass $M_{\rm GC}$, and the comoving volume horizon (with SNR of 15) at a given redshift $V$, Equation 1 of \citealt{Arca_Sedda_2020_deciHz_GW_discoveries} provides an order-of-magnitude LISA detectable merger rate estimate for IMRIs formed in stellar clusters:

\begin{multline}
    \Gamma_{\rm LISA} \simeq 0.012 n_{\rm rep} \left( \frac{p_{\rm IMBH}}{0.2} \right) \left( \frac{p_{\rm IMRI}}{0.5} \right) \left( \frac{f_{\rm com}}{0.1} \right) \\ \times \left( \frac{\rho_{\rm SFR}}{0.005 \, M_{\odot} \,  \rm{yr}^{-1} \, \rm{Mpc}^{-3}} \right) \left( \frac{M_{\rm GC}}{10^6 \, M_{\odot}} \right)^{-1} \left( \frac{V}{2.5 \, \rm{Gpc}^3} \right) \, \rm{yr}^{-1},
    \label{eq:IMRI_merger_rate}
\end{multline}
which is calibrated to a $1000 + 30 \, M_{\odot}$ IMRI system in a $10^6 \, M_{\odot}$ globular cluster, sensitive out to redshift $z \sim 0.2$ (equivalently, comoving volume $\sim 2.5$ Gpc$^3$) for a 4-year LISA mission with SNR of 15. Assuming a constant $\rho_{\rm SFR} = 0.05 \, M_{\odot} \, \rm{yr}^{-1} \, \rm{Mpc}^{-3}$ across the redshift range $z = 2-8$ \citep{Katz_Ricotti_2013_globular_cluster_fromation}, $f_{\rm COM} = 0.01$ \citep{Arca_Sedda_2019_MOCCA-SURVEY_IMBH_MW_GC}, $p_{\rm IMBH}  = 0.2$, and $p_{\rm IMRI} = 0.5$, this translates to a $1000 + 30 \, M_{\odot}$-like IMRI detection rate of $0.012 \, \rm{yr}^{-1}$ for LISA \citep{Arca_Sedda_2020_deciHz_GW_discoveries}. We consider a similar case study for a $1000 + 30 \, M_{\odot}$ IMRI system with an initial eccentricity of $e = 0.99$ and an initial observation time of 4 years before its merger. LILA-Pioneer also achieves an SNR of about 15 at $z \sim 0.2$ for this IMRI system, leading to a similar IMRI detection rate of $0.012 \, \rm{yr}^{-1}$. LILA-Horizon significantly boosts this rate due to its improved sensitivity by expanding the SNR $= 15$ detection horizon to $z \sim 4$, or a comoving volume of $\sim 1650$ Gpc$^3$, which translates to a detection rate of $660 \, \Gamma_{\rm LISA} \approx 8 \, \rm{yr}^{-1}$, based on Equation \ref{eq:IMRI_merger_rate}. While this calculation only provides a rough estimate, it illustrates that a deci-Hz detector like LILA can improve the IMRI detection rate of LISA by more than two orders of magnitude with optimistic prospects to detect a few $1000 + 30 \, M_{\odot}$-like IMRI systems annually. 

To highlight LILA's potential for discovering IMRI systems, Figure \ref{fig:mass_ratio_horizon} depicts detection redshift contours (with SNR threshold of 8) for black hole binary systems with different mass ratios and total masses, demonstrating how LILA-Horizon could confidently detect IMRIs with total masses of $\sim 10^5 \, M_{\odot}$ and mass ratios of $10^{-4}-1$ back to the seeding epoch $z > 10$. This provides another major pathway for discovering IMBHs and gaining insights into their astrophysical environments, such as dense stellar clusters where IMBHs are likely to form binary systems with other stellar remnants. IMRI systems such as tidal disruption of white dwarfs by IMBHs are also potential multi-messenger sources that could become extremely luminous above the Eddington limit \citep{Sesana_2008_WD_BH_multimessenger, Zalamea_2010_WD_BH_tidal_stripping_multi_messenger, MacLeod_2014_WD_BH_tidal, Yang_2026_WD_IMBH_tidal_dissipation}. There has been tentative observational evidence of white dwarf--IMBH binary systems through X-ray quasi-periodic oscillations \citep{Giustini_2020_X-ray_QPO} and the discovery of the ultra-long gamma-ray burst GRB 250702B \citep{Levan_2025_GRB250702B, Neights_2026_GRB250702B, Yuan_2026_GRB_250702B_IMBH_WD_tidal_disruption, Eyles-Ferris_2026_GRB250702B_IMBH_WD_tidal_disruption, Sato_2026_GRB250702B_IMBH_WD_tidal_disruption}, which further underscores the need for deci-Hz detectors like LILA to conclusively determine the nature of these transient events.

\subsection{Early Warning \& Localization Potential}
\label{subsec:early_warning}

\begin{figure}[hbt!]
  \centering
  \begin{subfigure}{0.5\linewidth}
    \centering
    \includegraphics[width=\linewidth]{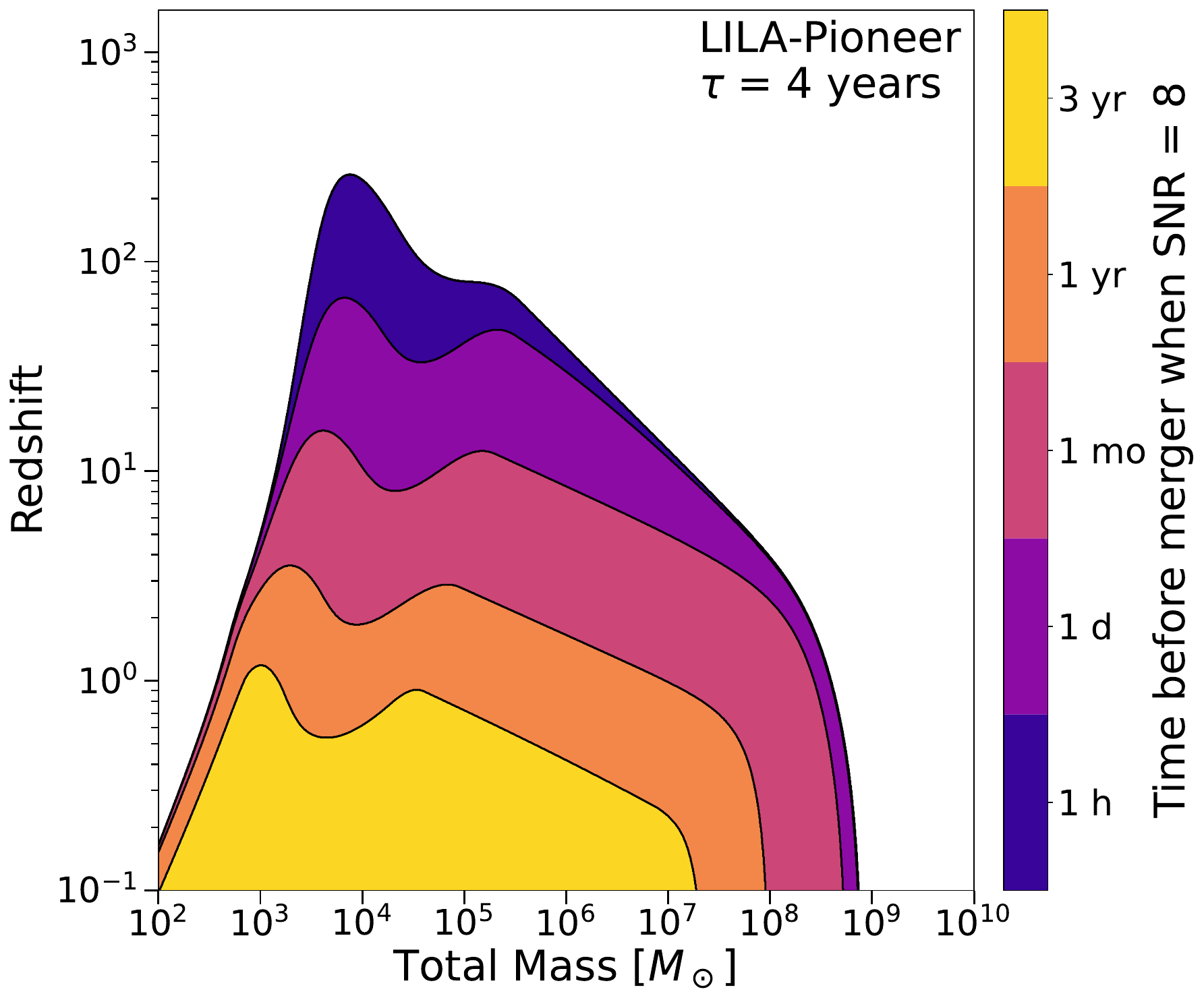}
  \end{subfigure}\hfill
  \begin{subfigure}{0.5\linewidth}
    \centering
    \includegraphics[width=\linewidth]{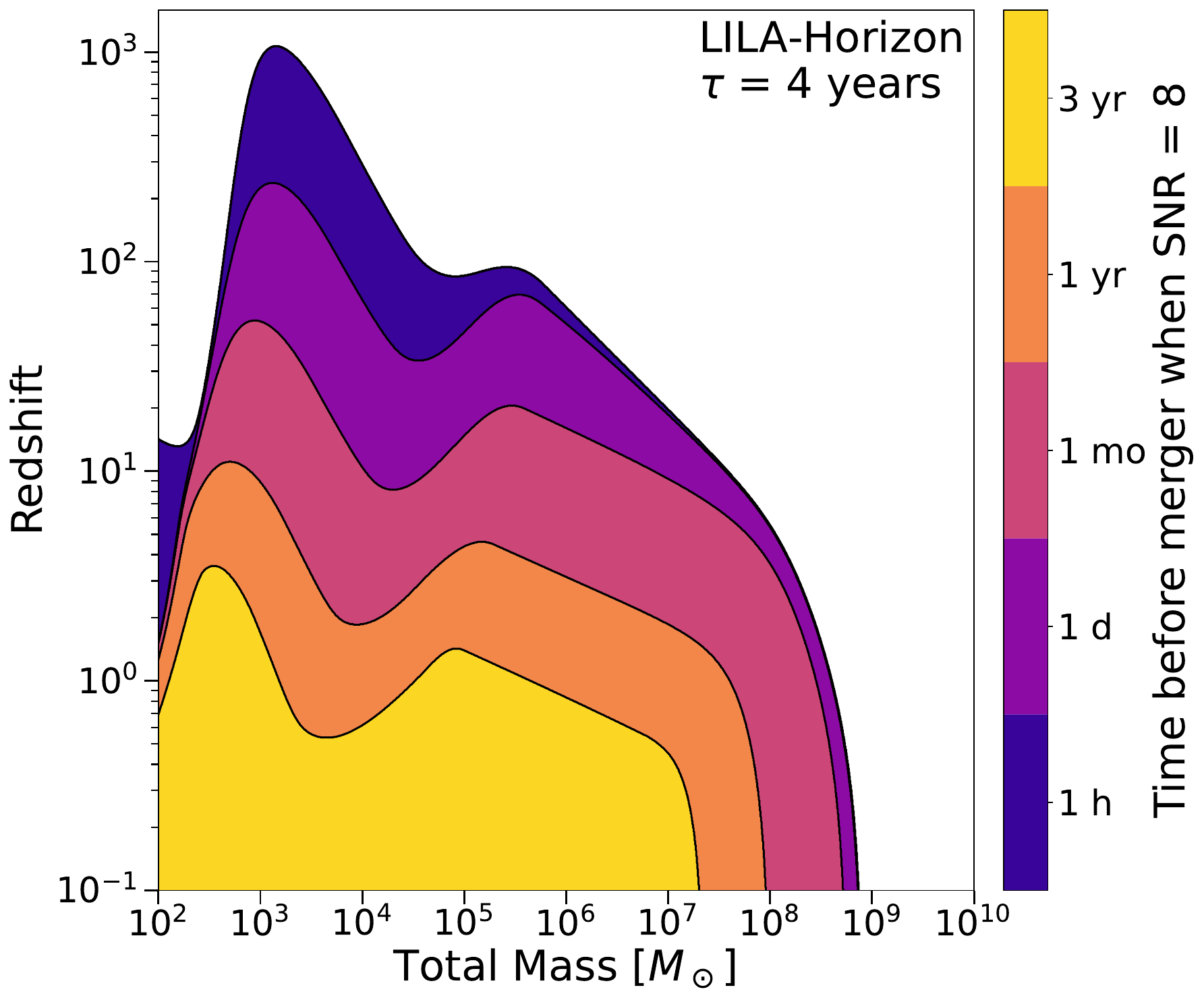}
  \end{subfigure}
  \caption{Detection horizon of LILA-Pioneer (left) and LILA-Horizon (right) with color contours denoting the time before merger for a black hole binary to be detected with SNR = 8, assuming non-spinning, equal-mass, circular black hole binaries.}
  \label{fig:time_to_snr8}
\end{figure}

As demonstrated in Figure \ref{fig:strain_evolution_circular}, IMBH binary mergers enter the deci-Hz GW frequency band up to months and possibly years in advance, allowing LILA to provide early warning for multi-messenger observations and future merger events detectable by ground-based detectors (Subsection \ref{subsec:multiband_synergy}). To highlight LILA's prospects for providing early warning for multi-messenger and multi-band follow-up, Figure \ref{fig:snr_vs_time} illustrates the cumulative SNR of binary systems at redshift $z = 1$ with different orbital parameters as a function of time before merger, assuming that these binaries are observed from 4 years before merger. With an SNR threshold of 8, equal-mass, circular black hole binary systems with masses of $\sim 10^3 - 10^5 \, M_{\odot}$ become detectable by both LILA-Pioneer and LILA-Horizon more than a year before merger. Eccentricity negatively affects the LILA SNR as the strain is weaker in the deci-Hz band relative to circular binary systems (Figure \ref{fig:strain_evolution_eccentricity}), but highly eccentric binary systems with masses of $\sim 10^3 - 10^5 \, M_{\odot}$ and an initial eccentricity of 0.9999 could still become detectable by both LILA-Pioneer and LILA-Horizon more than a month in advance. 

\begin{figure}[hbt!]
  \centering
  \begin{subfigure}{0.5\linewidth}
    \centering
    \includegraphics[width=\linewidth]{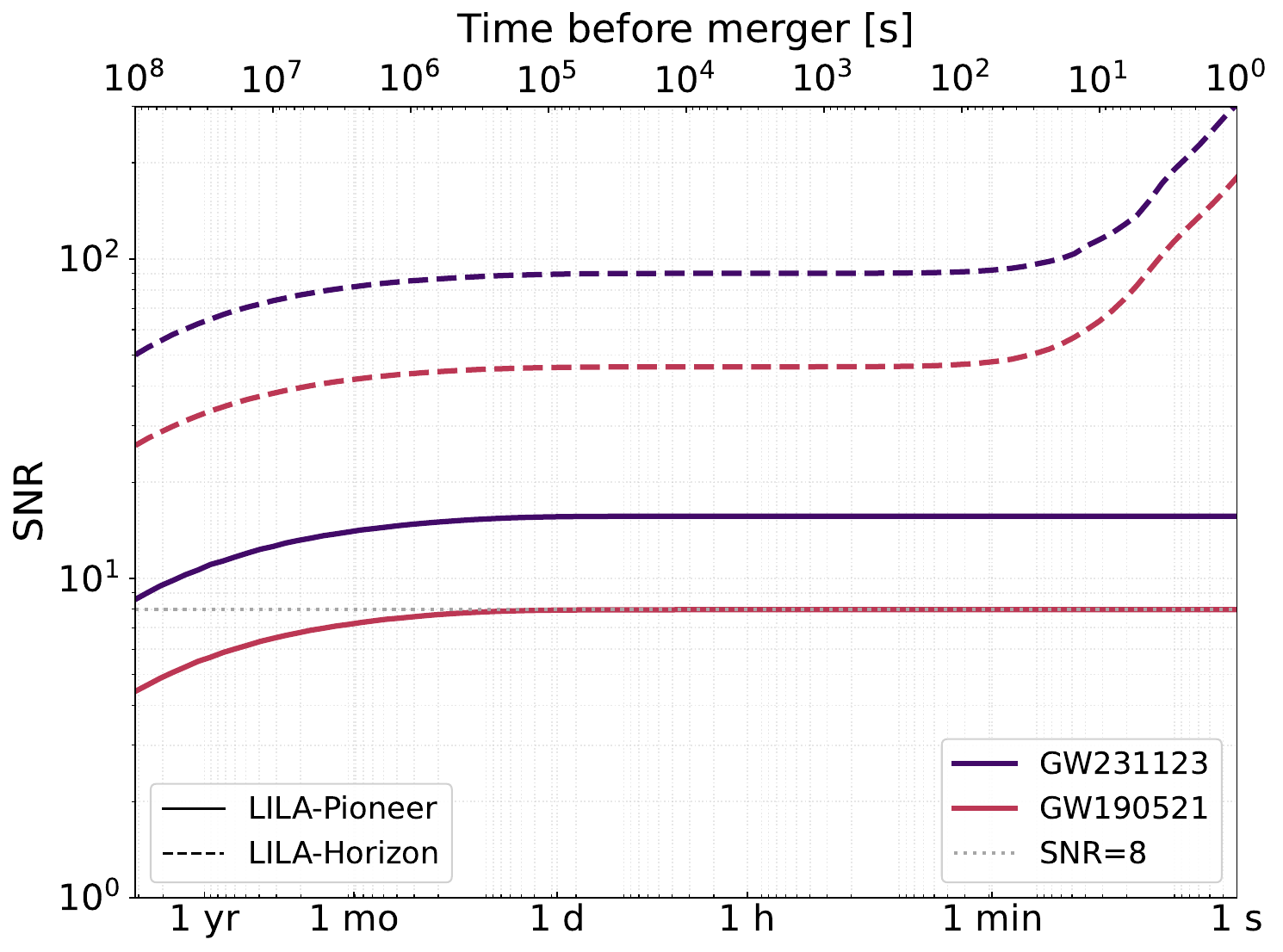}
  \end{subfigure}\hfill
  \begin{subfigure}{0.5\linewidth}
    \centering
    \includegraphics[width=\linewidth]{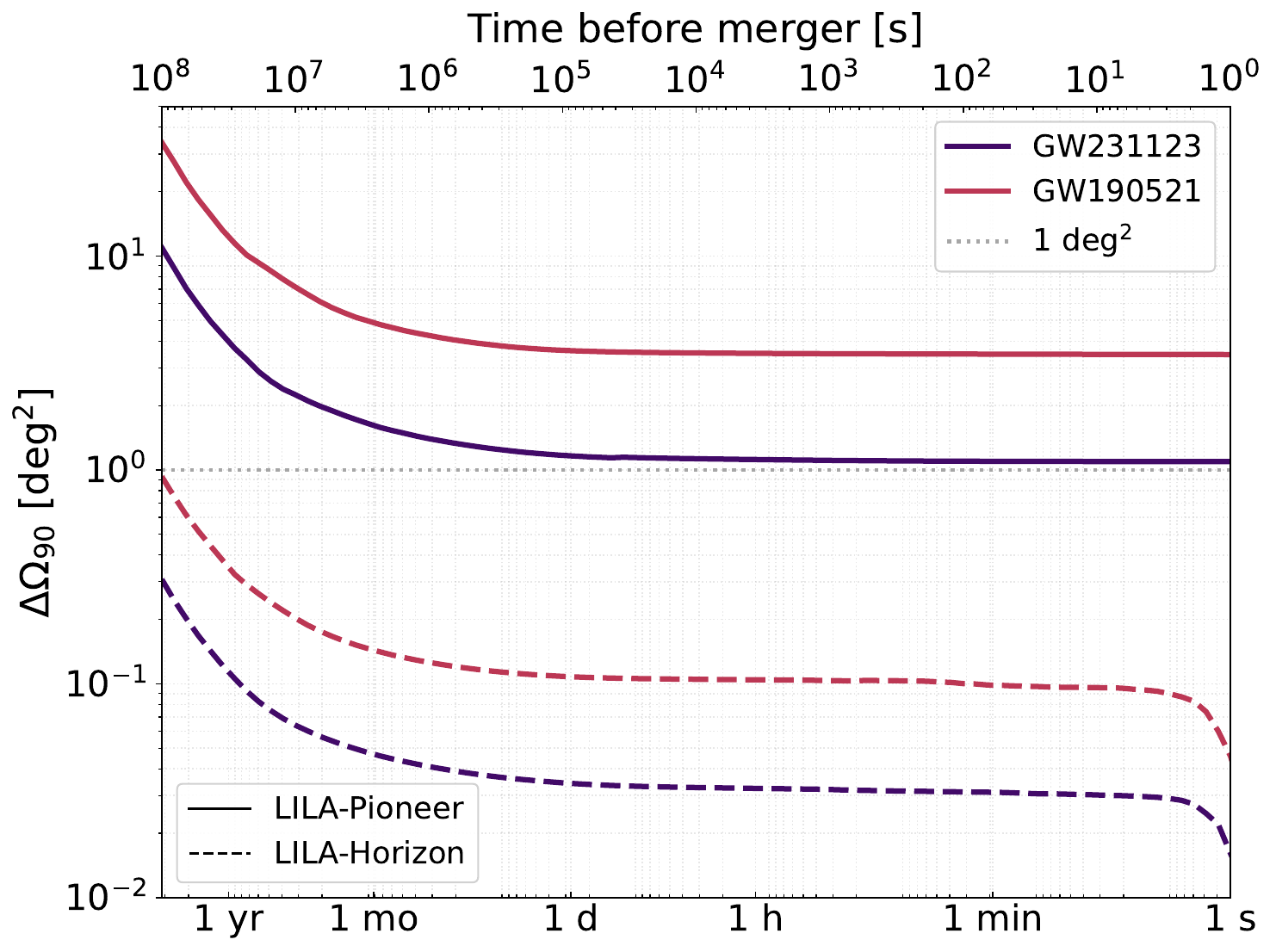}
  \end{subfigure}
  \caption{Cumulative SNR (left) and sky localization error (right) as a function of time before merger for GW190521 ($85 + 66 \, M_{\odot}, z = 0.82$) and GW231123 ($137 + 101 \, M_{\odot}, z = 0.39$), observed by LVK. The horizon dotted line in each panel denotes SNR = 8 (left) and 1 deg$^2$, respectively.}
  \label{fig:skylocerr_EW}
\end{figure}

IMRI systems with a primary mass of $10^4 \, M_{\odot}$ and mass ratios of $M_2/M_1 = 0.0005-0.01$ at redshift $z = 1$ are detectable by LILA-Horizon more than a year (about a month) in advance for an initial orbital eccentricity of 0 (0.9999). Figure \ref{fig:time_to_snr8} further illustrates the early warning detection horizon (with SNR = 8 threshold) of LILA. Non-spinning, equal-mass, circular IMBH binaries within the local Universe ($z < 1$) can be detected by LILA up to 3 years before merger, and IMBH binaries in the early Universe ($z \sim 10-20$) can be detected by LILA from days to about a month before merger, which is crucial for detailed characterization of the first population of massive black holes and their seeds. Additionally, as demonstrated in Figure \ref{fig:skylocerr_EW}, LILA-Pioneer and LILA-Horizon could localize events on the order of $10^{-2}$ to a few square degrees, in contrast to the current localization capability of LIGO on the order of tens of square degrees. Sub-square degree localization would be critical for follow-up observations, especially for multi-messenger sources, such as massive black holes in AGN and circumbinary disk environments.

\subsection{Multi-Band Synergy with Ground-Based Detectors in Probing $\mathit{100-1000 \, M_{\odot}}$ Black Holes}
\label{subsec:multiband_synergy}

\begin{figure}[hbt!]
    \centering
    \includegraphics[width=0.98\linewidth]{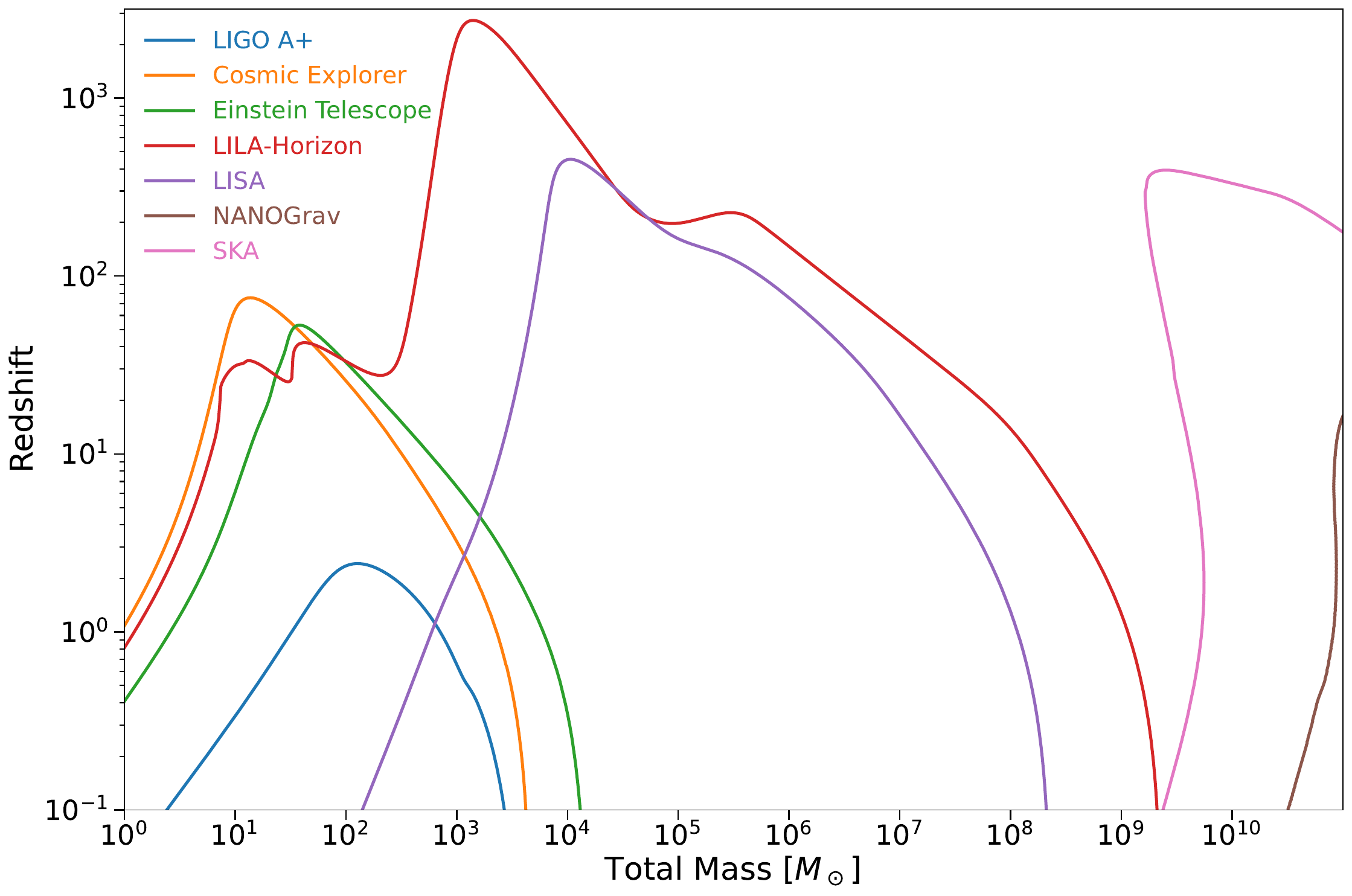}
    \caption{SNR = 8 detection horizon curves for LIGO A+ \citep{Barsotti_2018_Advanced_LIGO_sensitivity}, CE  \citep{Abbott_2017_Cosmic_Explorer}, ET \citep{Hild_2011_ET_sensitivity}, LILA-Horizon \citep{Creighton_2025_LILA_noise_sensitivity}, LISA \citep{Robson_2019_LISA_sensitivity}, 15-year NANOGrav \citep{NANOGrav_15yr_detector_characterization_noise}, and 15-year SKA \citep{Sesana_2008_GW_background_BH_binary_PTA}, assuming non-spinning, equal-mass, circular black hole binary systems initially observed 4 years before merger. The 4-year SNR is only applicable to LILA-Horizon, LISA, NANOGrav, and SKA because the binaries only become sensitive to ground detectors within a minute before their mergers.}
    \label{fig:snr_horizon_multiband}
\end{figure}

With dual-band capability \citep{Shapiro_2025_LILA_vibration_isolation}, LILA-Horizon can extend its detection reach down to light IMBHs with masses of $100-1000 \, M_{\odot}$ and achieve sensitivity comparable to that of future ground detectors like ET and CE for probing this class of black hole binaries. Figure \ref{fig:snr_horizon_multiband} illustrates that current and future GW detectors, both ground-based (LIGO, CE, ET) and space-based (LISA, LILA), and pulsar timing arrays (NANOGrav, SKA) provide complementary coverage across the GW spectrum to uncover the full mass range of black hole binaries across cosmic time throughout the observable Universe. Figure \ref{fig:strain_evolution_circular} shows the evolutionary track of a $100 + 100 \, M_{\odot}$ binary, which enters the deci-Hz band a year in advance and stays within the LILA-Horizon sensitive band for most of its time before reaching the $\gtrsim 10$ Hz band of ground detectors right before merging. Figure \ref{fig:snr_vs_time} shows that a $100 + 100 \, M_{\odot}$ binary inspiral could reach SNR = 8 from weeks to years before merger depending on the initial orbital eccentricity, demonstrating strong prospects for providing early warning for future merger events to be detected by ground detectors.

One motivation for this science case is GW231123 ($137 + 101 \, M_{\odot}$), the most massive black hole merger event discovered so far \citep{GW231123_total_225_Msol}. The initial black hole mass posteriors largely span the mass gap of $\sim 60-130 \, M_{\odot}$ arising from pair-instability supernovae, which do not leave behind a stellar remnant \citep{Farmer_2019_lower_pair_instability_mass_gap, Woosley_2021_pair_instability_mass_gap, Hendriks_2023_pulsational_pair_instability_SNe}. The existence of GW231123 is naturally explained by hierarchical mergers in dense stellar clusters \citep{GW231123_total_225_Msol, Li_Fan_2025_GW231123_hierarchical, Liu_Lai_2025_GW231123_hierarchical, Li_2025_GW231123_hierarchical_10_BH, Passenger_2025_GW231123_hierarchical, Paiella_2025_GW231123_star_clusters_stellar_evolution_hierarchical_mergers, Angeloni_GW231123_formation_channel} with the assistance of environments that keep the intermediate merger remnant from escaping because of recoil kicks, such as AGN disks \citep{Delfavero_2025_GW231123_hierarchical_AGN}. However, the inferred high spin in GW231123 is not likely to be a result of dynamically assembled black holes \citep{Stegmann_2025_GW231123_ancestors}, leading to alternative explanations, such as efficient accretion in an isolated binary \citep{van_Son_2020_pair_instability_super_eddington_isolated, Bartos_2026_GW231123_accretion}, binary evolution of pop III stars avoiding pair instability \citep{Tanikawa_2025_GW231123_pop_III_isolated}, collapse of high-mass rapidly-spinning helium stars \citep{Gottlieb_2025_GW231123_GRMHD, Croon_2026_GW231123_stellar_origin} through chemically homogeneous evolution \citep{Popa_deMink_2025_GW231123_chemically_homogeneous_evolution}, primordial black holes \citep{Yuan_2025_GW231123_primordial_BH, De_Luca_2025_GW231123_primordial_BH}, and cosmic strings \citep{Cuceu_2026_GW231123_cosmic_string}. Uncovering more systems like GW231123 would further shed light on the formation and evolutionary pathways of black hole binaries in the pair-instability mass gap. In addition to GW231123, LVK detectors have discovered a few mergers (such as GW190521; \citealt{GW190521}) and high-significance candidates (see extended gravitational-wave transient catalogs; \citealt{GWTC-3_IMBH_search, GWTC2.1_extended_catalog, GWTC4.0_catalog}) with black hole masses in the pair instability gap and final remnant masses above $100 \, M_{\odot}$ \citep{Ruiz-Rocha_2025_O3_lite_IMBH}. Some of the most massive merger detections and candidates are summarized in Table \ref{table:SNR}, with estimated SNR for future detectors such as CE, ET, LISA, and LILA-Horizon, assuming an initial observation time of 4 years before merger. All of the merger events in Table \ref{table:SNR} would be confidently detected by LILA-Horizon with an SNR $> 100$.

\begin{table}[ht]
\centering
\caption{Summary of primary ($M_1$), secondary ($M_2$), and final remnant ($M_{\rm f}$) masses from LIGO/Virgo black hole binary merger observations and estimated SNR for future detectors, assuming an initial observation time of 4 years before merger for an individual, non-spinning, circular binary system. The 4-year SNR is only applicable to LISA and LILA-Horizon because the binaries only become sensitive to ground detectors within a minute before their mergers. The sources are some of the most massive merger detections and high-significance candidates so far: GW190521 \citep{GW190521}, GW190426$\_$190642 \citep{GWTC2.1_extended_catalog}, GW190403$\_$051519 \citep{GWTC2.1_extended_catalog}, GW231123 \citep{GW231123_total_225_Msol}, GW231028 \citep{GWTC4.0_catalog}, GW230704$\_$212616 \citep{GWTC4.0_catalog}.}
\label{tab:snr_summary}
\begin{tabular}{l|ccccc|cccc}
\hline
\textbf{Events} &
\multicolumn{5}{c|}{\textbf{LIGO/Virgo Observations}} &
\multicolumn{4}{c}{\textbf{Simulated SNR}} \\
\cline{2-6}\cline{7-10}
 & $M_1 \, [M_{\odot}]$ & $M_2 \, [M_{\odot}]$ & $M_{\rm f} \, [M_{\odot}]$ & $z$ & SNR
 & CE & ET & LISA & LILA-Horizon \\
\hline
GW190521 & 85 & 66 & 142 & 0.82 & 14.2 & 643 & 254 & 2 & 525 \\
GW190426$\_$190642 & 106 & 76 & 173 & 0.73 & 8.7 & 820 & 319 & 2 & 631 \\
GW190403$\_$051519 & 85 & 20 & 102 & 1.18 & 7.6 & 278 & 110 & 0.5 & 227  \\
GW231123 & 137 & 101 & 222 & 0.39 & 20.7 & 1814 & 702 & 5 & 1363 \\
GW231028 & 95 & 58 & 144 & 0.67 & 21 & 758 & 301 & 2 & 630 \\
GW230704$\_$212616 & 89 & 49 & 132 & 1.1 & 8 & 449 & 176 & 1 & 354 \\
\hline
\end{tabular}
\label{table:SNR}
\end{table}

At the current sensitivity of LIGO, discovering GW231123-like events is exceptionally rare, as the merger rate density upper limit for $100 + 100 \, M_{\odot}$ binaries is empirically constrained to be about $0.06$ year$^{-1}$ Gpc$^{-3}$ \citep{GWTC-3_IMBH_search}. LILA-Horizon can confidently discover GW231123-like events out to $z \sim 5$ ($\sim 8$ Gpc) with an SNR $> 80$, which translates into an observational volume of about 6,400 Gpc$^3$, or about 400 events every year assuming the merger rate density upper limit quoted earlier. In addition, the empirical GW190521-like ($85 + 66 \, M_{\odot}$) merger rate is inferred to be $0.08^{+0.19}_{-0.07}$ year$^{-1}$ Gpc$^{-3}$. LILA-Horizon can confidently discover GW190521-like events out to $z \sim 5$ ($\sim 8$ Gpc) with an SNR $> 50$, which translates to an estimated GW190521-like merger rate of $504^{+1198}_{-441}$ year$^{-1}$ Gpc$^{-3}$. While this rough estimate only extrapolates from LIGO-Virgo detection rate to high redshift and does not account for the formation rate of GW231123 and GW190521-like binaries throughout cosmic time, it suggests a promising pathway towards detecting many more pair-instability mass gap events and lower-range IMBHs with LILA. The sensitivity improvement in next-generation ground detectors like CE and ET will further complement deci-Hz detectors like LILA in uncovering a larger fraction of this black hole binary demographic, allowing statistical analysis of the formation mechanisms of black holes found in the pair-instability gap and expanding the upper envelope of discovered black holes with stellar origin to masses $\gtrsim 250 \, M_{\odot}$. High-SNR detections with LILA and next-generation ground detectors, together with the multi-band synergy between these facilities, will also provide an internal consistency test of GR through independent waveform reconstruction and binary parameter estimation.

\subsection{Testing GR}
\label{subsec:GR_tests}
Figure \ref{fig:snr_horizon} shows that mergers of black hole masses of $\sim 10^4 \, M_{\odot}$ out to redshift $z \sim 10$ and $z \sim 100$ can be detected by LILA-Pioneer and LILA-Horizon with an SNR of 100, respectively. Such a high SNR detection would enable a wide range of novel strong-field tests of gravity. Recently, the GW250114 merger event has been detected by LIGO with an unprecedented SNR of 80, resulting in the experimental confirmation of Hawking's area theorem \citep{Hawking_1971_radiation_mergers} and the Kerr nature of black holes \citep{GW250114_Hawking_area_Kerr_nature}. This strong SNR signal allows detailed black hole spectroscopy analysis of the quasi-normal modes during the ringdown phase \citep{Yang_2025_GW250114_nonlinear_QNM, GW250114_spectroscopy_GR_test, Wang_2026_GW250114_nonlinear_ringdown}, enabling stringent verification of GR, the Kerr nature of black holes, and the presence of an event horizon \citep{, Chandra_2025_GW250114_strong_field_test, Akyuz_2025_GW250114_science, Lu_2025_GW250114_horizon_signatures, Grimaldi_2026_GW250114_plunge_merger_ringdown_GR_test}, as well as constraining possible deviations from GR and modified gravity theories \citep{Guo_2025_GW250114_modified_gravity, Andres-Carcasona_2025_GW250114_no_Love}. Hence, LILA detectors have strong potential to provide even more rigorous tests of GR and modified gravity theories by uncovering many more black hole merger events with similar or higher SNR than GW250114. Recently, the fourth observing run of LVK detectors has doubled the number of confident detections (defined to be measured in at least two detectors with false alarm rates $\leq 10^{-3} \, \textrm{yr}^{-1}$), allowing highly sensitive tests of GR with some of the most stringent constraints on deviations from post-Newtonian coefficients in GR \citep{GWTC4.0_GR_tests_I_overview, GWTC4.0_GR_tests_II_parametrized, GWTC4.0_GR_tests_III_remnants}. Alongside next-generation GW detectors, with a detection horizon towards most of the observable Universe (Subsection \ref{subsec:LILA_observational_horizon}), LILA will substantially expand the number of confident high-SNR events to continue pushing experimental tests of GR to their limits.

LILA's ability to test gravity is not limited to the strong-field regime. As shown in Figure \ref{fig:strain_evolution_circular} and \ref{fig:strain_evolution_eccentricity}, as the IMBH binary strain enters the deci-Hz band up to months and years before merger, LILA can potentially track hundreds to thousands of GW signals during the inspiral. Figure \ref{fig:snr_vs_time} demonstrates that many IMBH binaries and IMRI systems can be detected by LILA detectors with an SNR of 8 up to months or years in advance, depending on the individual masses, mass ratio, and initial orbital eccentricity, making LILA sensitive to deviations from GR in the post-Newtonian expansion of the waveforms, enabling tests of modified gravity theories and searches for signatures of new physics \citep{Nair_2019_GWTC-1_higher_curvature, Das_2024_non_Kerr_inspirals, ACA_2024_EMRI_gravity_test, Zi_2025_eccentric_EMRI_ultra-light_vector, Speri_2026_EMRI_scalar_charge, Seymour_2026_inspiral_test_GR}.



\section{Conclusion}
\label{sec:conclusion}

The proposed lunar-based LILA will take full advantage of the lack of seismic, atmospheric, and anthropogenic noise on the Moon to access the deci-Hz GW band and uncover a new population of GW sources with IMBH binaries and IMRI systems, potentially observing binary inspirals months to years before merger. The early warning capacity is essential for multi-messenger and multi-band follow-up, as well as detailed characterization of the astrophysical environments and formation pathways of black hole binaries. The initial pathfinder mission, LILA-Pioneer, could already uncover a large fraction of IMBHs in the very early Universe ($z > 10$), crucial to understanding the formation and evolution of the first massive black hole population. The follow-up LILA-Horizon detector will deliver transformative milli-Hz to kilo-Hz broadband GW sensitivity and open up new multi-band opportunities with next-generation ground-based detectors. The key takeaways of this work on characterizing the detection landscape and proposing science cases for LILA are the following:

\begin{itemize}
    \item \textbf{Discovering IMBH Binaries (Section \ref{subsec:LILA_observational_horizon}):} LILA is capable of detecting black hole binaries across a wide mass range, with its peak sensitivity in the deci-Hz GW band corresponding to IMBHs of $\sim 10^3-10^6 \, M_{\odot}$ (Figures \ref{fig:strain_evolution_circular}--\ref{fig:snr_horizon}).
    \item \textbf{Massive Black Hole Seeding $\&$ Growth Models (Section \ref{subsec:LILA_observational_horizon}):} With an observational period of 4 years, LILA can extend its IMBH detection horizon to the very early Universe beyond our current limit (JWST at $z \sim 14$), directly probing the first population of massive black holes in the seeding epoch ($z \sim 20-30$), with important implications for massive black hole formation and evolution through mapping the cosmological black hole mass and spin distributions throughout cosmic time (Figure \ref{fig:snr_horizon}).
    \item \textbf{Discovering Eccentric Binaries (Section \ref{subsec:eccentric}):} The milli-to-deci-Hz sensitivity band of LILA enables the discovery of black hole binaries some time before mergers with measurable eccentricity residuals retained from their formation (e.g., dynamical channels like dense stellar clusters), probing astrophysical environments and evolutionary pathways of black hole binaries (Figure \ref{fig:strain_evolution_eccentricity}).
    \item \textbf{Discovering IMRIs (Section \ref{subsec:IMRIs}):} LILA-Horizon could confidently detect IMRI systems with mass $\sim 10^5 \, M_{\odot}$ and mass ratio of $\sim 10^{-4}-10^{-2}$ towards the very early Universe ($z > 10$), providing another pathway for discovering IMBHs and their astrophysical environments (Figure \ref{fig:mass_ratio_horizon}). With high sensitivity in the deci-Hz band, LILA-Horizon could potentially detect $1000 + 30 \, M_{\odot}$-like IMRI systems with a rate of 8 yr$^{-1}$, two orders of magnitude above the LISA rate.
    \item \textbf{Early Warning Potential (Section \ref{subsec:early_warning}):} IMBH binaries within the local Universe ($z \lesssim 1$) can be detected by LILA more than a year before merger (Figure \ref{fig:snr_vs_time}), while IMBH binaries in the early Universe ($z \sim 10-20$) can be detected by LILA up to about a month before merger (Figure \ref{fig:time_to_snr8}). LILA also enables sky localization on the order of $10^{-2}$ to a few square degrees, crucial for multi-messenger follow-up (Figure \ref{fig:skylocerr_EW}). 
    \item \textbf{Multi-Band GW Astronomy (Section \ref{subsec:multiband_synergy}):} Current and future GW detectors, both ground-based (LIGO, CE, ET) and space-based (LISA, LILA), and pulsar timing arrays (NANOGrav, SKA) provide complementary coverage across the GW spectrum and the mass range of black hole binaries across cosmic time (Figure \ref{fig:snr_horizon_multiband}). LILA-Horizon will expand current LVK searches of GW231123-like events by improving the $100 + 100 \, M_{\odot}$ binary detection rate by two orders of magnitude, further probing black holes in the pair-instability mass gap (Table \ref{tab:snr_summary}) and hierarchical mergers. With next-generation ground detectors (CE, ET), LILA will also expand the upper envelope of discovered black holes with stellar origin to masses $\gtrsim 250 \, M_{\odot}$.
    \item \textbf{Testing GR (Section \ref{subsec:GR_tests}):} The ability to detect IMBH binaries and their mergers with high SNR $\gtrsim 100$ (Figure \ref{fig:snr_horizon}) will allow LILA to perform strong-field tests of gravity. The multi-band synergy between LILA and other future facilities like LISA, CE, and ET will also provide an internal consistency test of GR through independent waveform reconstruction and binary parameter estimation.
\end{itemize}
Future work could extend the LILA selection function presented in this paper to make astrophysical predictions, such as expected LILA detection rates and redshift evolution of black hole merger rates, using semi-analytical models (e.g., \citealt{Volonteri_2008, Barausse_2012, Pacucci_2015, Valiante_2018, Dayal_2019, Sassano_2021, Trinca_2023, Evans_2025, Ricarte_2025_SEROTINA_spin_evolution, Bonoli_2025_L-GalaxiesBH, Davari_2026_merger_rate_CAT}) and cosmological simulations (e.g., \citealt{Di_Matteo_2012, Vogelsberger_2014, Khandai_2015, Somerville_Dave_2015, Tremmel_2017, Griffin_2019, Nelson_2019, Habouzit_2021, Bhowmick_2024, Jeon_2025_JWST_SMBH_pathways}) with realistic black hole population models, seeding mechanisms, and evolutionary pathways. In addition, future forecasts should incorporate dynamical models of dense stellar clusters, since cluster dynamics can strongly shape stellar binary populations through disruption, hardening, and mass segregation (e.g., \citealt{Heggie_1975_binary_evolution_stellar_dynamics, Phillips_2026_binary_stream_cluster, Wu_2026_binary_stellar_clusters, O_Connor_2026_binary_population_GC, Bruce_2026_binary_evolution_GC}), potentially playing an important role in the formation of compact binaries with IMBHs in these environments and affecting merger rates of IMBH binaries and IMRI systems relevant for LILA (e.g., \citealt{Di_Carlo_2021_IMBH_star_clusters, Purohit_2024_IMBH_collisional_runaway, Lee_2025_IMBH_star_cluster_formation_evolution, Vergara_2025_IMBH_runaway_collisions, Rantala_2026_IMBH_formation_runaway_metallicity, Mestichelli_2026_IMBH_seeding_stellar_collisions, Paiella_2026_BH_growth_stellar_clusters}). Being sensitive to the IMBH mass range back to the seeding epoch of $z \sim 20-30$, LILA is uniquely positioned to bridge the evolutionary gap between the earliest black hole seeds and current SMBH demographics and complement ongoing observational searches for IMBHs in galaxy nuclei and globular clusters. Mapping out the black hole mass and spin distribution detectable with LILA across cosmic time will help inform future searches for signatures of massive black hole formation and guide the mission lifetime for LILA.

%
%

\ack{T. N. would like to acknowledge Michael D. Johnson and Erandi Chavez for providing valuable input and discussion on this work. This work used \texttt{numpy} \citep{Numpy}, \texttt{matplotlib} \citep{matplotlib}, \texttt{astropy} \citep{Astropy}, and \texttt{gwent} \citep{Kaiser_McWilliams_2011_sensitivity_GW_spectrum}.}

\funding{A. B. Y., R. N., and K. J.’s work was supported in part by the Lunar Labs Initiative at Vanderbilt University, which is funded by the Vanderbilt Scaling Grant from the Office of the Vice Provost for Research and Innovation and the John Templeton Foundation. T. N. was supported in part by the Astronomy Department at Harvard University. T. N. and A. R.'s work was supported in part by the Black Hole Initiative at Harvard University, which is funded by the Gordon and Betty Moore Foundation (Grant \#13526) and the John Templeton Foundation (Grant \#63445). A. R. acknowledges financial support from the Gordon and Betty Moore Foundation (Grant GBMF-12987) and the National Science Foundation (AST-2307887). The opinions expressed in this publication are those of the author(s) and do not necessarily reflect the views of these Foundations.}



\data{The data that support the findings of this study are available from the corresponding author, T. N., upon reasonable request.}


\bibliographystyle{aasjournal}
\bibliography{bh_cosmo, eht, ms}

@ARTICLE{King_2006,
       author = {{King}, A.~R. and {Pringle}, J.~E.},
        title = "{Growing supermassive black holes by chaotic accretion}",
      journal = {\mnras},
         year = "2006",
        month = "Nov",
       volume = {373},
       number = {1},
        pages = {L90-L92},
          doi = {10.1111/j.1745-3933.2006.00249.x},
archivePrefix = {arXiv},
       eprint = {astro-ph/0609598},
 primaryClass = {astro-ph},
       adsurl = {https://ui.adsabs.harvard.edu/abs/2006MNRAS.373L..90K}
}

@ARTICLE{Barausse_2012,
       author = {{Barausse}, Enrico},
        title = "{The evolution of massive black holes and their spins in their galactic hosts}",
      journal = {\mnras},
         year = "2012",
        month = "Jul",
       volume = {423},
       number = {3},
        pages = {2533-2557},
          doi = {10.1111/j.1365-2966.2012.21057.x},
archivePrefix = {arXiv},
       eprint = {1201.5888},
 primaryClass = {astro-ph.CO},
       adsurl = {https://ui.adsabs.harvard.edu/abs/2012MNRAS.423.2533B}
}

@ARTICLE{BL_2000,
       author = {{Barkana}, Rennan and {Loeb}, Abraham},
        title = "{High-Redshift Galaxies: Their Predicted Size and Surface Brightness
        Distributions and Their Gravitational Lensing Probability}",
      journal = {\apj},
         year = 2000,
        month = Mar,
       volume = {531},
        pages = {613-623},
          doi = {10.1086/308503},
archivePrefix = {arXiv},
       eprint = {astro-ph/9906398},
 primaryClass = {astro-ph},
       adsurl = {https://ui.adsabs.harvard.edu/#abs/2000ApJ...531..613B}
}

@ARTICLE{Pacucci_2022_search,
       author = {{Pacucci}, Fabio and {Loeb}, Abraham},
        title = "{The search for the farthest quasar: consequences for black hole growth and seed models}",
      journal = {\mnras},
         year = 2022,
        month = jan,
       volume = {509},
       number = {2},
        pages = {1885-1891},
          doi = {10.1093/mnras/stab3071},
archivePrefix = {arXiv},
       eprint = {2110.10176},
 primaryClass = {astro-ph.GA},
       adsurl = {https://ui.adsabs.harvard.edu/abs/2022MNRAS.509.1885P}
}

@ARTICLE{Fan_2023,
       author = {{Fan}, Xiaohui and {Ba{\~n}ados}, Eduardo and {Simcoe}, Robert A.},
        title = "{Quasars and the Intergalactic Medium at Cosmic Dawn}",
      journal = {\araa},
         year = 2023,
        month = aug,
       volume = {61},
        pages = {373-426},
          doi = {10.1146/annurev-astro-052920-102455},
archivePrefix = {arXiv},
       eprint = {2212.06907},
 primaryClass = {astro-ph.GA},
       adsurl = {https://ui.adsabs.harvard.edu/abs/2023ARA&A..61..373F}
}

@ARTICLE{Kocevski_2023,
       author = {{Kocevski}, Dale D. and {Onoue}, Masafusa and {Inayoshi}, Kohei and {Trump}, Jonathan R. and {Arrabal Haro}, Pablo and {Grazian}, Andrea and {Dickinson}, Mark and {Finkelstein}, Steven L. and {Kartaltepe}, Jeyhan S. and {Hirschmann}, Michaela and {Aird}, James and {Holwerda}, Benne W. and {Fujimoto}, Seiji and {Juneau}, St{\'e}phanie and {Amor{\'\i}n}, Ricardo O. and {Backhaus}, Bren E. and {Bagley}, Micaela B. and {Barro}, Guillermo and {Bell}, Eric F. and {Bisigello}, Laura and {Calabr{\`o}}, Antonello and {Cleri}, Nikko J. and {Cooper}, M.~C. and {Ding}, Xuheng and {Grogin}, Norman A. and {Ho}, Luis C. and {Hutchison}, Taylor A. and {Inoue}, Akio K. and {Jiang}, Linhua and {Jones}, Brenda and {Koekemoer}, Anton M. and {Li}, Wenxiu and {Li}, Zhengrong and {McGrath}, Elizabeth J. and {Molina}, Juan and {Papovich}, Casey and {P{\'e}rez-Gonz{\'a}lez}, Pablo G. and {Pirzkal}, Nor and {Wilkins}, Stephen M. and {Yang}, Guang and {Yung}, L.~Y. Aaron},
        title = "{Hidden Little Monsters: Spectroscopic Identification of Low-mass, Broad-line AGNs at z > 5 with CEERS}",
      journal = {\apjl},
         year = 2023,
        month = sep,
       volume = {954},
       number = {1},
          eid = {L4},
        pages = {L4},
          doi = {10.3847/2041-8213/ace5a0},
archivePrefix = {arXiv},
       eprint = {2302.00012},
 primaryClass = {astro-ph.GA},
       adsurl = {https://ui.adsabs.harvard.edu/abs/2023ApJ...954L...4K}
}

@ARTICLE{Harikane_2023,
       author = {{Harikane}, Yuichi and {Zhang}, Yechi and {Nakajima}, Kimihiko and {Ouchi}, Masami and {Isobe}, Yuki and {Ono}, Yoshiaki and {Hatano}, Shun and {Xu}, Yi and {Umeda}, Hiroya},
        title = "{A JWST/NIRSpec First Census of Broad-line AGNs at z = 4-7: Detection of 10 Faint AGNs with M $_{BH}$ {}10$^{6}$-{}10$^{8}$ M $_{{\ensuremath{\odot}}}$ and Their Host Galaxy Properties}",
      journal = {\apj},
         year = 2023,
        month = dec,
       volume = {959},
       number = {1},
          eid = {39},
        pages = {39},
          doi = {10.3847/1538-4357/ad029e},
archivePrefix = {arXiv},
       eprint = {2303.11946},
 primaryClass = {astro-ph.GA},
       adsurl = {https://ui.adsabs.harvard.edu/abs/2023ApJ...959...39H}
}

@ARTICLE{LISA_2017,
       author = {{Amaro-Seoane}, Pau and {Audley}, Heather and {Babak}, Stanislav and
        {Baker}, John and {Barausse}, Enrico and {Bender}, Peter and
        {Berti}, Emanuele and {Binetruy}, Pierre and {Born}, Michael and
        {Bortoluzzi}, Daniele and {Camp}, Jordan and {Caprini}, Chiara
        and {Cardoso}, Vitor and {Colpi}, Monica and {Conklin}, John and
        {Cornish}, Neil and {Cutler}, Curt and {Danzmann}, Karsten and
        {Dolesi}, Rita and {Ferraioli}, Luigi and {Ferroni}, Valerio and
        {Fitzsimons}, Ewan and {Gair}, Jonathan and {Gesa Bote}, Lluis
        and {Giardini}, Domenico and {Gibert}, Ferran and {Grimani},
        Catia and {Halloin}, Hubert and {Heinzel}, Gerhard and {Hertog},
        Thomas and {Hewitson}, Martin and {Holley-Bockelmann}, Kelly and
        {Hollington}, Daniel and {Hueller}, Mauro and {Inchauspe}, Henri
        and {Jetzer}, Philippe and {Karnesis}, Nikos and {Killow},
        Christian and {Klein}, Antoine and {Klipstein}, Bill and
        {Korsakova}, Natalia and {Larson}, Shane L and {Livas}, Jeffrey
        and {Lloro}, Ivan and {Man}, Nary and {Mance}, Davor and
        {Martino}, Joseph and {Mateos}, Ignacio and {McKenzie}, Kirk and
        {McWilliams}, Sean T and {Miller}, Cole and {Mueller}, Guido and
        {Nardini}, Germano and {Nelemans}, Gijs and {Nofrarias}, Miquel
        and {Petiteau}, Antoine and {Pivato}, Paolo and {Plagnol}, Eric
        and {Porter}, Ed and {Reiche}, Jens and {Robertson}, David and
        {Robertson}, Norna and {Rossi}, Elena and {Russano}, Giuliana
        and {Schutz}, Bernard and {Sesana}, Alberto and {Shoemaker},
        David and {Slutsky}, Jacob and {Sopuerta}, Carlos F. and
        {Sumner}, Tim and {Tamanini}, Nicola and {Thorpe}, Ira and
        {Troebs}, Michael and {Vallisneri}, Michele and {Vecchio},
        Alberto and {Vetrugno}, Daniele and {Vitale}, Stefano and
        {Volonteri}, Marta and {Wanner}, Gudrun and {Ward}, Harry and
        {Wass}, Peter and {Weber}, William and {Ziemer}, John and
        {Zweifel}, Peter},
        title = "{Laser Interferometer Space Antenna}",
      journal = {ArXiv e-prints},
         year = 2017,
        month = Feb,
          eid = {arXiv:1702.00786},
        pages = {arXiv:1702.00786},
archivePrefix = {arXiv},
       eprint = {1702.00786},
       adsurl = {https://ui.adsabs.harvard.edu/#abs/2017arXiv170200786A}
}

@ARTICLE{Chilingarian_2018,
       author = {{Chilingarian}, Igor V. and {Katkov}, Ivan Yu. and {Zolotukhin}, Ivan
        Yu. and {Grishin}, Kirill A. and {Beletsky}, Yuri and {Boutsia},
        Konstantina and {Osip}, David J.},
        title = "{A Population of Bona Fide Intermediate Mass Black Holes Identified as
        Low Luminosity Active Galactic Nuclei}",
      journal = {ArXiv e-prints},
         year = 2018,
        month = May,
archivePrefix = {arXiv},
       eprint = {1805.01467},
       adsurl = {https://ui.adsabs.harvard.edu/#abs/2018arXiv180501467C}
}

@ARTICLE{McConnell_Ma_2013,
   author = {{McConnell}, N.~J. and {Ma}, C.-P.},
    title = "{Revisiting the Scaling Relations of Black Hole Masses and Host Galaxy Properties}",
  journal = {\apj},
archivePrefix = "arXiv",
   eprint = {1211.2816},
     year = 2013,
    month = feb,
   volume = 764,
      eid = {184},
    pages = {184},
      doi = {10.1088/0004-637X/764/2/184},
   adsurl = {http://adsabs.harvard.edu/abs/2013ApJ...764..184M}
}

@ARTICLE{Dubois_2014,
       author = {{Dubois}, Yohan and {Volonteri}, Marta and {Silk}, Joseph},
        title = "{Black hole evolution - III. Statistical properties of mass growth and
        spin evolution using large-scale hydrodynamical cosmological
        simulations}",
      journal = {\mnras},
         year = 2014,
        month = May,
       volume = {440},
        pages = {1590-1606},
          doi = {10.1093/mnras/stu373},
       adsurl = {https://ui.adsabs.harvard.edu/#abs/2014MNRAS.440.1590D}
}

@ARTICLE{Ricarte_2018,
       author = {{Ricarte}, Angelo and {Natarajan}, Priyamvada},
        title = "{Exploring SMBH assembly with semi-analytic modelling}",
      journal = {\mnras},
         year = 2018,
        month = Feb,
       volume = {474},
        pages = {1995-2011},
          doi = {10.1093/mnras/stx2851},
       adsurl = {https://ui.adsabs.harvard.edu/#abs/2018MNRAS.474.1995R}
}

@ARTICLE{Planck_2018,
   author = {{Planck Collaboration} and {Aghanim}, N. and {Akrami}, Y. and 
	{Ashdown}, M. and {Aumont}, J. and {Baccigalupi}, C. and {Ballardini}, M. and 
	{Banday}, A.~J. and {Barreiro}, R.~B. and {Bartolo}, N. and 
	{Basak}, S. and {Battye}, R. and {Benabed}, K. and {Bernard}, J.-P. and 
	{Bersanelli}, M. and {Bielewicz}, P. and {Bock}, J.~J. and {Bond}, J.~R. and 
	{Borrill}, J. and {Bouchet}, F.~R. and {Boulanger}, F. and {Bucher}, M. and 
	{Burigana}, C. and {Butler}, R.~C. and {Calabrese}, E. and {Cardoso}, J.-F. and 
	{Carron}, J. and {Challinor}, A. and {Chiang}, H.~C. and {Chluba}, J. and 
	{Colombo}, L.~P.~L. and {Combet}, C. and {Contreras}, D. and 
	{Crill}, B.~P. and {Cuttaia}, F. and {de Bernardis}, P. and 
	{de Zotti}, G. and {Delabrouille}, J. and {Delouis}, J.-M. and 
	{Di Valentino}, E. and {Diego}, J.~M. and {Dor{\'e}}, O. and 
	{Douspis}, M. and {Ducout}, A. and {Dupac}, X. and {Dusini}, S. and 
	{Efstathiou}, G. and {Elsner}, F. and {En{\ss}lin}, T.~A. and 
	{Eriksen}, H.~K. and {Fantaye}, Y. and {Farhang}, M. and {Fergusson}, J. and 
	{Fernandez-Cobos}, R. and {Finelli}, F. and {Forastieri}, F. and 
	{Frailis}, M. and {Franceschi}, E. and {Frolov}, A. and {Galeotta}, S. and 
	{Galli}, S. and {Ganga}, K. and {G{\'e}nova-Santos}, R.~T. and 
	{Gerbino}, M. and {Ghosh}, T. and {Gonz{\'a}lez-Nuevo}, J. and 
	{G{\'o}rski}, K.~M. and {Gratton}, S. and {Gruppuso}, A. and 
	{Gudmundsson}, J.~E. and {Hamann}, J. and {Handley}, W. and 
	{Herranz}, D. and {Hivon}, E. and {Huang}, Z. and {Jaffe}, A.~H. and 
	{Jones}, W.~C. and {Karakci}, A. and {Keih{\"a}nen}, E. and 
	{Keskitalo}, R. and {Kiiveri}, K. and {Kim}, J. and {Kisner}, T.~S. and 
	{Knox}, L. and {Krachmalnicoff}, N. and {Kunz}, M. and {Kurki-Suonio}, H. and 
	{Lagache}, G. and {Lamarre}, J.-M. and {Lasenby}, A. and {Lattanzi}, M. and 
	{Lawrence}, C.~R. and {Le Jeune}, M. and {Lemos}, P. and {Lesgourgues}, J. and 
	{Levrier}, F. and {Lewis}, A. and {Liguori}, M. and {Lilje}, P.~B. and 
	{Lilley}, M. and {Lindholm}, V. and {L{\'o}pez-Caniego}, M. and 
	{Lubin}, P.~M. and {Ma}, Y.-Z. and {Mac{\'{\i}}as-P{\'e}rez}, J.~F. and 
	{Maggio}, G. and {Maino}, D. and {Mandolesi}, N. and {Mangilli}, A. and 
	{Marcos-Caballero}, A. and {Maris}, M. and {Martin}, P.~G. and 
	{Martinelli}, M. and {Mart{\'{\i}}nez-Gonz{\'a}lez}, E. and 
	{Matarrese}, S. and {Mauri}, N. and {McEwen}, J.~D. and {Meinhold}, P.~R. and 
	{Melchiorri}, A. and {Mennella}, A. and {Migliaccio}, M. and 
	{Millea}, M. and {Mitra}, S. and {Miville-Desch{\^e}nes}, M.-A. and 
	{Molinari}, D. and {Montier}, L. and {Morgante}, G. and {Moss}, A. and 
	{Natoli}, P. and {N{\o}rgaard-Nielsen}, H.~U. and {Pagano}, L. and 
	{Paoletti}, D. and {Partridge}, B. and {Patanchon}, G. and {Peiris}, H.~V. and 
	{Perrotta}, F. and {Pettorino}, V. and {Piacentini}, F. and 
	{Polastri}, L. and {Polenta}, G. and {Puget}, J.-L. and {Rachen}, J.~P. and 
	{Reinecke}, M. and {Remazeilles}, M. and {Renzi}, A. and {Rocha}, G. and 
	{Rosset}, C. and {Roudier}, G. and {Rubi{\~n}o-Mart{\'{\i}}n}, J.~A. and 
	{Ruiz-Granados}, B. and {Salvati}, L. and {Sandri}, M. and {Savelainen}, M. and 
	{Scott}, D. and {Shellard}, E.~P.~S. and {Sirignano}, C. and 
	{Sirri}, G. and {Spencer}, L.~D. and {Sunyaev}, R. and {Suur-Uski}, A.-S. and 
	{Tauber}, J.~A. and {Tavagnacco}, D. and {Tenti}, M. and {Toffolatti}, L. and 
	{Tomasi}, M. and {Trombetti}, T. and {Valenziano}, L. and {Valiviita}, J. and 
	{Van Tent}, B. and {Vibert}, L. and {Vielva}, P. and {Villa}, F. and 
	{Vittorio}, N. and {Wandelt}, B.~D. and {Wehus}, I.~K. and {White}, M. and 
	{White}, S.~D.~M. and {Zacchei}, A. and {Zonca}, A.},
    title = "{Planck 2018 results. VI. Cosmological parameters}",
  journal = {ArXiv e-prints},
archivePrefix = "arXiv",
   eprint = {1807.06209},
     year = 2018,
    month = jul,
   adsurl = {http://adsabs.harvard.edu/abs/2018arXiv180706209P}
}

@ARTICLE{Baldassare_2015,
       author = {{Baldassare}, Vivienne F. and {Reines}, Amy E. and {Gallo}, Elena and
        {Greene}, Jenny E.},
        title = "{A ̃50,000 M<SUB>☉</SUB> Solar Mass Black Hole in the Nucleus of RGG 118}",
      journal = {\apj},
         year = 2015,
        month = Aug,
       volume = {809},
          doi = {10.1088/2041-8205/809/1/L14},
       adsurl = {https://ui.adsabs.harvard.edu/#abs/2015ApJ...809L..14B}
}

@ARTICLE{Kormendy_Ho_2013,
       author = {{Kormendy}, John and {Ho}, Luis C.},
        title = "{Coevolution (Or Not) of Supermassive Black Holes and Host Galaxies}",
      journal = {Annual Review of Astronomy and Astrophysics},
         year = 2013,
        month = Aug,
       volume = {51},
        pages = {511-653},
          doi = {10.1146/annurev-astro-082708-101811},
       adsurl = {https://ui.adsabs.harvard.edu/#abs/2013ARA&A..51..511K}
}

@ARTICLE{Boekholt_2018,
       author = {{Boekholt}, T.~C.~N. and {Schleicher}, D.~R.~G. and {Fellhauer}, M. and
         {Klessen}, R.~S. and {Reinoso}, B. and {Stutz}, A.~M. and
         {Haemmerl{\'e}}, L.},
        title = "{Formation of massive seed black holes via collisions and accretion}",
      journal = {\mnras},
         year = 2018,
        month = may,
       volume = {476},
       number = {1},
        pages = {366-380},
          doi = {10.1093/mnras/sty208},
archivePrefix = {arXiv},
       eprint = {1801.05841},
 primaryClass = {astro-ph.GA},
       adsurl = {https://ui.adsabs.harvard.edu/abs/2018MNRAS.476..366B}
}

@ARTICLE{Bromm_Loeb_2003,
   author = {{Bromm}, V. and {Loeb}, A.},
    title = "{Formation of the First Supermassive Black Holes}",
  journal = {\apj},
   eprint = {astro-ph/0212400},
     year = 2003,
    month = oct,
   volume = 596,
    pages = {34-46},
      doi = {10.1086/377529},
   adsurl = {http://adsabs.harvard.edu/abs/2003ApJ...596...34B}
}

@ARTICLE{Volonteri_2008,
   author = {{Volonteri}, M. and {Lodato}, G. and {Natarajan}, P.},
    title = "{The evolution of massive black hole seeds}",
  journal = {\mnras},
archivePrefix = "arXiv",
   eprint = {0709.0529},
     year = 2008,
    month = jan,
   volume = 383,
    pages = {1079-1088},
      doi = {10.1111/j.1365-2966.2007.12589.x},
   adsurl = {http://adsabs.harvard.edu/abs/2008MNRAS.383.1079V}
}

@ARTICLE{Loeb_Rasio_1994,
   author = {{Loeb}, A. and {Rasio}, F.~A.},
    title = "{Collapse of primordial gas clouds and the formation of quasar black holes}",
  journal = {\apj},
   eprint = {astro-ph/9401026},
     year = 1994,
    month = sep,
   volume = 432,
    pages = {52-61},
      doi = {10.1086/174548},
   adsurl = {http://adsabs.harvard.edu/abs/1994ApJ...432...52L}
}

@ARTICLE{Ferrara_2014,
   author = {{Ferrara}, A. and {Salvadori}, S. and {Yue}, B. and {Schleicher}, D.
	},
    title = "{Initial mass function of intermediate-mass black hole seeds}",
  journal = {\mnras},
archivePrefix = "arXiv",
   eprint = {1406.6685},
     year = 2014,
    month = sep,
   volume = 443,
    pages = {2410-2425},
      doi = {10.1093/mnras/stu1280},
   adsurl = {http://adsabs.harvard.edu/abs/2014MNRAS.443.2410F}
}

@ARTICLE{BL01,
   author = {{Barkana}, R. and {Loeb}, A.},
    title = "{In the beginning: the first sources of light and the reionization of the universe}",
  journal = {\physrep},
   eprint = {astro-ph/0010468},
     year = 2001,
    month = jul,
   volume = 349,
    pages = {125-238},
      doi = {10.1016/S0370-1573(01)00019-9},
   adsurl = {http://adsabs.harvard.edu/abs/2001PhR...349..125B}
}

@ARTICLE{Nguyen_2019,
       author = {{Nguyen}, Dieu D. and {Seth}, Anil C. and {Neumayer}, Nadine and {Iguchi}, Satoru and {Cappellari}, Michelle and {Strader}, Jay and {Chomiuk}, Laura and {Tremou}, Evangelia and {Pacucci}, Fabio and {Nakanishi}, Kouichiro and {Bahramian}, Arash and {Nguyen}, Phuong M. and {den Brok}, Mark and {Ahn}, Christopher C. and {Voggel}, Karina T. and {Kacharov}, Nikolay and {Tsukui}, Takafumi and {Ly}, Cuc K. and {Dumont}, Antoine and {Pechetti}, Renuka},
        title = "{Improved Dynamical Constraints on the Masses of the Central Black Holes in Nearby Low-mass Early-type Galactic Nuclei and the First Black Hole Determination for NGC 205}",
      journal = {\apj},
         year = 2019,
        month = feb,
       volume = {872},
       number = {1},
          eid = {104},
        pages = {104},
          doi = {10.3847/1538-4357/aafe7a},
archivePrefix = {arXiv},
       eprint = {1901.05496},
 primaryClass = {astro-ph.GA},
       adsurl = {https://ui.adsabs.harvard.edu/abs/2019ApJ...872..104N}
}

@ARTICLE{Greene_2020_review,
       author = {{Greene}, Jenny E. and {Strader}, Jay and {Ho}, Luis C.},
        title = "{Intermediate-Mass Black Holes}",
      journal = {\araa},
         year = 2020,
        month = aug,
       volume = {58},
        pages = {257-312},
          doi = {10.1146/annurev-astro-032620-021835},
archivePrefix = {arXiv},
       eprint = {1911.09678},
 primaryClass = {astro-ph.GA},
       adsurl = {https://ui.adsabs.harvard.edu/abs/2020ARA&A..58..257G}
}

@ARTICLE{Pacucci_2018,
       author = {{Pacucci}, Fabio and {Loeb}, Abraham and {Mezcua}, Mar and
        {Mart{\'\i}n-Navarro}, Ignacio},
        title = "{Glimmering in the Dark: Modeling the Low-mass End of the M
        $_{{\ensuremath{\bullet}}}$─{\ensuremath{\sigma}} Relation and
        of the Quasar Luminosity Function}",
      journal = {\apj},
         year = 2018,
        month = Sep,
       volume = {864},
          eid = {L6},
        pages = {L6},
          doi = {10.3847/2041-8213/aad8b2},
archivePrefix = {arXiv},
       eprint = {1808.09452},
 primaryClass = {astro-ph.GA},
       adsurl = {https://ui.adsabs.harvard.edu/#abs/2018ApJ...864L...6P}
}

@ARTICLE{Madau_2014,
       author = {{Madau}, Piero and {Dickinson}, Mark},
        title = "{Cosmic Star-Formation History}",
      journal = {Annual Review of Astronomy and Astrophysics},
         year = 2014,
        month = Aug,
       volume = {52},
        pages = {415-486},
          doi = {10.1146/annurev-astro-081811-125615},
archivePrefix = {arXiv},
       eprint = {1403.0007},
 primaryClass = {astro-ph.CO},
       adsurl = {https://ui.adsabs.harvard.edu/#abs/2014ARA&A..52..415M}
}

@ARTICLE{Volonteri_2005,
   author = {{Volonteri}, M. and {Rees}, M.~J.},
    title = "{Rapid Growth of High-Redshift Black Holes}",
  journal = {\apj},
   eprint = {astro-ph/0506040},
     year = 2005,
    month = nov,
   volume = 633,
    pages = {624-629},
      doi = {10.1086/466521},
   adsurl = {http://adsabs.harvard.edu/abs/2005ApJ...633..624V}
}

@ARTICLE{Volonteri_2010,
   author = {{Volonteri}, M.},
    title = "{Formation of supermassive black holes}",
  journal = "{Astronomy and Astrophysics Review}",
archivePrefix = "arXiv",
   eprint = {1003.4404},
 primaryClass = "astro-ph.CO",
     year = 2010,
    month = jul,
   volume = 18,
    pages = {279-315},
      doi = {10.1007/s00159-010-0029-x},
   adsurl = {http://adsabs.harvard.edu/abs/2010A%26ARv..18..279V}
}

@ARTICLE{Pacucci_2015,
   author = {{Pacucci}, F. and {Ferrara}, A.},
    title = "{Simulating the growth of Intermediate Mass Black Holes}",
  journal = {\mnras},
archivePrefix = "arXiv",
   eprint = {1501.00989},
 primaryClass = "astro-ph.HE",
     year = 2015,
    month = mar,
   volume = 448,
    pages = {104-118},
      doi = {10.1093/mnras/stv018},
   adsurl = {http://adsabs.harvard.edu/abs/2015MNRAS.448..104P}
}

@ARTICLE{Madau_Rees_2001,
   author = {{Madau}, P. and {Rees}, M.~J.},
    title = "{Massive Black Holes as Population III Remnants}",
  journal = {\apjl},
   eprint = {astro-ph/0101223},
     year = 2001,
    month = apr,
   volume = 551,
    pages = {L27-L30},
      doi = {10.1086/319848},
   adsurl = {http://adsabs.harvard.edu/abs/2001ApJ...551L..27M}
}

@ARTICLE{Shapiro_2005,
       author = {{Shapiro}, Stuart L.},
        title = "{Spin, Accretion, and the Cosmological Growth of Supermassive Black Holes}",
      journal = {\apj},
         year = "2005",
        month = "Feb",
       volume = {620},
       number = {1},
        pages = {59-68},
          doi = {10.1086/427065},
archivePrefix = {arXiv},
       eprint = {astro-ph/0411156},
 primaryClass = {astro-ph},
       adsurl = {https://ui.adsabs.harvard.edu/abs/2005ApJ...620...59S}
}

@ARTICLE{Pacucci_2015_GW,
   author = {{Pacucci}, F. and {Ferrara}, A. and {Marassi}, S.},
    title = "{Gravitational waves from direct collapse black holes formation}",
  journal = {\mnras},
archivePrefix = "arXiv",
   eprint = {1502.04125},
     year = 2015,
    month = may,
   volume = 449,
    pages = {1076-1083},
      doi = {10.1093/mnras/stv317},
   adsurl = {http://adsabs.harvard.edu/abs/2015MNRAS.449.1076P}
}

@ARTICLE{Trinca_2023,
       author = {{Trinca}, Alessandro and {Schneider}, Raffaella and {Maiolino}, Roberto and {Valiante}, Rosa and {Graziani}, Luca and {Volonteri}, Marta},
        title = "{Seeking the growth of the first black hole seeds with JWST}",
      journal = {\mnras},
         year = 2023,
        month = mar,
       volume = {519},
       number = {3},
        pages = {4753-4764},
          doi = {10.1093/mnras/stac3768},
archivePrefix = {arXiv},
       eprint = {2211.01389},
 primaryClass = {astro-ph.GA},
       adsurl = {https://ui.adsabs.harvard.edu/abs/2023MNRAS.519.4753T}
}

@ARTICLE{Rodriguez_2016,
   author = {{Rodriguez}, C.~L. and {Chatterjee}, S. and {Rasio}, F.~A.},
    title = "{Binary black hole mergers from globular clusters: Masses, merger rates, and the impact of stellar evolution}",
  journal = {\prd},
archivePrefix = "arXiv",
   eprint = {1602.02444},
 primaryClass = "astro-ph.HE",
     year = 2016,
    month = apr,
   volume = 93,
   number = 8,
      eid = {084029},
    pages = {084029},
      doi = {10.1103/PhysRevD.93.084029},
   adsurl = {http://adsabs.harvard.edu/abs/2016PhRvD..93h4029R}
}

@ARTICLE{Pacucci_2023,
       author = {{Pacucci}, Fabio and {Nguyen}, Bao and {Carniani}, Stefano and {Maiolino}, Roberto and {Fan}, Xiaohui},
        title = "{JWST CEERS and JADES Active Galaxies at z = 4-7 Violate the Local M $_{{\textbullet}}$-M $_{{\ensuremath{\star}}}$ Relation at >3{\ensuremath{\sigma}}: Implications for Low-mass Black Holes and Seeding Models}",
      journal = {\apjl},
         year = 2023,
        month = nov,
       volume = {957},
       number = {1},
          eid = {L3},
        pages = {L3},
          doi = {10.3847/2041-8213/ad0158},
archivePrefix = {arXiv},
       eprint = {2308.12331},
 primaryClass = {astro-ph.GA},
       adsurl = {https://ui.adsabs.harvard.edu/abs/2023ApJ...957L...3P}
}

@ARTICLE{Maiolino_2024_JADES_BH_z4_11,
       author = {{Maiolino}, Roberto and {Scholtz}, Jan and {Curtis-Lake}, Emma and {Carniani}, Stefano and {Baker}, William and {de Graaff}, Anna and {Tacchella}, Sandro and {{\"U}bler}, Hannah and {D'Eugenio}, Francesco and {Witstok}, Joris and {Curti}, Mirko and {Arribas}, Santiago and {Bunker}, Andrew J. and {Charlot}, St{\'e}phane and {Chevallard}, Jacopo and {Eisenstein}, Daniel J. and {Egami}, Eiichi and {Ji}, Zhiyuan and {Jones}, Gareth C. and {Lyu}, Jianwei and {Rawle}, Tim and {Robertson}, Brant and {Rujopakarn}, Wiphu and {Perna}, Michele and {Sun}, Fengwu and {Venturi}, Giacomo and {Williams}, Christina C. and {Willott}, Chris},
        title = "{JADES: The diverse population of infant black holes at 4 < z < 11: Merging, tiny, poor, but mighty}",
      journal = {\aap},
         year = 2024,
        month = nov,
       volume = {691},
          eid = {A145},
        pages = {A145},
          doi = {10.1051/0004-6361/202347640},
archivePrefix = {arXiv},
       eprint = {2308.01230},
 primaryClass = {astro-ph.GA},
       adsurl = {https://ui.adsabs.harvard.edu/abs/2024A&A...691A.145M}
}

@ARTICLE{TRINITY_2,
       author = {{Zhang}, Haowen and {Behroozi}, Peter and {Volonteri}, Marta and {Silk}, Joseph and {Fan}, Xiaohui and {Aird}, James and {Yang}, Jinyi and {Hopkins}, Philip F.},
        title = "{TRINITY II: The luminosity-dependent bias of the supermassive black hole mass-galaxy mass relation for bright quasars at z = 6}",
      journal = {\mnras},
         year = 2023,
        month = jul,
       volume = {523},
       number = {1},
        pages = {L69-L74},
          doi = {10.1093/mnrasl/slad060},
archivePrefix = {arXiv},
       eprint = {2303.08150},
 primaryClass = {astro-ph.GA},
       adsurl = {https://ui.adsabs.harvard.edu/abs/2023MNRAS.523L..69Z}
}

@ARTICLE{Fragione_2023,
       author = {{Fragione}, Giacomo and {Pacucci}, Fabio},
        title = "{Constraining the Properties of Black Hole Seeds from the Farthest Quasars}",
      journal = {arXiv e-prints},
         year = 2023,
        month = aug,
          eid = {arXiv:2308.14986},
        pages = {arXiv:2308.14986},
          doi = {10.48550/arXiv.2308.14986},
archivePrefix = {arXiv},
       eprint = {2308.14986},
 primaryClass = {astro-ph.GA},
       adsurl = {https://ui.adsabs.harvard.edu/abs/2023arXiv230814986F}
}

@ARTICLE{Ellis_2024a,
       author = {{Ellis}, John and {Fairbairn}, Malcolm and {H{\"u}tsi}, Gert and {Raidal}, Juhan and {Urrutia}, Juan and {Vaskonen}, Ville and {Veerm{\"a}e}, Hardi},
        title = "{Gravitational waves from supermassive black hole binaries in light of the NANOGrav 15-year data}",
      journal = {\prd},
         year = 2024,
        month = jan,
       volume = {109},
       number = {2},
          eid = {L021302},
        pages = {L021302},
          doi = {10.1103/PhysRevD.109.L021302},
archivePrefix = {arXiv},
       eprint = {2306.17021},
 primaryClass = {astro-ph.CO},
       adsurl = {https://ui.adsabs.harvard.edu/abs/2024PhRvD.109b1302E}
}

@ARTICLE{Ellis_2024b,
       author = {{Ellis}, John and {Fairbairn}, Malcolm and {H{\"u}tsi}, Gert and {Urrutia}, Juan and {Vaskonen}, Ville and {Veerm{\"a}e}, Hardi},
        title = "{Consistency of JWST Black Hole Observations with NANOGrav Gravitational Wave Measurements}",
      journal = {arXiv e-prints},
         year = 2024,
        month = mar,
          eid = {arXiv:2403.19650},
        pages = {arXiv:2403.19650},
          doi = {10.48550/arXiv.2403.19650},
archivePrefix = {arXiv},
       eprint = {2403.19650},
 primaryClass = {astro-ph.CO},
       adsurl = {https://ui.adsabs.harvard.edu/abs/2024arXiv240319650E}
}

@ARTICLE{Pacucci_2024,
       author = {{Pacucci}, Fabio and {Loeb}, Abraham},
        title = "{The Redshift Evolution of the M $_{{\textbullet}}${\textendash}M $_{{\ensuremath{\star}}}$ Relation for JWST's Supermassive Black Holes at z > 4}",
      journal = {\apj},
         year = 2024,
        month = apr,
       volume = {964},
       number = {2},
          eid = {154},
        pages = {154},
          doi = {10.3847/1538-4357/ad3044},
archivePrefix = {arXiv},
       eprint = {2401.04159},
 primaryClass = {astro-ph.GA},
       adsurl = {https://ui.adsabs.harvard.edu/abs/2024ApJ...964..154P}
}

@ARTICLE{Li_2024,
       author = {{Li}, Junyao and {Silverman}, John D. and {Shen}, Yue and {Volonteri}, Marta and {Jahnke}, Knud and {Zhuang}, Ming-Yang and {Scoggins}, Matthew T. and {Ding}, Xuheng and {Harikane}, Yuichi and {Onoue}, Masafusa and {Tanaka}, Takumi S.},
        title = "{Tip of the iceberg: overmassive black holes at 4<z<7 found by JWST are not inconsistent with the local $\mathcal{M}_{\rm BH}$-$\mathcal{M}_\star$ relation}",
      journal = {arXiv e-prints},
         year = 2024,
        month = feb,
          eid = {arXiv:2403.00074},
        pages = {arXiv:2403.00074},
          doi = {10.48550/arXiv.2403.00074},
archivePrefix = {arXiv},
       eprint = {2403.00074},
 primaryClass = {astro-ph.GA},
       adsurl = {https://ui.adsabs.harvard.edu/abs/2024arXiv240300074L}
}

@ARTICLE{Bhowmick_2024,
       author = {{Bhowmick}, Aklant K. and {Blecha}, Laura and {Torrey}, Paul and {Kelley}, Luke Zoltan and {Weinberger}, Rainer and {Vogelsberger}, Mark and {Hernquist}, Lars and {Somerville}, Rachel S. and {Evans}, Analis Eolyn},
        title = "{Introducing the BRAHMA simulation suite: Signatures of low mass black hole seeding models in cosmological simulations}",
      journal = {arXiv e-prints},
         year = 2024,
        month = feb,
          eid = {arXiv:2402.03626},
        pages = {arXiv:2402.03626},
          doi = {10.48550/arXiv.2402.03626},
archivePrefix = {arXiv},
       eprint = {2402.03626},
 primaryClass = {astro-ph.GA},
       adsurl = {https://ui.adsabs.harvard.edu/abs/2024arXiv240203626B}
}

@ARTICLE{Habouzit_2021,
       author = {{Habouzit}, M{\'e}lanie and {Li}, Yuan and {Somerville}, Rachel S. and {Genel}, Shy and {Pillepich}, Annalisa and {Volonteri}, Marta and {Dav{\'e}}, Romeel and {Rosas-Guevara}, Yetli and {McAlpine}, Stuart and {Peirani}, S{\'e}bastien and {Hernquist}, Lars and {Angl{\'e}s-Alc{\'a}zar}, Daniel and {Reines}, Amy and {Bower}, Richard and {Dubois}, Yohan and {Nelson}, Dylan and {Pichon}, Christophe and {Vogelsberger}, Mark},
        title = "{Supermassive black holes in cosmological simulations I: M$_{BH}$ - M$_{{\ensuremath{\star}}}$ relation and black hole mass function}",
      journal = {\mnras},
         year = 2021,
        month = may,
       volume = {503},
       number = {2},
        pages = {1940-1975},
          doi = {10.1093/mnras/stab496},
archivePrefix = {arXiv},
       eprint = {2006.10094},
 primaryClass = {astro-ph.GA},
       adsurl = {https://ui.adsabs.harvard.edu/abs/2021MNRAS.503.1940H}
}

@ARTICLE{Jeon_2025_JWST_SMBH_pathways,
       author = {{Jeon}, Junehyoung and {Bromm}, Volker and {Liu}, Boyuan and {Finkelstein}, Steven L.},
        title = "{Physical Pathways for JWST-observed Supermassive Black Holes in the Early Universe}",
      journal = {\apj},
         year = 2025,
        month = feb,
       volume = {979},
       number = {2},
          eid = {127},
        pages = {127},
          doi = {10.3847/1538-4357/ad9f3a},
archivePrefix = {arXiv},
       eprint = {2402.18773},
 primaryClass = {astro-ph.GA},
       adsurl = {https://ui.adsabs.harvard.edu/abs/2025ApJ...979..127J}
}

@ARTICLE{Natarajan_2024_UHZ1,
       author = {{Natarajan}, Priyamvada and {Pacucci}, Fabio and {Ricarte}, Angelo and {Bogd{\'a}n}, {\'A}kos and {Goulding}, Andy D. and {Cappelluti}, Nico},
        title = "{First Detection of an Overmassive Black Hole Galaxy UHZ1: Evidence for Heavy Black Hole Seed Formation from Direct Collapse}",
      journal = {\apjl},
         year = 2024,
        month = jan,
       volume = {960},
       number = {1},
          eid = {L1},
        pages = {L1},
          doi = {10.3847/2041-8213/ad0e76},
archivePrefix = {arXiv},
       eprint = {2308.02654},
 primaryClass = {astro-ph.HE},
       adsurl = {https://ui.adsabs.harvard.edu/abs/2024ApJ...960L...1N}
}

@ARTICLE{Fei_2025_JWST_GLIMPSE_IMBH_direct_collapse,
       author = {{Fei}, Qinyue and {Fujimoto}, Seiji and {Naidu}, Rohan P. and {Chisholm}, John and {Atek}, Hakim and {Brammer}, Gabriel and {Asada}, Yoshihisa and {Bromm}, Volker and {Furtak}, Lukas J. and {Greene}, Jenny E. and {Hsiao}, Tiger Yu-Yang and {Jeon}, Junehyoung and {Kokorev}, Vasily and {Matthee}, Jorryt and {Natarajan}, Priyamvada and {Richard}, Johan and {Saldana-Lopez}, Alberto and {Schaerer}, Daniel and {Volonteri}, Marta and {Zitrin}, Adi},
        title = "{A GLIMPSE of Intermediate Mass Black holes in the epoch of reionization: Witnessing the Descendants of Direct Collapse?}",
      journal = {arXiv e-prints},
         year = 2025,
        month = sep,
          eid = {arXiv:2509.20452},
        pages = {arXiv:2509.20452},
          doi = {10.48550/arXiv.2509.20452},
archivePrefix = {arXiv},
       eprint = {2509.20452},
 primaryClass = {astro-ph.GA},
       adsurl = {https://ui.adsabs.harvard.edu/abs/2025arXiv250920452F}
}

@ARTICLE{Geris_2026_JADES_high_z_BH,
       author = {{Geris}, Sophia and {Maiolino}, Roberto and {Isobe}, Yuki and {Scholtz}, Jan and {D'Eugenio}, Francesco and {Ji}, Xihan and {Juod{\v{z}}balis}, Ignas and {Simmonds}, Charlotte and {Dayal}, Pratika and {Trinca}, Alessandro and {Schneider}, Raffaella and {Arribas}, Santiago and {Bhatawdekar}, Rachana and {Bunker}, Andrew J. and {Carniani}, Stefano and {Charlot}, St{\'e}phane and {Chevallard}, Jacopo and {Curtis-Lake}, Emma and {Johnson}, Benjamin D. and {Parlanti}, Eleonora and {Rinaldi}, Pierluigi and {Robertson}, Brant and {Tacchella}, Sandro and {{\"U}bler}, Hannah and {Venturi}, Giacomo and {Williams}, Christina C. and {Witstok}, Joris},
        title = "{JADES reveals a large population of low-mass black holes at high redshift}",
      journal = {\mnras},
         year = 2026,
        month = jan,
       volume = {545},
       number = {1},
          eid = {staf1979},
        pages = {staf1979},
          doi = {10.1093/mnras/staf1979},
archivePrefix = {arXiv},
       eprint = {2506.22147},
 primaryClass = {astro-ph.GA},
       adsurl = {https://ui.adsabs.harvard.edu/abs/2026MNRAS.545S1979G}
}

@ARTICLE{NANOGrav_15_year,
       author = {{Agazie}, Gabriella and {Anumarlapudi}, Akash and {Archibald}, Anne M. and {Arzoumanian}, Zaven and {Baker}, Paul T. and {B{\'e}csy}, Bence and {Blecha}, Laura and {Brazier}, Adam and {Brook}, Paul R. and {Burke-Spolaor}, Sarah and {Burnette}, Rand and {Case}, Robin and {Charisi}, Maria and {Chatterjee}, Shami and {Chatziioannou}, Katerina and {Cheeseboro}, Belinda D. and {Chen}, Siyuan and {Cohen}, Tyler and {Cordes}, James M. and {Cornish}, Neil J. and {Crawford}, Fronefield and {Cromartie}, H. Thankful and {Crowter}, Kathryn and {Cutler}, Curt J. and {Decesar}, Megan E. and {Degan}, Dallas and {Demorest}, Paul B. and {Deng}, Heling and {Dolch}, Timothy and {Drachler}, Brendan and {Ellis}, Justin A. and {Ferrara}, Elizabeth C. and {Fiore}, William and {Fonseca}, Emmanuel and {Freedman}, Gabriel E. and {Garver-Daniels}, Nate and {Gentile}, Peter A. and {Gersbach}, Kyle A. and {Glaser}, Joseph and {Good}, Deborah C. and {G{\"u}ltekin}, Kayhan and {Hazboun}, Jeffrey S. and {Hourihane}, Sophie and {Islo}, Kristina and {Jennings}, Ross J. and {Johnson}, Aaron D. and {Jones}, Megan L. and {Kaiser}, Andrew R. and {Kaplan}, David L. and {Kelley}, Luke Zoltan and {Kerr}, Matthew and {Key}, Joey S. and {Klein}, Tonia C. and {Laal}, Nima and {Lam}, Michael T. and {Lamb}, William G. and {Lazio}, T. Joseph W. and {Lewandowska}, Natalia and {Littenberg}, Tyson B. and {Liu}, Tingting and {Lommen}, Andrea and {Lorimer}, Duncan R. and {Luo}, Jing and {Lynch}, Ryan S. and {Ma}, Chung-Pei and {Madison}, Dustin R. and {Mattson}, Margaret A. and {McEwen}, Alexander and {McKee}, James W. and {McLaughlin}, Maura A. and {McMann}, Natasha and {Meyers}, Bradley W. and {Meyers}, Patrick M. and {Mingarelli}, Chiara M.~F. and {Mitridate}, Andrea and {Natarajan}, Priyamvada and {Ng}, Cherry and {Nice}, David J. and {Ocker}, Stella Koch and {Olum}, Ken D. and {Pennucci}, Timothy T. and {Perera}, Benetge B.~P. and {Petrov}, Polina and {Pol}, Nihan S. and {Radovan}, Henri A. and {Ransom}, Scott M. and {Ray}, Paul S. and {Romano}, Joseph D. and {Sardesai}, Shashwat C. and {Schmiedekamp}, Ann and {Schmiedekamp}, Carl and {Schmitz}, Kai and {Schult}, Levi and {Shapiro-Albert}, Brent J. and {Siemens}, Xavier and {Simon}, Joseph and {Siwek}, Magdalena S. and {Stairs}, Ingrid H. and {Stinebring}, Daniel R. and {Stovall}, Kevin and {Sun}, Jerry P. and {Susobhanan}, Abhimanyu and {Swiggum}, Joseph K. and {Taylor}, Jacob and {Taylor}, Stephen R. and {Turner}, Jacob E. and {Unal}, Caner and {Vallisneri}, Michele and {van Haasteren}, Rutger and {Vigeland}, Sarah J. and {Wahl}, Haley M. and {Wang}, Qiaohong and {Witt}, Caitlin A. and {Young}, Olivia and {Nanograv Collaboration}},
        title = "{The NANOGrav 15 yr Data Set: Evidence for a Gravitational-wave Background}",
      journal = {\apjl},
         year = 2023,
        month = jul,
       volume = {951},
       number = {1},
          eid = {L8},
        pages = {L8},
          doi = {10.3847/2041-8213/acdac6},
archivePrefix = {arXiv},
       eprint = {2306.16213},
 primaryClass = {astro-ph.HE},
       adsurl = {https://ui.adsabs.harvard.edu/abs/2023ApJ...951L...8A}
}

@ARTICLE{NANOGrav_15_year_BH_binary,
       author = {{Agazie}, Gabriella and {Anumarlapudi}, Akash and {Archibald}, Anne M. and {Baker}, Paul T. and {B{\'e}csy}, Bence and {Blecha}, Laura and {Bonilla}, Alexander and {Brazier}, Adam and {Brook}, Paul R. and {Burke-Spolaor}, Sarah and {Burnette}, Rand and {Case}, Robin and {Casey-Clyde}, J. Andrew and {Charisi}, Maria and {Chatterjee}, Shami and {Chatziioannou}, Katerina and {Cheeseboro}, Belinda D. and {Chen}, Siyuan and {Cohen}, Tyler and {Cordes}, James M. and {Cornish}, Neil J. and {Crawford}, Fronefield and {Cromartie}, H. Thankful and {Crowter}, Kathryn and {Cutler}, Curt J. and {D'Orazio}, Daniel J. and {Decesar}, Megan E. and {Degan}, Dallas and {Demorest}, Paul B. and {Deng}, Heling and {Dolch}, Timothy and {Drachler}, Brendan and {Ferrara}, Elizabeth C. and {Fiore}, William and {Fonseca}, Emmanuel and {Freedman}, Gabriel E. and {Gardiner}, Emiko and {Garver-Daniels}, Nate and {Gentile}, Peter A. and {Gersbach}, Kyle A. and {Glaser}, Joseph and {Good}, Deborah C. and {G{\"u}ltekin}, Kayhan and {Hazboun}, Jeffrey S. and {Hourihane}, Sophie and {Islo}, Kristina and {Jennings}, Ross J. and {Johnson}, Aaron and {Jones}, Megan L. and {Kaiser}, Andrew R. and {Kaplan}, David L. and {Kelley}, Luke Zoltan and {Kerr}, Matthew and {Key}, Joey S. and {Laal}, Nima and {Lam}, Michael T. and {Lamb}, William G. and {Lazio}, T. Joseph W. and {Lewandowska}, Natalia and {Littenberg}, Tyson B. and {Liu}, Tingting and {Luo}, Jing and {Lynch}, Ryan S. and {Ma}, Chung-Pei and {Madison}, Dustin R. and {McEwen}, Alexander and {McKee}, James W. and {McLaughlin}, Maura A. and {McMann}, Natasha and {Meyers}, Bradley W. and {Meyers}, Patrick M. and {Mingarelli}, Chiara M.~F. and {Mitridate}, Andrea and {Natarajan}, Priyamvada and {Ng}, Cherry and {Nice}, David J. and {Ocker}, Stella Koch and {Olum}, Ken D. and {Pennucci}, Timothy T. and {Perera}, Benetge B.~P. and {Petrov}, Polina and {Pol}, Nihan S. and {Radovan}, Henri A. and {Ransom}, Scott M. and {Ray}, Paul S. and {Romano}, Joseph D. and {Runnoe}, Jessie C. and {Sardesai}, Shashwat C. and {Schmiedekamp}, Ann and {Schmiedekamp}, Carl and {Schmitz}, Kai and {Schult}, Levi and {Shapiro-Albert}, Brent J. and {Siemens}, Xavier and {Simon}, Joseph and {Siwek}, Magdalena S. and {Stairs}, Ingrid H. and {Stinebring}, Daniel R. and {Stovall}, Kevin and {Sun}, Jerry P. and {Susobhanan}, Abhimanyu and {Swiggum}, Joseph K. and {Taylor}, Jacob and {Taylor}, Stephen R. and {Turner}, Jacob E. and {Unal}, Caner and {Vallisneri}, Michele and {Vigeland}, Sarah J. and {Wachter}, Jeremy M. and {Wahl}, Haley M. and {Wang}, Qiaohong and {Witt}, Caitlin A. and {Wright}, David and {Young}, Olivia and {Nanograv Collaboration}},
        title = "{The NANOGrav 15 yr Data Set: Constraints on Supermassive Black Hole Binaries from the Gravitational-wave Background}",
      journal = {\apjl},
         year = 2023,
        month = aug,
       volume = {952},
       number = {2},
          eid = {L37},
        pages = {L37},
          doi = {10.3847/2041-8213/ace18b},
archivePrefix = {arXiv},
       eprint = {2306.16220},
 primaryClass = {astro-ph.HE},
       adsurl = {https://ui.adsabs.harvard.edu/abs/2023ApJ...952L..37A}
}

@ARTICLE{Ruiter_2010,
       author = {{Ruiter}, Ashley J. and {Belczynski}, Krzysztof and {Benacquista}, Matthew and {Larson}, Shane L. and {Williams}, Gabriel},
        title = "{The LISA Gravitational Wave Foreground: A Study of Double White Dwarfs}",
      journal = {\apj},
         year = 2010,
        month = jul,
       volume = {717},
       number = {2},
        pages = {1006-1021},
          doi = {10.1088/0004-637X/717/2/1006},
archivePrefix = {arXiv},
       eprint = {0705.3272},
 primaryClass = {astro-ph},
       adsurl = {https://ui.adsabs.harvard.edu/abs/2010ApJ...717.1006R}
}

@ARTICLE{Ricarte_2023_jet_spin_MAD,
       author = {{Ricarte}, Angelo and {Narayan}, Ramesh and {Curd}, Brandon},
        title = "{Recipes for Jet Feedback and Spin Evolution of Black Holes with Strongly Magnetized Super-Eddington Accretion Disks}",
      journal = {\apjl},
         year = 2023,
        month = sep,
       volume = {954},
       number = {1},
          eid = {L22},
        pages = {L22},
          doi = {10.3847/2041-8213/aceda5},
archivePrefix = {arXiv},
       eprint = {2307.04621},
 primaryClass = {astro-ph.HE},
       adsurl = {https://ui.adsabs.harvard.edu/abs/2023ApJ...954L..22R}
}

@ARTICLE{Jeon_2025_BHMF_high_z,
       author = {{Jeon}, Junehyoung and {Liu}, Boyuan and {Taylor}, Anthony J. and {Kokorev}, Vasily and {Chisholm}, John and {Kocevski}, Dale D. and {Finkelstein}, Steven L. and {Bromm}, Volker},
        title = "{The Emerging Black Hole Mass Function in the High-redshift Universe}",
      journal = {\apj},
         year = 2025,
        month = jul,
       volume = {988},
       number = {1},
          eid = {110},
        pages = {110},
          doi = {10.3847/1538-4357/ade2e1},
archivePrefix = {arXiv},
       eprint = {2503.14703},
 primaryClass = {astro-ph.GA},
       adsurl = {https://ui.adsabs.harvard.edu/abs/2025ApJ...988..110J}
}

@ARTICLE{Mathee_2024_JWST_LRD,
       author = {{Matthee}, Jorryt and {Naidu}, Rohan P. and {Brammer}, Gabriel and {Chisholm}, John and {Eilers}, Anna-Christina and {Goulding}, Andy and {Greene}, Jenny and {Kashino}, Daichi and {Labbe}, Ivo and {Lilly}, Simon J. and {Mackenzie}, Ruari and {Oesch}, Pascal A. and {Weibel}, Andrea and {Wuyts}, Stijn and {Xiao}, Mengyuan and {Bordoloi}, Rongmon and {Bouwens}, Rychard and {van Dokkum}, Pieter and {Illingworth}, Garth and {Kramarenko}, Ivan and {Maseda}, Michael V. and {Mason}, Charlotte and {Meyer}, Romain A. and {Nelson}, Erica J. and {Reddy}, Naveen A. and {Shivaei}, Irene and {Simcoe}, Robert A. and {Yue}, Minghao},
        title = "{Little Red Dots: An Abundant Population of Faint Active Galactic Nuclei at z {\ensuremath{\sim}} 5 Revealed by the EIGER and FRESCO JWST Surveys}",
      journal = {\apj},
         year = 2024,
        month = mar,
       volume = {963},
       number = {2},
          eid = {129},
        pages = {129},
          doi = {10.3847/1538-4357/ad2345},
archivePrefix = {arXiv},
       eprint = {2306.05448},
 primaryClass = {astro-ph.GA},
       adsurl = {https://ui.adsabs.harvard.edu/abs/2024ApJ...963..129M}
}

@ARTICLE{Taylor_2025_JWST_AGN_BHMF,
       author = {{Taylor}, Anthony J. and {Finkelstein}, Steven L. and {Kocevski}, Dale D. and {Jeon}, Junehyoung and {Bromm}, Volker and {Amor{\'\i}n}, Ricardo O. and {Arrabal Haro}, Pablo and {Backhaus}, Bren E. and {Bagley}, Micaela B. and {Banados}, Eduardo and {Bhatawdekar}, Rachana and {Brooks}, Madisyn and {Calabr{\`o}}, Antonello and {Ch{\'a}vez Ortiz}, {\'O}scar A. and {Cheng}, Yingjie and {Cleri}, Nikko J. and {Cole}, Justin W. and {Davis}, Kelcey and {Dickinson}, Mark and {Donnan}, Callum and {Dunlop}, James S. and {Ellis}, Richard S. and {Fern{\'a}ndez}, Vital and {Fontana}, Adriano and {Fujimoto}, Seiji and {Giavalisco}, Mauro and {Grazian}, Andrea and {Guo}, Jingsong and {Hathi}, Nimish P. and {Holwerda}, Benne W. and {Hirschmann}, Michaela and {Inayoshi}, Kohei and {Kartaltepe}, Jeyhan S. and {Khusanova}, Yana and {Koekemoer}, Anton M. and {Kokorev}, Vasily and {Larson}, Rebecca L. and {Leung}, Gene C.~K. and {Lucas}, Ray A. and {McLeod}, Derek J. and {Napolitano}, Lorenzo and {Onoue}, Masafusa and {Pacucci}, Fabio and {Papovich}, Casey and {P{\'e}rez-Gonz{\'a}lez}, Pablo G. and {Pirzkal}, Nor and {Somerville}, Rachel S. and {Trump}, Jonathan R. and {Wilkins}, Stephen M. and {Yung}, L.~Y. Aaron and {Zhang}, Haowen},
        title = "{Broad-line AGNs at 3.5 < z < 6: The Black Hole Mass Function and a Connection with Little Red Dots}",
      journal = {\apj},
         year = 2025,
        month = jun,
       volume = {986},
       number = {2},
          eid = {165},
        pages = {165},
          doi = {10.3847/1538-4357/add15b},
archivePrefix = {arXiv},
       eprint = {2409.06772},
 primaryClass = {astro-ph.GA},
       adsurl = {https://ui.adsabs.harvard.edu/abs/2025ApJ...986..165T}
}

@ARTICLE{Jeon_2026_LRD_progenitor_DCBH,
       author = {{Jeon}, Junehyoung and {Liu}, Boyuan and {Bromm}, Volker and {Fujimoto}, Seiji and {Taylor}, Anthony J. and {Kokorev}, Vasily and {Larson}, Rebecca L. and {Chisholm}, John and {Finkelstein}, Steven L. and {Kocevski}, Dale D.},
        title = "{Little Red Dots and Their Progenitors from Direct Collapse Black Holes}",
      journal = {\apj},
         year = 2026,
        month = feb,
       volume = {998},
       number = {1},
          eid = {148},
        pages = {148},
          doi = {10.3847/1538-4357/ae3725},
archivePrefix = {arXiv},
       eprint = {2508.14155},
 primaryClass = {astro-ph.GA},
       adsurl = {https://ui.adsabs.harvard.edu/abs/2026ApJ...998..148J}
}

@ARTICLE{Izquierdo-Villalba_2024_NANOGrav,
       author = {{Izquierdo-Villalba}, David and {Sesana}, Alberto and {Colpi}, Monica and {Spinoso}, Daniele and {Bonetti}, Matteo and {Bonoli}, Silvia and {Valiante}, Rosa},
        title = "{Connecting low-redshift LISA massive black hole mergers to the nHz stochastic gravitational wave background}",
      journal = {\aap},
         year = 2024,
        month = jun,
       volume = {686},
          eid = {A183},
        pages = {A183},
          doi = {10.1051/0004-6361/202449293},
archivePrefix = {arXiv},
       eprint = {2401.10983},
 primaryClass = {astro-ph.GA},
       adsurl = {https://ui.adsabs.harvard.edu/abs/2024A&A...686A.183I}
}

@article{Creighton_2025_LILA_noise_sensitivity,
  title={Fundamental Noise and Gravitational-Wave Sensitivity of the Laser Interferometer Lunar Antenna (LILA)},
  author={Creighton, Teviet and Lognonn{\'e}, Philippe and Panning, Mark P and Trippe, James and Quetschke, Volker and Jani, Karan},
  journal={arXiv preprint arXiv:2508.18437},
  year={2025}
}

@article{Jani_2025_LILA,
  title={Laser Interferometer Lunar Antenna (LILA): Advancing the US Priorities in Gravitational-wave and Lunar Science},
  author={Jani, Karan and Abernathy, Matthew and Berti, Emanuele and Boschi, Valerio and Chakrabarti, Sukanya and Cocoros, Alice and Conklin, John W and Creighton, Teviet and Dell'Agnello, Simone and Diels, Jean-Claude and others},
  journal={arXiv preprint arXiv:2508.11631},
  year={2025}
}

@article{Ricarte_2025_SEROTINA_spin_evolution,
  title={Multimessenger Probes of Supermassive Black Hole Spin Evolution},
  author={Ricarte, Angelo and Natarajan, Priyamvada and Narayan, Ramesh and Palumbo, Daniel CM},
  journal={The Astrophysical Journal},
  volume={980},
  number={1},
  pages={136},
  year={2025},
  publisher={IOP Publishing}
}

@ARTICLE{Ricarte_Natarajan_2018_SEROTINA_SMBH_assembly,
       author = {{Ricarte}, Angelo and {Natarajan}, Priyamvada},
        title = "{Exploring SMBH assembly with semi-analytic modelling}",
      journal = {\mnras},
         year = 2018,
        month = feb,
       volume = {474},
       number = {2},
        pages = {1995-2011},
          doi = {10.1093/mnras/stx2851},
archivePrefix = {arXiv},
       eprint = {1710.11532},
 primaryClass = {astro-ph.HE},
       adsurl = {https://ui.adsabs.harvard.edu/abs/2018MNRAS.474.1995R}
}

@ARTICLE{Astropy,
       author = {{Astropy Collaboration} and {Robitaille}, Thomas P. and {Tollerud}, Erik J. and {Greenfield}, Perry and {Droettboom}, Michael and {Bray}, Erik and {Aldcroft}, Tom and {Davis}, Matt and {Ginsburg}, Adam and {Price-Whelan}, Adrian M. and {Kerzendorf}, Wolfgang E. and {Conley}, Alexander and {Crighton}, Neil and {Barbary}, Kyle and {Muna}, Demitri and {Ferguson}, Henry and {Grollier}, Fr{\'e}d{\'e}ric and {Parikh}, Madhura M. and {Nair}, Prasanth H. and {Unther}, Hans M. and {Deil}, Christoph and {Woillez}, Julien and {Conseil}, Simon and {Kramer}, Roban and {Turner}, James E.~H. and {Singer}, Leo and {Fox}, Ryan and {Weaver}, Benjamin A. and {Zabalza}, Victor and {Edwards}, Zachary I. and {Azalee Bostroem}, K. and {Burke}, D.~J. and {Casey}, Andrew R. and {Crawford}, Steven M. and {Dencheva}, Nadia and {Ely}, Justin and {Jenness}, Tim and {Labrie}, Kathleen and {Lim}, Pey Lian and {Pierfederici}, Francesco and {Pontzen}, Andrew and {Ptak}, Andy and {Refsdal}, Brian and {Servillat}, Mathieu and {Streicher}, Ole},
        title = "{Astropy: A community Python package for astronomy}",
      journal = {\aap},
         year = 2013,
        month = oct,
       volume = {558},
          eid = {A33},
        pages = {A33},
          doi = {10.1051/0004-6361/201322068},
archivePrefix = {arXiv},
       eprint = {1307.6212},
 primaryClass = {astro-ph.IM},
       adsurl = {https://ui.adsabs.harvard.edu/abs/2013A&A...558A..33A}
}

@ARTICLE{Numpy,
       author = {{Harris}, Charles R. and {Millman}, K. Jarrod and {van der Walt}, St{\'e}fan J. and {Gommers}, Ralf and {Virtanen}, Pauli and {Cournapeau}, David and {Wieser}, Eric and {Taylor}, Julian and {Berg}, Sebastian and {Smith}, Nathaniel J. and {Kern}, Robert and {Picus}, Matti and {Hoyer}, Stephan and {van Kerkwijk}, Marten H. and {Brett}, Matthew and {Haldane}, Allan and {del R{\'\i}o}, Jaime Fern{\'a}ndez and {Wiebe}, Mark and {Peterson}, Pearu and {G{\'e}rard-Marchant}, Pierre and {Sheppard}, Kevin and {Reddy}, Tyler and {Weckesser}, Warren and {Abbasi}, Hameer and {Gohlke}, Christoph and {Oliphant}, Travis E.},
        title = "{Array programming with NumPy}",
      journal = {\nat},
         year = 2020,
        month = sep,
       volume = {585},
       number = {7825},
        pages = {357-362},
          doi = {10.1038/s41586-020-2649-2},
archivePrefix = {arXiv},
       eprint = {2006.10256},
 primaryClass = {cs.MS},
       adsurl = {https://ui.adsabs.harvard.edu/abs/2020Natur.585..357H}
}

@ARTICLE{GW150914,
       author = {{Abbott}, B.~P. and {Abbott}, R. and {Abbott}, T.~D. and {Abernathy}, M.~R. and {Acernese}, F. and {Ackley}, K. and {Adams}, C. and {Adams}, T. and {Addesso}, P. and {Adhikari}, R.~X. and {Adya}, V.~B. and {Affeldt}, C. and {Agathos}, M. and {Agatsuma}, K. and {Aggarwal}, N. and {Aguiar}, O.~D. and {Aiello}, L. and {Ain}, A. and {Ajith}, P. and {Allen}, B. and {Allocca}, A. and {Altin}, P.~A. and {Anderson}, S.~B. and {Anderson}, W.~G. and {Arai}, K. and {Arain}, M.~A. and {Araya}, M.~C. and {Arceneaux}, C.~C. and {Areeda}, J.~S. and {Arnaud}, N. and {Arun}, K.~G. and {Ascenzi}, S. and {Ashton}, G. and {Ast}, M. and {Aston}, S.~M. and {Astone}, P. and {Aufmuth}, P. and {Aulbert}, C. and {Babak}, S. and {Bacon}, P. and {Bader}, M.~K.~M. and {Baker}, P.~T. and {Baldaccini}, F. and {Ballardin}, G. and {Ballmer}, S.~W. and {Barayoga}, J.~C. and {Barclay}, S.~E. and {Barish}, B.~C. and {Barker}, D. and {Barone}, F. and {Barr}, B. and {Barsotti}, L. and {Barsuglia}, M. and {Barta}, D. and {Bartlett}, J. and {Barton}, M.~A. and {Bartos}, I. and {Bassiri}, R. and {Basti}, A. and {Batch}, J.~C. and {Baune}, C. and {Bavigadda}, V. and {Bazzan}, M. and {Behnke}, B. and {Bejger}, M. and {Belczynski}, C. and {Bell}, A.~S. and {Bell}, C.~J. and {Berger}, B.~K. and {Bergman}, J. and {Bergmann}, G. and {Berry}, C.~P.~L. and {Bersanetti}, D. and {Bertolini}, A. and {Betzwieser}, J. and {Bhagwat}, S. and {Bhandare}, R. and {Bilenko}, I.~A. and {Billingsley}, G. and {Birch}, J. and {Birney}, I.~A. and {Birnholtz}, O. and {Biscans}, S. and {Bisht}, A. and {Bitossi}, M. and {Biwer}, C. and {Bizouard}, M.~A. and {Blackburn}, J.~K. and {Blair}, C.~D. and {Blair}, D.~G. and {Blair}, R.~M. and {Bloemen}, S. and {Bock}, O. and {Bodiya}, T.~P. and {Boer}, M. and {Bogaert}, G. and {Bogan}, C. and {Bohe}, A. and {Bojtos}, P. and {Bond}, C. and {Bondu}, F. and {Bonnand}, R. and {Boom}, B.~A. and {Bork}, R. and {Boschi}, V. and {Bose}, S. and {Bouffanais}, Y. and {Bozzi}, A. and {Bradaschia}, C. and {Brady}, P.~R. and {Braginsky}, V.~B. and {Branchesi}, M. and {Brau}, J.~E. and {Briant}, T. and {Brillet}, A. and {Brinkmann}, M. and {Brisson}, V. and {Brockill}, P. and {Brooks}, A.~F. and {Brown}, D.~A. and {Brown}, D.~D. and {Brown}, N.~M. and {Buchanan}, C.~C. and {Buikema}, A. and {Bulik}, T. and {Bulten}, H.~J. and {Buonanno}, A. and {Buskulic}, D. and {Buy}, C. and {Byer}, R.~L. and {Cabero}, M. and {Cadonati}, L. and {Cagnoli}, G. and {Cahillane}, C. and {Bustillo}, J. Calder{\'o}n and {Callister}, T. and {Calloni}, E. and {Camp}, J.~B. and {Cannon}, K.~C. and {Cao}, J. and {Capano}, C.~D. and {Capocasa}, E. and {Carbognani}, F. and {Caride}, S. and {Diaz}, J. Casanueva and {Casentini}, C. and {Caudill}, S. and {Cavagli{\`a}}, M. and {Cavalier}, F. and {Cavalieri}, R. and {Cella}, G. and {Cepeda}, C.~B. and {Baiardi}, L. Cerboni and {Cerretani}, G. and {Cesarini}, E. and {Chakraborty}, R. and {Chalermsongsak}, T. and {Chamberlin}, S.~J. and {Chan}, M. and {Chao}, S. and {Charlton}, P. and {Chassande-Mottin}, E. and {Chen}, H.~Y. and {Chen}, Y. and {Cheng}, C. and {Chincarini}, A. and {Chiummo}, A. and {Cho}, H.~S. and {Cho}, M. and {Chow}, J.~H. and {Christensen}, N. and {Chu}, Q. and {Chua}, S. and {Chung}, S. and {Ciani}, G. and {Clara}, F. and {Clark}, J.~A. and {Cleva}, F. and {Coccia}, E. and {Cohadon}, P.-F. and {Colla}, A. and {Collette}, C.~G. and {Cominsky}, L. and {Constancio}, M. and {Conte}, A. and {Conti}, L. and {Cook}, D. and {Corbitt}, T.~R. and {Cornish}, N. and {Corsi}, A. and {Cortese}, S. and {Costa}, C.~A. and {Coughlin}, M.~W. and {Coughlin}, S.~B. and {Coulon}, J.-P. and {Countryman}, S.~T. and {Couvares}, P. and {Cowan}, E.~E. and {Coward}, D.~M. and {Cowart}, M.~J.},
        title = "{Observation of Gravitational Waves from a Binary Black Hole Merger}",
      journal = {\prl},
         year = 2016,
        month = feb,
       volume = {116},
       number = {6},
          eid = {061102},
        pages = {061102},
          doi = {10.1103/PhysRevLett.116.061102},
archivePrefix = {arXiv},
       eprint = {1602.03837},
 primaryClass = {gr-qc},
       adsurl = {https://ui.adsabs.harvard.edu/abs/2016PhRvL.116f1102A}
}

@ARTICLE{Ricarte_2019_SEROTINA_undetected_highz_BH,
       author = {{Ricarte}, Angelo and {Pacucci}, Fabio and {Cappelluti}, Nico and {Natarajan}, Priyamvada},
        title = "{The clustering of undetected high-redshift black holes and their signatures in cosmic backgrounds}",
      journal = {\mnras},
         year = 2019,
        month = oct,
       volume = {489},
       number = {1},
        pages = {1006-1022},
          doi = {10.1093/mnras/stz1891},
archivePrefix = {arXiv},
       eprint = {1907.03675},
 primaryClass = {astro-ph.GA},
       adsurl = {https://ui.adsabs.harvard.edu/abs/2019MNRAS.489.1006R}
}

@ARTICLE{Sesana_2005,
       author = {{Sesana}, Alberto and {Haardt}, Francesco and {Madau}, Piero and {Volonteri}, Marta},
        title = "{The Gravitational Wave Signal from Massive Black Hole Binaries and Its Contribution to the LISA Data Stream}",
      journal = {\apj},
         year = 2005,
        month = apr,
       volume = {623},
       number = {1},
        pages = {23-30},
          doi = {10.1086/428492},
archivePrefix = {arXiv},
       eprint = {astro-ph/0409255},
 primaryClass = {astro-ph},
       adsurl = {https://ui.adsabs.harvard.edu/abs/2005ApJ...623...23S}
}

@ARTICLE{Amaro-Seoane_2023_LISA_astrophysics,
       author = {{Amaro-Seoane}, Pau and {Andrews}, Jeff and {Arca Sedda}, Manuel and {Askar}, Abbas and {Baghi}, Quentin and {Balasov}, Razvan and {Bartos}, Imre and {Bavera}, Simone S. and {Bellovary}, Jillian and {Berry}, Christopher P.~L. and {Berti}, Emanuele and {Bianchi}, Stefano and {Blecha}, Laura and {Blondin}, St{\'e}phane and {Bogdanovi{\'c}}, Tamara and {Boissier}, Samuel and {Bonetti}, Matteo and {Bonoli}, Silvia and {Bortolas}, Elisa and {Breivik}, Katelyn and {Capelo}, Pedro R. and {Caramete}, Laurentiu and {Cattorini}, Federico and {Charisi}, Maria and {Chaty}, Sylvain and {Chen}, Xian and {Chru{\'s}li{\'n}ska}, Martyna and {Chua}, Alvin J.~K. and {Church}, Ross and {Colpi}, Monica and {D'Orazio}, Daniel and {Danielski}, Camilla and {Davies}, Melvyn B. and {Dayal}, Pratika and {De Rosa}, Alessandra and {Derdzinski}, Andrea and {Destounis}, Kyriakos and {Dotti}, Massimo and {Du{\c{t}}an}, Ioana and {Dvorkin}, Irina and {Fabj}, Gaia and {Foglizzo}, Thierry and {Ford}, Saavik and {Fouvry}, Jean-Baptiste and {Franchini}, Alessia and {Fragos}, Tassos and {Fryer}, Chris and {Gaspari}, Massimo and {Gerosa}, Davide and {Graziani}, Luca and {Groot}, Paul and {Habouzit}, Melanie and {Haggard}, Daryl and {Haiman}, Zoltan and {Han}, Wen-Biao and {Istrate}, Alina and {Johansson}, Peter H. and {Khan}, Fazeel Mahmood and {Kimpson}, Tomas and {Kokkotas}, Kostas and {Kong}, Albert and {Korol}, Valeriya and {Kremer}, Kyle and {Kupfer}, Thomas and {Lamberts}, Astrid and {Larson}, Shane and {Lau}, Mike and {Liu}, Dongliang and {Lloyd-Ronning}, Nicole and {Lodato}, Giuseppe and {Lupi}, Alessandro and {Ma}, Chung-Pei and {Maccarone}, Tomas and {Mandel}, Ilya and {Mangiagli}, Alberto and {Mapelli}, Michela and {Mathis}, St{\'e}phane and {Mayer}, Lucio and {McGee}, Sean and {McKernan}, Berry and {Miller}, M. Coleman and {Mota}, David F. and {Mumpower}, Matthew and {Nasim}, Syeda S. and {Nelemans}, Gijs and {Noble}, Scott and {Pacucci}, Fabio and {Panessa}, Francesca and {Paschalidis}, Vasileios and {Pfister}, Hugo and {Porquet}, Delphine and {Quenby}, John and {Ricarte}, Angelo and {R{\"o}pke}, Friedrich K. and {Regan}, John and {Rosswog}, Stephan and {Ruiter}, Ashley and {Ruiz}, Milton and {Runnoe}, Jessie and {Schneider}, Raffaella and {Schnittman}, Jeremy and {Secunda}, Amy and {Sesana}, Alberto and {Seto}, Naoki and {Shao}, Lijing and {Shapiro}, Stuart and {Sopuerta}, Carlos and {Stone}, Nicholas C. and {Suvorov}, Arthur and {Tamanini}, Nicola and {Tamfal}, Tomas and {Tauris}, Thomas and {Temmink}, Karel and {Tomsick}, John and {Toonen}, Silvia and {Torres-Orjuela}, Alejandro and {Toscani}, Martina and {Tsokaros}, Antonios and {Unal}, Caner and {V{\'a}zquez-Aceves}, Ver{\'o}nica and {Valiante}, Rosa and {van Putten}, Maurice and {van Roestel}, Jan and {Vignali}, Christian and {Volonteri}, Marta and {Wu}, Kinwah and {Younsi}, Ziri and {Yu}, Shenghua and {Zane}, Silvia and {Zwick}, Lorenz and {Antonini}, Fabio and {Baibhav}, Vishal and {Barausse}, Enrico and {Bonilla Rivera}, Alexander and {Branchesi}, Marica and {Branduardi-Raymont}, Graziella and {Burdge}, Kevin and {Chakraborty}, Srija and {Cuadra}, Jorge and {Dage}, Kristen and {Davis}, Benjamin and {de Mink}, Selma E. and {Decarli}, Roberto and {Doneva}, Daniela and {Escoffier}, Stephanie and {Gandhi}, Poshak and {Haardt}, Francesco and {Lousto}, Carlos O. and {Nissanke}, Samaya and {Nordhaus}, Jason and {O'Shaughnessy}, Richard and {Portegies Zwart}, Simon and {Pound}, Adam and {Schussler}, Fabian and {Sergijenko}, Olga and {Spallicci}, Alessandro and {Vernieri}, Daniele and {Vigna-G{\'o}mez}, Alejandro},
        title = "{Astrophysics with the Laser Interferometer Space Antenna}",
      journal = {Living Reviews in Relativity},
         year = 2023,
        month = dec,
       volume = {26},
       number = {1},
          eid = {2},
        pages = {2},
          doi = {10.1007/s41114-022-00041-y},
archivePrefix = {arXiv},
       eprint = {2203.06016},
 primaryClass = {gr-qc},
       adsurl = {https://ui.adsabs.harvard.edu/abs/2023LRR....26....2A}
}

@ARTICLE{LIGO_O3_population,
       author = {{Abbott}, R. and {Abbott}, T.~D. and {Acernese}, F. and {Ackley}, K. and {Adams}, C. and {Adhikari}, N. and {Adhikari}, R.~X. and {Adya}, V.~B. and {Affeldt}, C. and {Agarwal}, D. and {Agathos}, M. and {Agatsuma}, K. and {Aggarwal}, N. and {Aguiar}, O.~D. and {Aiello}, L. and {Ain}, A. and {Ajith}, P. and {Akutsu}, T. and {de Alarc{\'o}n}, P.~F. and {Akcay}, S. and {Albanesi}, S. and {Allocca}, A. and {Altin}, P.~A. and {Amato}, A. and {Anand}, C. and {Anand}, S. and {Ananyeva}, A. and {Anderson}, S.~B. and {Anderson}, W.~G. and {Ando}, M. and {Andrade}, T. and {Andres}, N. and {Andri{\'c}}, T. and {Angelova}, S.~V. and {Ansoldi}, S. and {Antelis}, J.~M. and {Antier}, S. and {Antonini}, F. and {Appert}, S. and {Arai}, Koji and {Arai}, Koya and {Arai}, Y. and {Araki}, S. and {Araya}, A. and {Araya}, M.~C. and {Areeda}, J.~S. and {Ar{\`e}ne}, M. and {Aritomi}, N. and {Arnaud}, N. and {Arogeti}, M. and {Aronson}, S.~M. and {Arun}, K.~G. and {Asada}, H. and {Asali}, Y. and {Ashton}, G. and {Aso}, Y. and {Assiduo}, M. and {Aston}, S.~M. and {Astone}, P. and {Aubin}, F. and {Austin}, C. and {Babak}, S. and {Badaracco}, F. and {Bader}, M.~K.~M. and {Badger}, C. and {Bae}, S. and {Bae}, Y. and {Baer}, A.~M. and {Bagnasco}, S. and {Bai}, Y. and {Baiotti}, L. and {Baird}, J. and {Bajpai}, R. and {Ball}, M. and {Ballardin}, G. and {Ballmer}, S.~W. and {Balsamo}, A. and {Baltus}, G. and {Banagiri}, S. and {Bankar}, D. and {Barayoga}, J.~C. and {Barbieri}, C. and {Barish}, B.~C. and {Barker}, D. and {Barneo}, P. and {Barone}, F. and {Barr}, B. and {Barsotti}, L. and {Barsuglia}, M. and {Barta}, D. and {Bartlett}, J. and {Barton}, M.~A. and {Bartos}, I. and {Bassiri}, R. and {Basti}, A. and {Bawaj}, M. and {Bayley}, J.~C. and {Baylor}, A.~C. and {Bazzan}, M. and {B{\'e}csy}, B. and {Bedakihale}, V.~M. and {Bejger}, M. and {Belahcene}, I. and {Benedetto}, V. and {Beniwal}, D. and {Bennett}, T.~F. and {Bentley}, J.~D. and {Benyaala}, M. and {Bergamin}, F. and {Berger}, B.~K. and {Bernuzzi}, S. and {Berry}, C.~P.~L. and {Bersanetti}, D. and {Bertolini}, A. and {Betzwieser}, J. and {Beveridge}, D. and {Bhandare}, R. and {Bhardwaj}, U. and {Bhattacharjee}, D. and {Bhaumik}, S. and {Bilenko}, I.~A. and {Billingsley}, G. and {Bini}, S. and {Birney}, R. and {Birnholtz}, O. and {Biscans}, S. and {Bischi}, M. and {Biscoveanu}, S. and {Bisht}, A. and {Biswas}, B. and {Bitossi}, M. and {Bizouard}, M.-A. and {Blackburn}, J.~K. and {Blair}, C.~D. and {Blair}, D.~G. and {Blair}, R.~M. and {Bobba}, F. and {Bode}, N. and {Boer}, M. and {Bogaert}, G. and {Boldrini}, M. and {Bonavena}, L.~D. and {Bondu}, F. and {Bonilla}, E. and {Bonnand}, R. and {Booker}, P. and {Boom}, B.~A. and {Bork}, R. and {Boschi}, V. and {Bose}, N. and {Bose}, S. and {Bossilkov}, V. and {Boudart}, V. and {Bouffanais}, Y. and {Bozzi}, A. and {Bradaschia}, C. and {Brady}, P.~R. and {Bramley}, A. and {Branch}, A. and {Branchesi}, M. and {Brandt}, J. and {Brau}, J.~E. and {Breschi}, M. and {Briant}, T. and {Briggs}, J.~H. and {Brillet}, A. and {Brinkmann}, M. and {Brockill}, P. and {Brooks}, A.~F. and {Brooks}, J. and {Brown}, D.~D. and {Brunett}, S. and {Bruno}, G. and {Bruntz}, R. and {Bryant}, J. and {Bulik}, T. and {Bulten}, H.~J. and {Buonanno}, A. and {Buscicchio}, R. and {Buskulic}, D. and {Buy}, C. and {Byer}, R.~L. and {Cadonati}, L. and {Cagnoli}, G. and {Cahillane}, C. and {Bustillo}, J. Calder{\'o}n and {Callaghan}, J.~D. and {Callister}, T.~A. and {Calloni}, E. and {Cameron}, J. and {Camp}, J.~B. and {Canepa}, M. and {Canevarolo}, S. and {Cannavacciuolo}, M. and {Cannon}, K.~C. and {Cao}, H. and {Cao}, Z. and {Capocasa}, E. and {Capote}, E. and {Carapella}, G.},
        title = "{Population of Merging Compact Binaries Inferred Using Gravitational Waves through GWTC-3}",
      journal = {Physical Review X},
         year = 2023,
        month = jan,
       volume = {13},
       number = {1},
          eid = {011048},
        pages = {011048},
          doi = {10.1103/PhysRevX.13.011048},
archivePrefix = {arXiv},
       eprint = {2111.03634},
 primaryClass = {astro-ph.HE},
       adsurl = {https://ui.adsabs.harvard.edu/abs/2023PhRvX..13a1048A}
}

@ARTICLE{GW150914_GR,
       author = {{Abbott}, B.~P. and {Abbott}, R. and {Abbott}, T.~D. and {Abernathy}, M.~R. and {Acernese}, F. and {Ackley}, K. and {Adams}, C. and {Adams}, T. and {Addesso}, P. and {Adhikari}, R.~X. and {Adya}, V.~B. and {Affeldt}, C. and {Agathos}, M. and {Agatsuma}, K. and {Aggarwal}, N. and {Aguiar}, O.~D. and {Aiello}, L. and {Ain}, A. and {Ajith}, P. and {Allen}, B. and {Allocca}, A. and {Altin}, P.~A. and {Anderson}, S.~B. and {Anderson}, W.~G. and {Arai}, K. and {Araya}, M.~C. and {Arceneaux}, C.~C. and {Areeda}, J.~S. and {Arnaud}, N. and {Arun}, K.~G. and {Ascenzi}, S. and {Ashton}, G. and {Ast}, M. and {Aston}, S.~M. and {Astone}, P. and {Aufmuth}, P. and {Aulbert}, C. and {Babak}, S. and {Bacon}, P. and {Bader}, M.~K.~M. and {Baker}, P.~T. and {Baldaccini}, F. and {Ballardin}, G. and {Ballmer}, S.~W. and {Barayoga}, J.~C. and {Barclay}, S.~E. and {Barish}, B.~C. and {Barker}, D. and {Barone}, F. and {Barr}, B. and {Barsotti}, L. and {Barsuglia}, M. and {Barta}, D. and {Bartlett}, J. and {Bartos}, I. and {Bassiri}, R. and {Basti}, A. and {Batch}, J.~C. and {Baune}, C. and {Bavigadda}, V. and {Bazzan}, M. and {Behnke}, B. and {Bejger}, M. and {Bell}, A.~S. and {Bell}, C.~J. and {Berger}, B.~K. and {Bergman}, J. and {Bergmann}, G. and {Berry}, C.~P.~L. and {Bersanetti}, D. and {Bertolini}, A. and {Betzwieser}, J. and {Bhagwat}, S. and {Bhandare}, R. and {Bilenko}, I.~A. and {Billingsley}, G. and {Birch}, J. and {Birney}, I.~A. and {Birnholtz}, O. and {Biscans}, S. and {Bisht}, A. and {Bitossi}, M. and {Biwer}, C. and {Bizouard}, M.~A. and {Blackburn}, J.~K. and {Blair}, C.~D. and {Blair}, D.~G. and {Blair}, R.~M. and {Bloemen}, S. and {Bock}, O. and {Bodiya}, T.~P. and {Boer}, M. and {Bogaert}, G. and {Bogan}, C. and {Bohe}, A. and {Bojtos}, P. and {Bond}, C. and {Bondu}, F. and {Bonnand}, R. and {Boom}, B.~A. and {Bork}, R. and {Boschi}, V. and {Bose}, S. and {Bouffanais}, Y. and {Bozzi}, A. and {Bradaschia}, C. and {Brady}, P.~R. and {Braginsky}, V.~B. and {Branchesi}, M. and {Brau}, J.~E. and {Briant}, T. and {Brillet}, A. and {Brinkmann}, M. and {Brisson}, V. and {Brockill}, P. and {Brooks}, A.~F. and {Brown}, D.~A. and {Brown}, D.~D. and {Brown}, N.~M. and {Buchanan}, C.~C. and {Buikema}, A. and {Bulik}, T. and {Bulten}, H.~J. and {Buonanno}, A. and {Buskulic}, D. and {Buy}, C. and {Byer}, R.~L. and {Cadonati}, L. and {Cagnoli}, G. and {Cahillane}, C. and {Calder{\'o}n Bustillo}, J. and {Callister}, T. and {Calloni}, E. and {Camp}, J.~B. and {Cannon}, K.~C. and {Cao}, J. and {Capano}, C.~D. and {Capocasa}, E. and {Carbognani}, F. and {Caride}, S. and {Casanueva Diaz}, J. and {Casentini}, C. and {Caudill}, S. and {Cavagli{\`a}}, M. and {Cavalier}, F. and {Cavalieri}, R. and {Cella}, G. and {Cepeda}, C.~B. and {Cerboni Baiardi}, L. and {Cerretani}, G. and {Cesarini}, E. and {Chakraborty}, R. and {Chalermsongsak}, T. and {Chamberlin}, S.~J. and {Chan}, M. and {Chao}, S. and {Charlton}, P. and {Chassande-Mottin}, E. and {Chen}, H.~Y. and {Chen}, Y. and {Cheng}, C. and {Chincarini}, A. and {Chiummo}, A. and {Cho}, H.~S. and {Cho}, M. and {Chow}, J.~H. and {Christensen}, N. and {Chu}, Q. and {Chua}, S. and {Chung}, S. and {Ciani}, G. and {Clara}, F. and {Clark}, J.~A. and {Cleva}, F. and {Coccia}, E. and {Cohadon}, P.-F. and {Colla}, A. and {Collette}, C.~G. and {Cominsky}, L. and {Constancio}, M. and {Conte}, A. and {Conti}, L. and {Cook}, D. and {Corbitt}, T.~R. and {Cornish}, N. and {Corsi}, A. and {Cortese}, S. and {Costa}, C.~A. and {Coughlin}, M.~W. and {Coughlin}, S.~B. and {Coulon}, J.-P. and {Countryman}, S.~T. and {Couvares}, P. and {Cowan}, E.~E. and {Coward}, D.~M. and {Cowart}, M.~J. and {Coyne}, D.~C. and {Coyne}, R. and {Craig}, K. and {Creighton}, J.~D.~E.},
        title = "{Tests of General Relativity with GW150914}",
      journal = {\prl},
         year = 2016,
        month = may,
       volume = {116},
       number = {22},
          eid = {221101},
        pages = {221101},
          doi = {10.1103/PhysRevLett.116.221101},
archivePrefix = {arXiv},
       eprint = {1602.03841},
 primaryClass = {gr-qc},
       adsurl = {https://ui.adsabs.harvard.edu/abs/2016PhRvL.116v1101A}
}

@ARTICLE{LIGO_GWTC3_GR,
       author = {{Abbott}, R. and {Abe}, H. and {Acernese}, F. and {Ackley}, K. and {Adhikari}, N. and {Adhikari}, R.~X. and {Adkins}, V.~K. and {Adya}, V.~B. and {Affeldt}, C. and {Agarwal}, D. and {Agathos}, M. and {Agatsuma}, K. and {Aggarwal}, N. and {Aguiar}, O.~D. and {Aiello}, L. and {Ain}, A. and {Ajith}, P. and {Akutsu}, T. and {de Alarc{\'o}n}, P.~F. and {Albanesi}, S. and {Alfaidi}, R.~A. and {Allocca}, A. and {Altin}, P.~A. and {Amato}, A. and {Anand}, C. and {Anand}, S. and {Ananyeva}, A. and {Anderson}, S.~B. and {Anderson}, W.~G. and {Ando}, M. and {Andrade}, T. and {Andres}, N. and {Andr{\'e}s-Carcasona}, M. and {Andri{\'c}}, T. and {Angelova}, S.~V. and {Ansoldi}, S. and {Antelis}, J.~M. and {Antier}, S. and {Apostolatos}, T. and {Appavuravther}, E.~Z. and {Appert}, S. and {Apple}, S.~K. and {Arai}, K. and {Araya}, A. and {Araya}, M.~C. and {Areeda}, J.~S. and {Ar{\`e}ne}, M. and {Aritomi}, N. and {Arnaud}, N. and {Arogeti}, M. and {Aronson}, S.~M. and {Arun}, K.~G. and {Asada}, H. and {Asali}, Y. and {Ashton}, G. and {Aso}, Y. and {Assiduo}, M. and {Melo}, S. Assis De Souza and {Aston}, S.~M. and {Astone}, P. and {Aubin}, F. and {Aultoneal}, K. and {Austin}, C. and {Babak}, S. and {Badaracco}, F. and {Bader}, M.~K.~M. and {Badger}, C. and {Bae}, S. and {Bae}, Y. and {Baer}, A.~M. and {Bagnasco}, S. and {Bai}, Y. and {Baird}, J. and {Bajpai}, R. and {Baka}, T. and {Ball}, M. and {Ballardin}, G. and {Ballmer}, S.~W. and {Balsamo}, A. and {Baltus}, G. and {Banagiri}, S. and {Banerjee}, B. and {Bankar}, D. and {Barayoga}, J.~C. and {Barbieri}, C. and {Barish}, B.~C. and {Barker}, D. and {Barneo}, P. and {Barone}, F. and {Barr}, B. and {Barsotti}, L. and {Barsuglia}, M. and {Barta}, D. and {Bartlett}, J. and {Barton}, M.~A. and {Bartos}, I. and {Basak}, S. and {Bassiri}, R. and {Basti}, A. and {Bawaj}, M. and {Bayley}, J.~C. and {Bazzan}, M. and {Becher}, B.~R. and {B{\'e}csy}, B. and {Bedakihale}, V.~M. and {Beirnaert}, F. and {Bejger}, M. and {Belahcene}, I. and {Benedetto}, V. and {Beniwal}, D. and {Benjamin}, M.~G. and {Bennett}, T.~F. and {Bentley}, J.~D. and {Benyaala}, M. and {Bera}, S. and {Berbel}, M. and {Bergamin}, F. and {Berger}, B.~K. and {Bernuzzi}, S. and {Berry}, C.~P.~L. and {Bersanetti}, D. and {Bertolini}, A. and {Betzwieser}, J. and {Beveridge}, D. and {Bhandare}, R. and {Bhandari}, A.~V. and {Bhardwaj}, U. and {Bhatt}, R. and {Bhattacharjee}, D. and {Bhaumik}, S. and {Bianchi}, A. and {Bilenko}, I.~A. and {Billingsley}, G. and {Bini}, S. and {Birney}, R. and {Birnholtz}, O. and {Biscans}, S. and {Bischi}, M. and {Biscoveanu}, S. and {Bisht}, A. and {Biswas}, B. and {Bitossi}, M. and {Bizouard}, M.-A. and {Blackburn}, J.~K. and {Blair}, C.~D. and {Blair}, D.~G. and {Blair}, R.~M. and {Bobba}, F. and {Bode}, N. and {Bo{\"e}r}, M. and {Bogaert}, G. and {Boldrini}, M. and {Bolingbroke}, G.~N. and {Bonavena}, L.~D. and {Bondu}, F. and {Bonilla}, E. and {Bonnand}, R. and {Booker}, P. and {Boom}, B.~A. and {Bork}, R. and {Boschi}, V. and {Bose}, N. and {Bose}, S. and {Bossilkov}, V. and {Boudart}, V. and {Bouffanais}, Y. and {Bozzi}, A. and {Bradaschia}, C. and {Brady}, P.~R. and {Bramley}, A. and {Branch}, A. and {Branchesi}, M. and {Brau}, J.~E. and {Breschi}, M. and {Briant}, T. and {Briggs}, J.~H. and {Brillet}, A. and {Brinkmann}, M. and {Brockill}, P. and {Brooks}, A.~F. and {Brooks}, J. and {Brown}, D.~D. and {Brunett}, S. and {Bruno}, G. and {Bruntz}, R. and {Bryant}, J. and {Bucci}, F. and {Bulik}, T. and {Bulten}, H.~J. and {Buonanno}, A. and {Burtnyk}, K. and {Buscicchio}, R. and {Buskulic}, D. and {Buy}, C. and {Byer}, R.~L. and {Davies}, G.~S. Cabourn and {Cabras}, G. and {Cabrita}, R. and {Cadonati}, L. and {Caesar}, M.},
        title = "{Tests of general relativity with GWTC-3}",
      journal = {\prd},
         year = 2025,
        month = oct,
       volume = {112},
       number = {8},
          eid = {084080},
        pages = {084080},
          doi = {10.1103/PhysRevD.112.084080},
archivePrefix = {arXiv},
       eprint = {2112.06861},
 primaryClass = {gr-qc},
       adsurl = {https://ui.adsabs.harvard.edu/abs/2025PhRvD.112h4080A}
}

@ARTICLE{Broekgaarden_2022,
       author = {{Broekgaarden}, Floor S. and {Berger}, Edo and {Stevenson}, Simon and {Justham}, Stephen and {Mandel}, Ilya and {Chru{\'s}li{\'n}ska}, Martyna and {van Son}, Lieke A.~C. and {Wagg}, Tom and {Vigna-G{\'o}mez}, Alejandro and {de Mink}, Selma E. and {Chattopadhyay}, Debatri and {Neijssel}, Coenraad J.},
        title = "{Impact of massive binary star and cosmic evolution on gravitational wave observations - II. Double compact object rates and properties}",
      journal = {\mnras},
         year = 2022,
        month = nov,
       volume = {516},
       number = {4},
        pages = {5737-5761},
          doi = {10.1093/mnras/stac1677},
archivePrefix = {arXiv},
       eprint = {2112.05763},
 primaryClass = {astro-ph.HE},
       adsurl = {https://ui.adsabs.harvard.edu/abs/2022MNRAS.516.5737B}
}

@ARTICLE{Fishbach_2018_BH_merger_rate_redshift,
       author = {{Fishbach}, Maya and {Holz}, Daniel E. and {Farr}, Will M.},
        title = "{Does the Black Hole Merger Rate Evolve with Redshift?}",
      journal = {\apjl},
         year = 2018,
        month = aug,
       volume = {863},
       number = {2},
          eid = {L41},
        pages = {L41},
          doi = {10.3847/2041-8213/aad800},
archivePrefix = {arXiv},
       eprint = {1805.10270},
 primaryClass = {astro-ph.HE},
       adsurl = {https://ui.adsabs.harvard.edu/abs/2018ApJ...863L..41F}
}

@ARTICLE{Fishbach_2020_GW190521_straddling_binary,
       author = {{Fishbach}, Maya and {Holz}, Daniel E.},
        title = "{Minding the Gap: GW190521 as a Straddling Binary}",
      journal = {\apjl},
         year = 2020,
        month = dec,
       volume = {904},
       number = {2},
          eid = {L26},
        pages = {L26},
          doi = {10.3847/2041-8213/abc827},
archivePrefix = {arXiv},
       eprint = {2009.05472},
 primaryClass = {astro-ph.HE},
       adsurl = {https://ui.adsabs.harvard.edu/abs/2020ApJ...904L..26F}
}

@ARTICLE{van_Son_2022_BH_merger_rate_evolution,
       author = {{van Son}, L.~A.~C. and {de Mink}, S.~E. and {Callister}, T. and {Justham}, S. and {Renzo}, M. and {Wagg}, T. and {Broekgaarden}, F.~S. and {Kummer}, F. and {Pakmor}, R. and {Mandel}, I.},
        title = "{The Redshift Evolution of the Binary Black Hole Merger Rate: A Weighty Matter}",
      journal = {\apj},
         year = 2022,
        month = may,
       volume = {931},
       number = {1},
          eid = {17},
        pages = {17},
          doi = {10.3847/1538-4357/ac64a3},
archivePrefix = {arXiv},
       eprint = {2110.01634},
 primaryClass = {astro-ph.HE},
       adsurl = {https://ui.adsabs.harvard.edu/abs/2022ApJ...931...17V}
}

@ARTICLE{van_Son_2022_mass_transfer_NSBH_gap,
       author = {{van Son}, L.~A.~C. and {de Mink}, S.~E. and {Renzo}, M. and {Justham}, S. and {Zapartas}, E. and {Breivik}, K. and {Callister}, T. and {Farr}, W.~M. and {Conroy}, C.},
        title = "{No Peaks without Valleys: The Stable Mass Transfer Channel for Gravitational-wave Sources in Light of the Neutron Star-Black Hole Mass Gap}",
      journal = {\apj},
         year = 2022,
        month = dec,
       volume = {940},
       number = {2},
          eid = {184},
        pages = {184},
          doi = {10.3847/1538-4357/ac9b0a},
archivePrefix = {arXiv},
       eprint = {2209.13609},
 primaryClass = {astro-ph.HE},
       adsurl = {https://ui.adsabs.harvard.edu/abs/2022ApJ...940..184V}
}

@ARTICLE{Sathyaprakash_Schutz_2009_GW_review,
       author = {{Sathyaprakash}, B.~S. and {Schutz}, Bernard F.},
        title = "{Physics, Astrophysics and Cosmology with Gravitational Waves}",
      journal = {Living Reviews in Relativity},
         year = 2009,
        month = dec,
       volume = {12},
       number = {1},
          eid = {2},
        pages = {2},
          doi = {10.12942/lrr-2009-2},
archivePrefix = {arXiv},
       eprint = {0903.0338},
 primaryClass = {gr-qc},
       adsurl = {https://ui.adsabs.harvard.edu/abs/2009LRR....12....2S}
}

@ARTICLE{Lyke_2020_SDSS_quasar_16th_DR,
       author = {{Lyke}, Brad W. and {Higley}, Alexandra N. and {McLane}, J.~N. and {Schurhammer}, Danielle P. and {Myers}, Adam D. and {Ross}, Ashley J. and {Dawson}, Kyle and {Chabanier}, Sol{\`e}ne and {Martini}, Paul and {Busca}, Nicol{\'a}s G. and {Mas des Bourboux}, H{\'e}lion du and {Salvato}, Mara and {Streblyanska}, Alina and {Zarrouk}, Pauline and {Burtin}, Etienne and {Anderson}, Scott F. and {Bautista}, Julian and {Bizyaev}, Dmitry and {Brandt}, W.~N. and {Brinkmann}, Jonathan and {Brownstein}, Joel R. and {Comparat}, Johan and {Green}, Paul and {de la Macorra}, Axel and {Mu{\~n}oz Guti{\'e}rrez}, Andrea and {Hou}, Jiamin and {Newman}, Jeffrey A. and {Palanque-Delabrouille}, Nathalie and {P{\^a}ris}, Isabelle and {Percival}, Will J. and {Petitjean}, Patrick and {Rich}, James and {Rossi}, Graziano and {Schneider}, Donald P. and {Smith}, Alexander and {Vivek}, M. and {Weaver}, Benjamin Alan},
        title = "{The Sloan Digital Sky Survey Quasar Catalog: Sixteenth Data Release}",
      journal = {\apjs},
         year = 2020,
        month = sep,
       volume = {250},
       number = {1},
          eid = {8},
        pages = {8},
          doi = {10.3847/1538-4365/aba623},
archivePrefix = {arXiv},
       eprint = {2007.09001},
 primaryClass = {astro-ph.GA},
       adsurl = {https://ui.adsabs.harvard.edu/abs/2020ApJS..250....8L}
}

@article{Peters_1964_GW_binary,
  title = {Gravitational Radiation and the Motion of Two Point Masses},
  author = {Peters, P. C.},
  journal = {Phys. Rev.},
  volume = {136},
  issue = {4B},
  pages = {B1224--B1232},
  numpages = {0},
  year = {1964},
  month = {Nov},
  publisher = {American Physical Society},
  doi = {10.1103/PhysRev.136.B1224},
  url = {https://link.aps.org/doi/10.1103/PhysRev.136.B1224}
}

@article{Peters_Mathews_1963_GW,
  title = {Gravitational Radiation from Point Masses in a Keplerian Orbit},
  author = {Peters, P. C. and Mathews, J.},
  journal = {Phys. Rev.},
  volume = {131},
  issue = {1},
  pages = {435--440},
  numpages = {0},
  year = {1963},
  month = {Jul},
  publisher = {American Physical Society},
  doi = {10.1103/PhysRev.131.435},
  url = {https://link.aps.org/doi/10.1103/PhysRev.131.435}
}

@ARTICLE{Huerta_2015_eccentric_SMBHB_SNR,
       author = {{Huerta}, E.~A. and {McWilliams}, Sean T. and {Gair}, Jonathan R. and {Taylor}, Stephen R.},
        title = "{Detection of eccentric supermassive black hole binaries with pulsar timing arrays: Signal-to-noise ratio calculations}",
      journal = {\prd},
         year = 2015,
        month = sep,
       volume = {92},
       number = {6},
          eid = {063010},
        pages = {063010},
          doi = {10.1103/PhysRevD.92.063010},
archivePrefix = {arXiv},
       eprint = {1504.00928},
 primaryClass = {gr-qc},
       adsurl = {https://ui.adsabs.harvard.edu/abs/2015PhRvD..92f3010H}
}

@ARTICLE{Enoki_Nagashima_GW_eccentric_SMBHB,
       author = {{Enoki}, M. and {Nagashima}, M.},
        title = "{The Effect of Orbital Eccentricity on Gravitational Wave Background Radiation from Supermassive Black Hole Binaries}",
      journal = {Progress of Theoretical Physics},
         year = 2007,
        month = feb,
       volume = {117},
       number = {2},
        pages = {241-256},
          doi = {10.1143/PTP.117.241},
archivePrefix = {arXiv},
       eprint = {astro-ph/0609377},
 primaryClass = {astro-ph},
       adsurl = {https://ui.adsabs.harvard.edu/abs/2007PThPh.117..241E}
}

@ARTICLE{Armitage_Natarajan_2005_SMBHB_eccentricity,
       author = {{Armitage}, Philip J. and {Natarajan}, Priyamvada},
        title = "{Eccentricity of Supermassive Black Hole Binaries Coalescing from Gas-rich Mergers}",
      journal = {\apj},
         year = 2005,
        month = dec,
       volume = {634},
       number = {2},
        pages = {921-927},
          doi = {10.1086/497108},
archivePrefix = {arXiv},
       eprint = {astro-ph/0508493},
 primaryClass = {astro-ph},
       adsurl = {https://ui.adsabs.harvard.edu/abs/2005ApJ...634..921A}
}

@ARTICLE{Iwasawa_2011_SMBHB_eccentric_evolution,
       author = {{Iwasawa}, Masaki and {An}, Sangyong and {Matsubayashi}, Tatsushi and {Funato}, Yoko and {Makino}, Junichiro},
        title = "{Eccentric Evolution of Supermassive Black Hole Binaries}",
      journal = {\apjl},
         year = 2011,
        month = apr,
       volume = {731},
       number = {1},
          eid = {L9},
        pages = {L9},
          doi = {10.1088/2041-8205/731/1/L9},
archivePrefix = {arXiv},
       eprint = {1011.4017},
 primaryClass = {astro-ph.GA},
       adsurl = {https://ui.adsabs.harvard.edu/abs/2011ApJ...731L...9I}
}

@ARTICLE{Sesana_2010_eccentric_SMBHB_stellar_envi,
       author = {{Sesana}, Alberto},
        title = "{Self Consistent Model for the Evolution of Eccentric Massive Black Hole Binaries in Stellar Environments: Implications for Gravitational Wave Observations}",
      journal = {\apj},
         year = 2010,
        month = aug,
       volume = {719},
       number = {1},
        pages = {851-864},
          doi = {10.1088/0004-637X/719/1/851},
archivePrefix = {arXiv},
       eprint = {1006.0730},
 primaryClass = {astro-ph.CO},
       adsurl = {https://ui.adsabs.harvard.edu/abs/2010ApJ...719..851S}
}

@ARTICLE{Marsh_2011_WDB_LISA,
       author = {{Marsh}, T.~R.},
        title = "{Double white dwarfs and LISA}",
      journal = {Classical and Quantum Gravity},
         year = 2011,
        month = may,
       volume = {28},
       number = {9},
          eid = {094019},
        pages = {094019},
          doi = {10.1088/0264-9381/28/9/094019},
archivePrefix = {arXiv},
       eprint = {1101.4970},
 primaryClass = {astro-ph.SR},
       adsurl = {https://ui.adsabs.harvard.edu/abs/2011CQGra..28i4019M}
}

@ARTICLE{Stroeer_Vecchio_2006_LISA_verification_binary,
       author = {{Stroeer}, A. and {Vecchio}, A.},
        title = "{The LISA verification binaries}",
      journal = {Classical and Quantum Gravity},
         year = 2006,
        month = oct,
       volume = {23},
       number = {19},
        pages = {S809-S817},
          doi = {10.1088/0264-9381/23/19/S19},
archivePrefix = {arXiv},
       eprint = {astro-ph/0605227},
 primaryClass = {astro-ph},
       adsurl = {https://ui.adsabs.harvard.edu/abs/2006CQGra..23S.809S}
}

@ARTICLE{Korol_2017_WDB_detectability_Gaia_LSST_LISA,
       author = {{Korol}, Valeriya and {Rossi}, Elena M. and {Groot}, Paul J. and {Nelemans}, Gijs and {Toonen}, Silvia and {Brown}, Anthony G.~A.},
        title = "{Prospects for detection of detached double white dwarf binaries with Gaia, LSST and LISA}",
      journal = {\mnras},
         year = 2017,
        month = sep,
       volume = {470},
       number = {2},
        pages = {1894-1910},
          doi = {10.1093/mnras/stx1285},
archivePrefix = {arXiv},
       eprint = {1703.02555},
 primaryClass = {astro-ph.HE},
       adsurl = {https://ui.adsabs.harvard.edu/abs/2017MNRAS.470.1894K}
}

@ARTICLE{Korol_2018_LISA_detectability_WD,
       author = {{Korol}, Valeria and {Koop}, Orlin and {Rossi}, Elena M.},
        title = "{Detectability of Double White Dwarfs in the Local Group with LISA}",
      journal = {\apjl},
         year = 2018,
        month = oct,
       volume = {866},
       number = {2},
          eid = {L20},
        pages = {L20},
          doi = {10.3847/2041-8213/aae587},
archivePrefix = {arXiv},
       eprint = {1808.05959},
 primaryClass = {astro-ph.HE},
       adsurl = {https://ui.adsabs.harvard.edu/abs/2018ApJ...866L..20K}
}

@ARTICLE{Karnesis_2021_compact_binary_background,
       author = {{Karnesis}, Nikolaos and {Babak}, Stanislav and {Pieroni}, Mauro and {Cornish}, Neil and {Littenberg}, Tyson},
        title = "{Characterization of the stochastic signal originating from compact binary populations as measured by LISA}",
      journal = {\prd},
         year = 2021,
        month = aug,
       volume = {104},
       number = {4},
          eid = {043019},
        pages = {043019},
          doi = {10.1103/PhysRevD.104.043019},
archivePrefix = {arXiv},
       eprint = {2103.14598},
 primaryClass = {astro-ph.IM},
       adsurl = {https://ui.adsabs.harvard.edu/abs/2021PhRvD.104d3019K}
}

@ARTICLE{Nelemans_2004_AMCVn_LISA,
       author = {{Nelemans}, G. and {Yungelson}, L.~R. and {Portegies Zwart}, S.~F.},
        title = "{Short-period AM CVn systems as optical, X-ray and gravitational-wave sources}",
      journal = {\mnras},
         year = 2004,
        month = mar,
       volume = {349},
       number = {1},
        pages = {181-192},
          doi = {10.1111/j.1365-2966.2004.07479.x},
archivePrefix = {arXiv},
       eprint = {astro-ph/0312193},
 primaryClass = {astro-ph},
       adsurl = {https://ui.adsabs.harvard.edu/abs/2004MNRAS.349..181N}
}

@ARTICLE{Korol_2020_WDB_galactic_population,
       author = {{Korol}, V. and {Toonen}, S. and {Klein}, A. and {Belokurov}, V. and {Vincenzo}, F. and {Buscicchio}, R. and {Gerosa}, D. and {Moore}, C.~J. and {Roebber}, E. and {Rossi}, E.~M. and {Vecchio}, A.},
        title = "{Populations of double white dwarfs in Milky Way satellites and their detectability with LISA}",
      journal = {\aap},
         year = 2020,
        month = jun,
       volume = {638},
          eid = {A153},
        pages = {A153},
          doi = {10.1051/0004-6361/202037764},
archivePrefix = {arXiv},
       eprint = {2002.10462},
 primaryClass = {astro-ph.GA},
       adsurl = {https://ui.adsabs.harvard.edu/abs/2020A&A...638A.153K}
}

\end{document}